\documentclass[a4paper,fleqn,usenatbib]{mnras}

\usepackage[T1]{fontenc}
\usepackage{ae,aecompl}

\usepackage{graphicx}	
\usepackage{amsmath}	
\usepackage{amssymb}	
\usepackage{epstopdf}
\usepackage{morefloats}
\usepackage{hyperref}
\usepackage{booktabs}
\usepackage{color}
\usepackage{ulem}

\title[Bars in LSB discs]{Characterising bars in low surface brightness disc galaxies}

\author[W.\ Peters \& R.\ Kuzio de Naray]{
Wesley Peters$^{1}$\thanks{E-mail: peters@astro.gsu.edu (WP); kuzio@astro.gsu.edu (RKD)} and
Rachel Kuzio de Naray$^{1}$\footnotemark[1]
\\
$^{1}$Department of Physics \& Astronomy, Georgia State University, PO Box 5060, Atlanta, GA 30302-5060, USA\\
}

\date{Accepted XXX. Received YYY; in original form ZZZ}

\pubyear{2017}

\begin{document}
\label{firstpage}
\pagerange{\pageref{firstpage}--\pageref{lastpage}}
\maketitle

\begin{abstract}
In this paper, we use \textit{B}-band, \textit{I}-band, and 3.6 $\mu$m azimuthal light profiles of four low surface brightness (LSB) galaxies (UGC~628, F568\nobreakdash-1, F568\nobreakdash-3, F563\nobreakdash-V2) to characterise three bar parameters: length, strength, and corotation radius. We employ three techniques to measure the radius of the bars, including a new method using the azimuthal light profiles. We find comparable bar radii between the \textit{I}-band and 3.6 $\mu$m for all four galaxies when using our azimuthal light profile method, and that our bar lengths are \textcolor{black}{comparable to} those in high surface brightness galaxies (HSBs). In addition, we find the bar strengths for our galaxies to be smaller than those for HSBs. Finally, we use Fourier transforms of the \textit{B}-band, \textit{I}-band, and 3.6 $\mu$m images to characterise the bars as either `fast' or `slow' by measuring the corotation radius via phase profiles. When using the \textit{B} and \textit{I}-band phase crossings, we find three of our galaxies have faster than expected relative bar pattern speeds \textcolor{black}{for galaxies expected to be embedded in centrally-dense cold dark matter haloes.} When using the \textit{B}-band and 3.6 $\mu$m phase crossings, we find more ambiguous results, although the relative bar pattern speeds are still faster than expected. Since we find a very slow bar in F563\nobreakdash-V2, we are confident that we are able to differentiate between fast and slow bars. \textcolor{black}{Finally, we} find no relation between bar strength and relative bar pattern speed when comparing our LSBs to HSBs. 
\end{abstract}


\section{Introduction}
\label{sec:introduction}

Low surface brightness galaxies (LSBs) are incredibly faint galaxies, typically defined as having central surface brightnesses $\mu_{0}(B) \geq 22.0$\ mag arcsec$^{-2}$. Due to this, LSBs are very hard to detect, and have been biased against in large-scale surveys, even though they may make up more than half of all galaxies \citep{mcgaugh1995b, bothun1997}. Although they are faint, LSBs are not simply just small and featureless galaxies, but come in a whole suite of morphologies comparable to `normal' high surface brightness galaxies (HSBs) \citep{mcgaugh1995a}. As LSBs have bluer colours, lower star formation, and lower metallicities than HSBs but similar total masses, they must have taken a different evolutionary path \citep{vanderhulst1993, vanzee1997, vandenhoek2000, kuzio2004}. Finally, LSBs are thought to be dark matter dominated at all radii \citep{deblok1996a, deblok1996b, mcgaugh2000, swaters2003}, making these galaxies vital to our understanding of galaxy formation and evolution. 

Although LSBs are comparatively not as well understood as HSBs, their dark matter domination is very well studied. Traditionally, understanding the nature of dark matter in LSBs has been approached with the examination of HI/H$\alpha$\ rotation curves and decomposition of baryonic and dark matter mass profiles \citep[e.g.][]{swaters2000, deblok2001b, mcgaugh2001, deblok2002, kuzio2006, kuzio2008}. In addition, mock observations of simulated LSBs have also been used to test predictions from cold dark matter simulations \citep{deblok2003, kuzio2009, kuzio2011b, pineda2017}. 

While there has historically been disagreement over the reliability of observations, such as concerns over HI beam smearing \citep[e.g.][]{vandenbosch2000, vandenbosch2001, blais2004}, H$\alpha$\ long-slit rotation curve resolution \citep[e.g.][]{swaters2003}, or, more recently, unaccounted for gas dynamics \citep[e.g.][]{pineda2017}, these have largely been accounted for. The result is that we are now left with observations that conflict with expectations from cold dark matter simulations \citep[e.g.][]{mcgaugh2001, deblok2002, spekkens2005, kuzio2006}. Despite this, pseudoisothermal dark matter haloes still match observations rather well, suggesting LSBs are embedded in massive dark matter haloes.

A subset of LSBs have been historically avoided, however, in these kinematic studies: those with bars. This is unfortunate as these galaxies may allow for a different approach to exploring dark matter. Barred LSBs have been avoided for two major reasons. Firstly, non-circular motions due to bars are not very well modeled or constrained, with both minimal \citep{eymeren2009} and significant \citep{spekkens2007,sellwood2010} effects on rotation curves being detected. Secondly, as a consequence of their implied dark matter domination, LSBs should be stable against bar formation, making these \textcolor{black}{features} very rare.

Through visual classifications, \citet{mihos1997} placed the bar fraction at $\sim$4$\%$ for LSBs. They also used numerical simulations to show that global instabilities are unlikely to form in LSB discs unless perturbed. \textcolor{black}{This bar fraction was recently updated to $\sim$8$\%$\ by \citet{honey2016} using optical SDSS images of LSB galaxies compiled from various catalogs.} \citet{sodi2017b} used a large sample of SDSS LSB galaxies classified from Galaxy Zoo 2 and found the bar fraction to be near 20$\%$. Each of these fractions are far lower than the $\sim$60$\%$\ found in HSBs when using NIR imaging \citep{eskridge2000,marinova2007,menendez2007}. This lower bar fraction likely stems from the higher gas content of LSBs \citep{sodi2017a}.

There has been recent work on exploring bars in LSBs, mostly with numerical simulations, to probe any correlations between bar morphology and host galaxy. \citet{mayer2004} used high resolution simulations of gas dominated discs and found that bars could form if the dark matter halo had a low concentration and the stellar component was higher than typically observed in LSBs. \citet{chequers2016} used hydrodynamical numerical simulations of UGC~628 to estimate the bar pattern speed and found it to suggest the galaxy was not dark matter dominated in the inner bar region. \citet{sodi2017b} observationally found a strong dependance on bar length with disc surface brightness, as well as finding weaker bars in comparison to HSBs. 

Given the current lack of data for barred LSBs, our goal is to quantify bars in LSBs. We hope to place these into morphological context with bars in HSBs, as well as infer properties of the dark matter haloes barred LSBs are embedded in. To accomplish this, we use optical \textit{B} and \textit{I}-band images of four barred LSB galaxies and use these new data in combination with archival Spitzer 3.6 $\mu$m data to quantify bar length and strength beyond a visual classification or description. In addition, we use a photometric technique from \citet{puerari1997} to characterise the relative bar pattern speeds of our sample.

The paper is organized as follows. In Sec.~\ref{sec:sam_dat} we discuss our sample selection, observations, and data reduction. In Sec.~\ref{sec:measurements} we detail how we construct our azimuthal light profiles and how we measure the bar length, strength, and speed. In Sec.~\ref{sec:results} we present the results for each individual galaxy. In Sec.~\ref{sec:discussion} we place our results into context with results from the literature as well as into context with previous dark matter studies. Finally, in Sec.~\ref{sec:conclusions} we list our conclusions from this work.

\section{Sample and Data}
\label{sec:sam_dat}

In this section we detail how we constructed our sample of barred LSB galaxies. We also discuss our acquisition of new broadband \textit{B} and \textit{I}-band images of our targets, and the retrieval of Spitzer 3.6 $\mu$m archival data. Finally, we detail our data reduction process, including how we deproject our galaxy images.

\subsection{Sample}
\label{ssec:sample}

As no single catalog of barred LSBs exists currently, we have begun assembling one from multiple sources. We have examined the catalogs of \citet{schombert1992} and \citet{impey1996} to visually identify barred LSBs. \citet{schombert1992} and \citet{impey1996} both list large numbers of LSBs that give us a starting point for a sample selection: \citet{schombert1992} lists 198 LSBs identified visually from the Second Palomar Sky Survery \citep{reid1991}, whereas \citet{impey1996} lists 693 LSBs identified from a combination of visual and machine scan searches. A few galaxies from \citet{schombert1992} have Spitzer 3.6 $\mu$m images \citep{schombert2014}, giving total stellar masses for these galaxies. The galaxies found in \citet{impey1996}, however, have very rarely been observed again. Because we are interested in the traditional LSB (i.e. dark matter dominated at all radii) from these catalogs, we avoid selecting bulge-dominated galaxies \citep[e.g.][]{beijersbergen1999} and those that only have LSB outer discs.

We also search the samples of \citet{swaters2000}, \citet{deblok2001b}, \citet{mcgaugh2001}, \citet{deblok2002}, \citet{kuzio2006}, and \citet{kuzio2008}. These studies provide rotation curves and mass models for numerous LSBs. Some of these galaxies overlap with the sample of \citet{schombert1992}. 

Finally, we also have selected UGC~628, which has had its bar pattern speed measured via Fabry-Perot kinematics \citep{chemin2009} and numerical modeling \citep{chequers2016}.

For the work presented in this paper, we have selected four LSBs of varied morphology and that have previous kinematic mass modeling: UGC~628, F568\nobreakdash-1, F568\nobreakdash-3, and F563\nobreakdash-V2. UGC~628 and F568\nobreakdash-1 display clear grand-design structure. However, these types of LSBs are rather rare. The other two, F568\nobreakdash-3 and F563\nobreakdash-V2, are more indicative of typical LSB structure, with F568\nobreakdash-3 having a clear disc but very messy spiral structure, and F563\nobreakdash-V2 having rather tenuous disc and arm structure.

Properties of our sample are listed in Table~\ref{sample}. The disc inclinations and position angles (P.A.) listed here are derived from our data (with F563\nobreakdash-V2 an exception) and discussed later in the text. The distance for UGC~628 is taken from \citet{deblok2002}, with the remaining distances taken from \citet{deblok2001b}.

\begin{table*}
  \centering
  \caption{Barred LSBs used in this study. Disc inclinations and P.A.s are derived from our data using \texttt{ELLIPSE}, except for F563\nobreakdash-V2 where we take the disc parameters from \citet{deblok2001b}. The distance for UGC~628 was taken from \citet{deblok2002}, with the rest taken from \citet{deblok2001b}.}
  \label{sample}
  \begin{tabular}{lccccc}
    \hline
    Galaxy        &   R.A.     &    Dec.   &      Inc.    &     P.A.      &     D   \\
                  & (J2000)    &  (J2000)  &   ($\degr$)  &   ($\degr$)   &   (Mpc) \\
    \hline
    UGC~628       & 01:00:51.9 & +19:28:33 & 58.2$\pm$0.7 & -42.8$\pm$0.9 &    65   \\
    F563\nobreakdash-V2       & 08:53:03.8 & +18:26:09 & 29           & -32           &    61   \\
    F568\nobreakdash-1        & 10:26:06.3 & +22:26:01 & 24.9$\pm$3.8 & -86.0$\pm$7.9 &    85   \\
    F568\nobreakdash-3        & 10:27:20.2 & +22:14:24 & 39.6$\pm$1.8 & -11.4$\pm$2.2 &    77   \\
    \hline
  \end{tabular}
\end{table*}

\subsection{Data}
\label{ssec:data}

We have obtained broadband \textit{B} and \textit{I}-band images of our targets using the ARCTIC imager on the 3.5-m telescope at Apache Point Observatory\footnote{Based on observations obtained with the Apache Point Observatory 3.5-meter telescope, which is owned and operated by the Astrophysical Research Consortium.}. ARCTIC has a field of view of 7.5$\arcmin \times$7.5$\arcmin$. We used ARCTIC in single read out mode with 2$\times$2 binning, giving a plate scale of 0.228 arcsec/pix.

ARCTIC observations for UGC~628 were taken on 2016 August 7. ARCTIC observations of F568\nobreakdash-1 and F563\nobreakdash-V2 were obtained on 2017 February 24. Observations for F568\nobreakdash-3 were taken on 2017 February 28, along with additional observations for F568\nobreakdash-1 and F563\nobreakdash-V2. We observed UGC~628, F568\nobreakdash-1, and F563\nobreakdash-V2 with 3$\times$600 sec exposures in both \textit{B} and \textit{I}. We observed F568\nobreakdash-3 with 2$\times$600 sec exposures in both bands. We dithered 15 arcsec between each exposure in both bands to correct for bad pixels and cosmic rays.

We also obtained fully reduced 3.6 $\mu$m images of our sample from the Infrared Science Archive (IRSA) published in \citet{schombert2014}. Observations had a maximum exposure time of 100 sec, and the final images have a pixel scale of 0.60 arcsec/pix.

\subsection{Data Reduction}
\label{ssec:reduc}

The Spitzer 3.6 $\mu$m images are fully reduced and flux calibrated. We converted the units of the image from MJy/sr to counts/pix using data from the header files in order to be compared with our optical images. The sky-background was determined by using the average value of six 50$\times$50 pixel star-free boxes on each image to determine a mean sky which was then subtracted from the image \citep[see][]{schombert2014}. 

\begin{figure*}
  \includegraphics[scale=1.0]{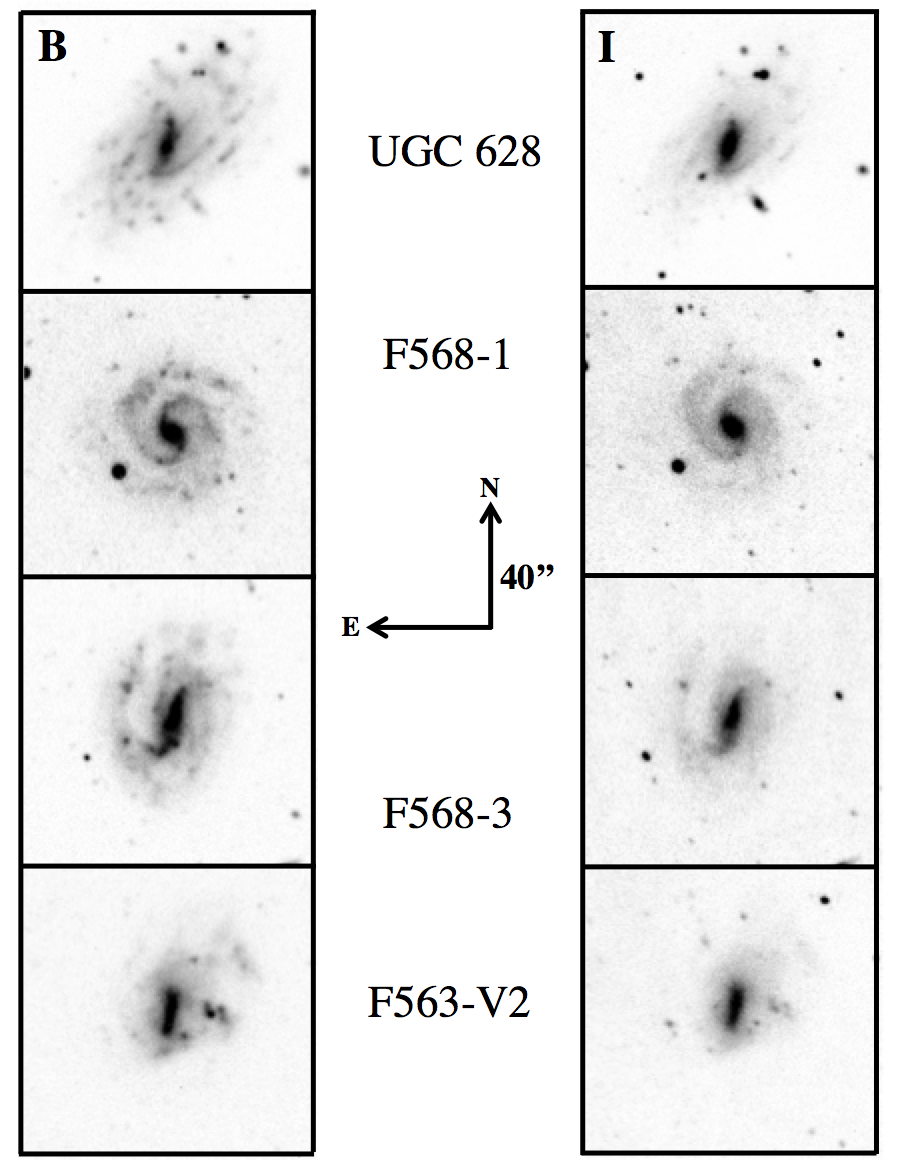}
  \caption{On-sky images of our target galaxies. \textit{B}-band images are in the left column and \textit{I}-band images are in the right column, all were taken with ARCTIC on the APO 3.5m. Directional arrows in the centre are both 40$\arcsec$\ long and apply to each image. Images are all scaled linearily in order to show the best contrast between the bar, spiral arms, and disc.}
  \label{gal_ims}
\end{figure*}

The ARCTIC data were reduced in \texttt{IRAF}\footnote{IRAF is distributed by the National Optical Astronomy Observatory, which is operated by the Association of Universities for Research in Astronomy (AURA) under a cooperative agreement with the National Science Foundation.} using standard packages and routines. The images were bias subtracted, dark subtracted, flat-fielded, and fringe corrected. The sky-subtraction was performed using the method outlined above, with six 100$\times$100 pixels boxes. The \textit{I}-band images were corrected for the fringe pattern present in each image. Our final, reduced and sky-subtracted \textit{B} and \textit{I}-band images are shown in Fig.~\ref{gal_ims}.

After the data were reduced, we deprojected all the data (including the Spitzer 3.6 $\mu$m images) so that the galaxies appear face-on. We have made the assumption that our disc galaxies are intrinsically circular. With this assumption, we used the \texttt{ELLIPSE} task in \texttt{IRAF} to fit elliptical isophotes to our galaxy images to determine the eccentricity ($\epsilon = 1 - b/a$) and position angle (P.A.) of the outer disc of our galaxies. We first ran \texttt{ELLIPSE} with all parameters free in order to determine the galaxy centre. We then re-ran \texttt{ELLIPSE} with the centre fixed to derive $\epsilon$\ and P.A. for the outer disc. We also confirmed that the centres found by \texttt{ELLIPSE} were the same location in the galaxy for all three bands. Finally, we used \texttt{GEOTRAN} to rotate the image by the disc position angle so that the major axis of the galaxy is aligned with the y-axis, and then `stretch' the minor axis (now aligned with the x-axis) by the axis ratio (b/a). If done properly, the deprojected galaxy images should have roughly circular outer isophotes. 

In order to ensure our images are deprojected in the same manner for each band, we derive inclinations and P.A. in the three bands for each galaxy and use the average as the final value. We use these final values to deproject each image. We list our derived inclinations and P.A. for our sample in Table~\ref{sample}, with the exception of F563\nobreakdash-V2 where we took the disc parameters from \citet{deblok2001b}. \textcolor{black}{The reason for this is as follows: the sharp change in the light profile between the bright bar and faint disc, as well as the rather tenuous disc structure, causes \texttt{IRAF}'s \texttt{ELLIPSE} task to fail when attempting to fit elliptical isophotes to the images. Due to this, we adopted literature values for our deprojection for this galaxy.} Because F568\nobreakdash-1 is nearly face on ($i \sim$25$^{\circ}$), the errors on the P.A. are relatively large. But because it is so face-on, the rotation angle does not affect the deprojection to a large degree.


\section{Measurements}
\label{sec:measurements}

For our four galaxies, we want to measure the bar lengths, strengths, and corotation radii. In this section we discuss the techniques that we use for measuring each of these bar properties.

\subsection{Bar Length}
\label{ssec:bar_len}

The length of a bar is rather straightforward, simply how far it extends from the centre of the galaxy. And yet, there are many ways of measuring the bar radius, since defining the end of a bar is more complex than expected. We use three in this work: the behaviour of azimuthal light profiles, Fourier analysis, and the behaviour of elliptical isophotes. \textcolor{black}{These techniques each provide an objective (and quantitative) measure of the bar length.} Each technique is described in detail below. Becase bars are stellar features, we only use the \textit{I}-band and 3.6 $\mu$m images to measure the bar length. \textcolor{black}{We compare the results of each technique to confirm that each is physically meaningful (i.e. does not include a significant portion of a spiral arm).}

\subsubsection{Azimuthal Light Profiles}
\label{sssec:az_profs}

Determining how light is distributed in a galaxy is very useful for studying properties of galaxies, including bars. For example, we can examine how the light behaves as a function of azimuthal angle and radius (i.e. an azimuthal light profile).

\citet{ohta1990} describe how azimuthal light profiles appear for different morphological features. In the very inner radii, the light profiles should be fairly constant due to the nucleus or a bulge. If a bar is in the galaxy, then there should be two `humps' present in the light profiles, separated by 180 degrees in azimuthal angle. These humps should also remain roughly constant in angle for all radii within the bar. Once outside the bar region, these humps should change depending on the morphology of the galaxy. If there are two spiral arms present, these humps should roughly remain 180 degrees out of phase but begin moving to different angles, as the arms are not at constant azimuthal angle. Because spiral arms and discs are typically fainter than bars, the peak intensities of these humps will also decrease dramatically outside the bar region. Using these changes in the humps, it should be possible to measure the radius of the bar. 

For consistency, we construct the azimuthal light profiles in the same manner for all of our galaxies. We use 120 azimuthal divisions for the \textit{B} and \textit{I}-bands and 60 divisions for 3.6 $\mu$m (i.e. 3$^{\circ}$\ and 6$^{\circ}$\ wide, respectively). We use a radial spacing of every 2 pixels for the \textit{B} and \textit{I}-bands and every 1 pixel for 3.6 $\mu$m (i.e. a spacing of 0.46$\arcsec$\ and 0.6$\arcsec$\ respectively). We begin our profiles at 3$\arcsec$\ for each galaxy, as starting too far in results in some bins containing no pixels. We then sum up the intensity of all pixels within each radial and azimuthal bin and divide by the number of pixels to return intensity per pix$^{2}$. In order to account for local variations that may be present, we use a nearest-neighbor smoothing at the end.

To measure the bar radius, we fit Gaussians to each hump in order to measure the azimuthal angle and peak intensity. By examining both the location and height of a hump at each radial slice relative to the other hump in a given radius, which should be 180 degrees out of phase and of similar intensity, we can track its behaviour across all radii. Where we find changes similar to what was described above and in \citet{ohta1990} we call the bar radius.

\subsubsection{Fourier Analysis}
\label{sssec:four_bar}

An alternative method for measuring the bar length is through a Fourier analysis \citep[e.g.][]{ohta1990,aguerri1998,aguerri2000,aguerri2009,puerari1997}. This involves decomposing our azimuthal light profiles via a Fourier transform, given as:

\begin{equation}
\mathcal{F}(r) = \int_{-\pi}^{\pi} I_{r}(\theta) \exp{(-2i \theta)} d\theta
\end{equation}
where $I_{r}(\theta)$\ are the azimuthal light profiles, and $\theta$\ is the azimuthal angle.

The Fourier coefficients are:
\begin{equation}
  A_{m}(r) = \frac{1}{\pi} \int_{0}^{2\pi} I_{r}(\theta) \cos{(m\theta)}d\theta
\end{equation}
\begin{equation}
  B_{m}(r) = \frac{1}{\pi} \int_{0}^{2\pi} I_{r}(\theta) \sin{(m\theta)}d\theta
\end{equation}
and the amplitudes are:
\begin{equation}
  I_{0}(r) = \frac{A_{0}(r)}{2}
\end{equation}
\begin{equation}
  I_{m}(r) = \sqrt{A_{m}^{2}(r) + B_{m}^{2}(r)}
\end{equation}

These amplitudes can be used both to confirm the presence of a bar, as well as measure the bar length. Within the bar region, the $m = 2$\ and $m = 4$\ amplitudes are strong compared with the odd modes \citep{ohta1990}. In order to measure the bar length, \citet{ohta1990} and \citet{aguerri2000} use the bar and interbar Fourier intensities ($I_{b}$\ and $I_{ib}$\ respectively), defined as: 
\begin{equation}
  I_{b} = I_{0} + I_{2} + I_{4} + I_{6}
\end{equation}
\begin{equation}
  I_{ib} = I_{0} - I_{2} + I_{4} - I_{6}
\end{equation}
From \citet{aguerri2000}, the bar region is defined as: 
\begin{equation}
  \frac{I_{b}}{I_{ib}} > \frac{1}{2}\left[ \left( \frac{I_{b}}{I_{ib}} \right)_{max} - \left( \frac{I_{b}}{I_{ib}} \right)_{min} \right] + \left( \frac{I_{b}}{I_{ib}} \right)_{min}
\end{equation}
and the last radius at which this is satisfied (within the bar region) is taken as the bar radius. More simply, \citet{ohta1990} define the bar region as ($I_{b}/I_{ib}) >$ 2. Because \citet{aguerri2000} claim Equation 8 takes the behaviour of the Fourier intensity profiles better into account than simply ($I_{b}/I_{ib}) >$ 2, we use their definition as the bar radius from the Fourier intensities.

\subsubsection{Elliptical Isophotes}
\label{sssec:iso_bar}

Perhaps a more familiar means of finding the bar radius is to fit elliptical isophotes to the light distribution of a galaxy disc and use either the radius of maximum ellipticty or the radius where the ellipticity changes discontinuously as an indication that the end of the bar has been reached \citep[e.g.,][]{wozniak1995,aguerri2000}. Because \citet{aguerri2000} find that choosing the radius of discontinuity often leads to overestimating the bar length, we use the radius at which the ellipticity reaches its maximum value as the bar length.

\subsection{Bar Strength}
\label{ssec:bar_strength}

The second bar parameter we want to measure is the strength. The strength of a bar can be thought of as a tracer of the underlying gravitational potential of the bar, best traced by NIR imaging since bars are stellar features. While originally only a visual classification based on how bright or large the bar appeared \citep{devaucouleurs1959}, with strong bars simply classified as SB and weaker bars as SAB, bar strength is now a quantifiable parameter. This parameter can be measured via multiple methods, such as the torques present in the galaxy \citep{buta2001,laurikainen2002}, or the ellipticity of isophotes \citep{martin1995}. Here, we will follow \citet{aguerri2000} and define the bar strength as

\begin{equation}
  S_{b} = \frac{1}{r_{bar}} \int_{0}^{r_{bar}} \frac{I_{2}}{I_{0}}dr
\end{equation}
where $r_{bar}$ is the radius of the bar, and $I_{2}$\ and $I_{0}$\ are the $m = 2$\ and $m = 0$\ Fourier amplitudes shown in Eq.~5. However, as we begin our azimuthal light profiles at 3$\arcsec$ and \textit{not} the very centre, we can only report a lower limit on the bar strengths for our galaxies. We will use the behaviour of our relative Fourier amplitudes as an indicator of how well our lower limits approximate the true strength. Similar to our bar radius measurements, we obtain bar strengths in only the \textit{I}-band and 3.6 $\mu$m. In order to get a more accurate indicator of the bar strength, we also remake our azimuthal light profiles starting at 1.5$\arcsec$\ and increase the azimuthal spacing from 3$\degr$\ to 6$\degr$\ for \textit{I} and from 6$\degr$\ to 12$\degr$\ for 3.6 $\mu$m. We do this to probe down closer to the centres of our galaxies to see if this increases the bar strength significantly. These new profiles are not used further in the analysis.

\subsection{Corotation Radius}
\label{ssec:corot_radii}

Finally, the third parameter we want to measure is the corotation radius. The corotation radius ($\mathrm{R_{CR}}$) of the bar is where orbital speeds are equal to the pattern speed of the bar. As will be discussed in Sec.~\ref{ssec:dis_speed}, $\mathrm{R_{CR}}$\ is useful for determining relative bar pattern speeds and in turn inferring properties of the dark matter halo. There are numerous ways of measuring or inferring $\mathrm{R_{CR}}$, including: identifying $\mathrm{R_{CR}}$\ with photometric rings \citep[i.e.][]{buta1986,perez2012}, modeling galaxies based on their luminosity distributions or velocity fields \citep[i.e.][]{salo1999,rautiainen2004,rautiainen2008}, phase intersections of multi-band photometry \citep{puerari1997}, and many others.

For this work, we will use the \citet[][hereafter PD97]{puerari1997} method to measure $\mathrm{R_{CR}}$. PD97 expanded upon the results of \citet{beckman1990}, by showing how phase crossings of \textit{B} and \textit{I}-band images can be used as an indicator of $\mathrm{R_{CR}}$. Based on spiral density wave theory \citep{lin1964}, spiral arms act as density waves that trigger star formation by compressing gas. As orbits are faster within corotation and slower outside, this means that the shock front triggering star formation is present on different sides of the arm on either side of $\mathrm{R_{CR}}$. Since the shock can be traced by newer O and B stars and the underlying density wave can be traced by the older stellar population, the phase intersection between the \textit{B} and \textit{I}-bands will occur at corotation. 

This method has been used to determine the corotation radii for HSBs, and has found consistent results with both direct measurements of the pattern speed and numerical simulations of real galaxies \citep{aguerri1998,vera2001,sierra2015}. In addition, \citet{martinez2014} expanded the use of this method to HI, CO, 24 $\mu$m, and FUV with success.

We calculate the phase as:

\begin{equation}
\Theta(r) = \arctan{\left( \frac{\mathrm{Re}(\mathcal{F}(r))}{\mathrm{Im}(\mathcal{F}(r))} \right)}
\end{equation}
where $\mathrm{Re}(\mathcal{F}(r))$\ and $\mathrm{Im}(\mathcal{F}(r))$\ are the real and imaginary parts of the Fourier transform shown in Eq.~1 respectively. In order to use the PD97 method, we require images in photometric bands separated by a large wavelength range, specifically one band for the newer stellar population and one for the older. We use the \textit{B}-band to trace the newer stars, and the \textit{I}-band and 3.6 $\mu$m to trace the older stars.

R$_{\mathrm{CR}}$\ is the intersection between the \textit{B}-band phase profile and either the \textit{I}-band or 3.6 $\mu$m phase profile. We take the first intersection after the bar radius to be R$_{\mathrm{CR}}$. Since we are using the \textit{I}-band and 3.6 $\mu$m as tracers of the older stellar populations, both phase profiles should intersect the \textit{B}-band at roughly the same radius. This is because even though the \textit{I}-band and 3.6 $\mu$m are separated by a large wavelength range, they both trace the older stars.

\section{Results}
\label{sec:results}

In this section we detail the results for each galaxy individually. Final bar lengths, strengths, and corotation radii are shown in Table~\ref{res_table}. In general, we fnd that all four of our galaxies show very strong $m = 2$\ amplitudes, and some show strong $m = 4$\ and $m = 6$\ amplitudes. This leaves little doubt that our galaxies are barred (see Sec.~\ref{sssec:four_bar}). In fact, the strength of our $m = 2$\ and $m = 4$\ amplitudes are comparable to those of HSBs from \citet{elmegreen1985}.

\begin{table*}
  \centering
  \caption{Bar radii ($\mathrm{R_{bar}}$), corotation radii ($\mathrm{R_{CR}}$), relative bar pattern speeds ($\mathcal{R}$), and lower limits on bar strengths ($S_{b}$) for our sample. Radii are in arcsec. Relative bar pattern speeds and bar strengths are dimensionless. Bar radii are those derived from our azimuthal light profile method (see Sec.~\ref{sssec:az_profs}).}
  \label{res_table}
  \begin{tabular}{lccccccccc}
    \hline
    & \multicolumn{4}{c}{\textit{I}} & & \multicolumn{4}{c}{3.6 $\mu$m} \\
    \cmidrule(lr){2-5} \cmidrule(lr){7-10}
    Galaxy & R$_{\mathrm{bar}}$ & R$_{\mathrm{CR}}$ & $\mathcal{R}$ & S$_{b}$ & & R$_{\mathrm{bar}}$ & R$_{\mathrm{CR}}$ & $\mathcal{R}$ & S$_{b}$ \\
    \hline
    UGC~628 & 11.21$\pm$0.92 & 13.96$\pm$0.46 & 1.25$\pm$0.11 & 0.26 & & 11.40$\pm$1.20 & 13.96$\pm$0.60 & 1.22$\pm$0.14 & 0.22 \\
    F568\nobreakdash-1  &  4.37$\pm$0.46 &  5.86$\pm$0.46 & 1.34$\pm$0.18 & 0.13 & & 4.80$\pm$0.60 &  7.06$\pm$0.60 & 1.47$\pm$0.22 & 0.14 \\
    F568\nobreakdash-3  &  8.93$\pm$0.92 & 10.06$\pm$0.46 & 1.13$\pm$0.13 & 0.19 & & 9.60$\pm$1.20 & 13.86$\pm$0.60 & 1.44$\pm$0.19 & 0.19 \\
    F563\nobreakdash-V2 &  6.65$\pm$0.46 & 15.86$\pm$0.46 & 2.38$\pm$0.18 & 0.29 & & 7.20$\pm$0.60 & 15.86$\pm$0.60 & 2.20$\pm$0.20 & 0.26 \\
    \hline
  \end{tabular}
\end{table*}

\subsection{UGC~628}
\label{ssec:ugc628_res}

As seen in Fig.~\ref{gal_ims}, UGC~628 displays a clear bar and dual spiral arm morphology. We also can see that there is a clear difference in appearance between the \textit{B} and \textit{I}-band images. We see the spiral arms more clearly in the \textit{B}-band image, as well as numerous HII regions that appear along the arms. In the \textit{I}-band image we can see the bar is both visually larger and fatter.


\begin{figure}
  \centering
  \includegraphics[scale=0.35]{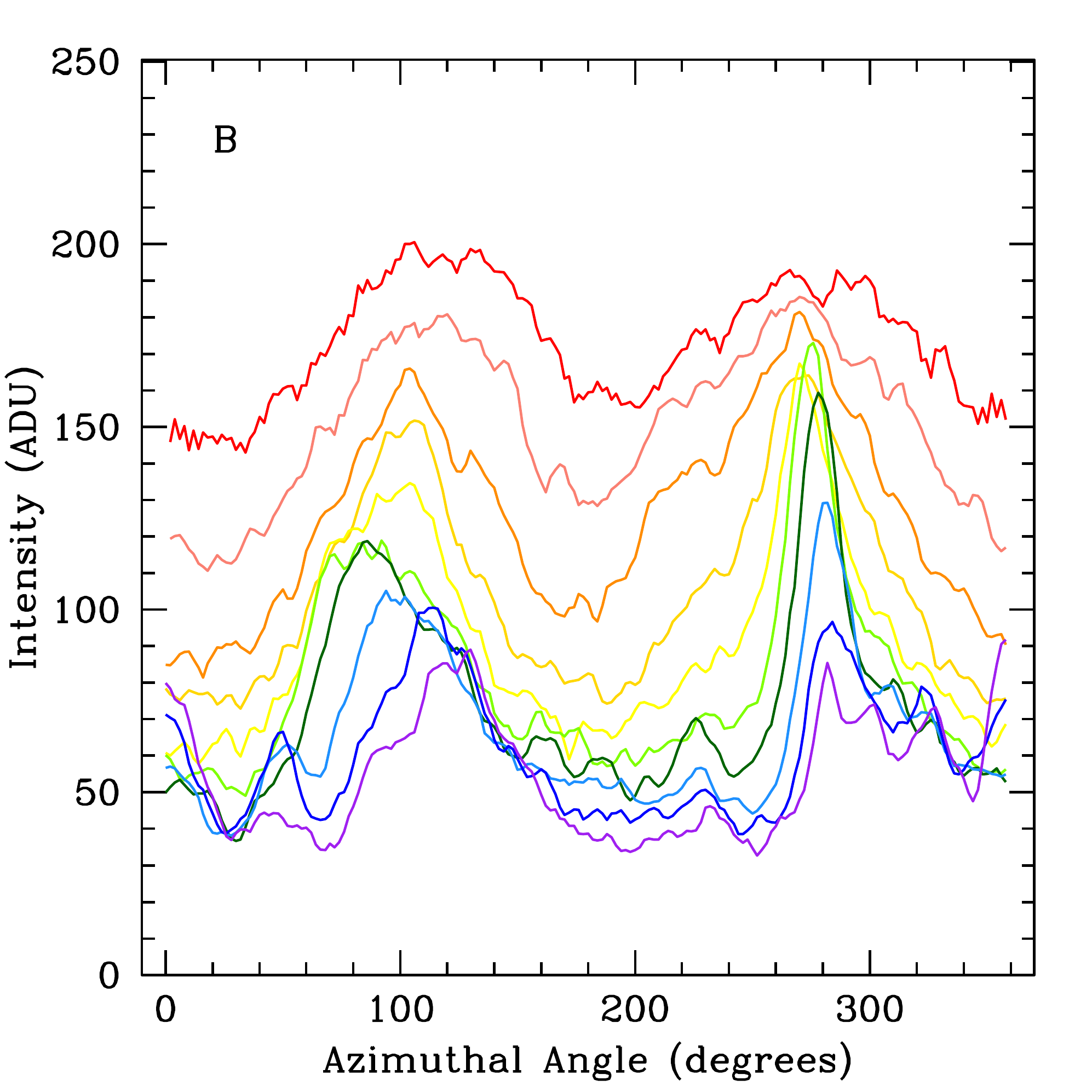}
  \includegraphics[scale=0.35]{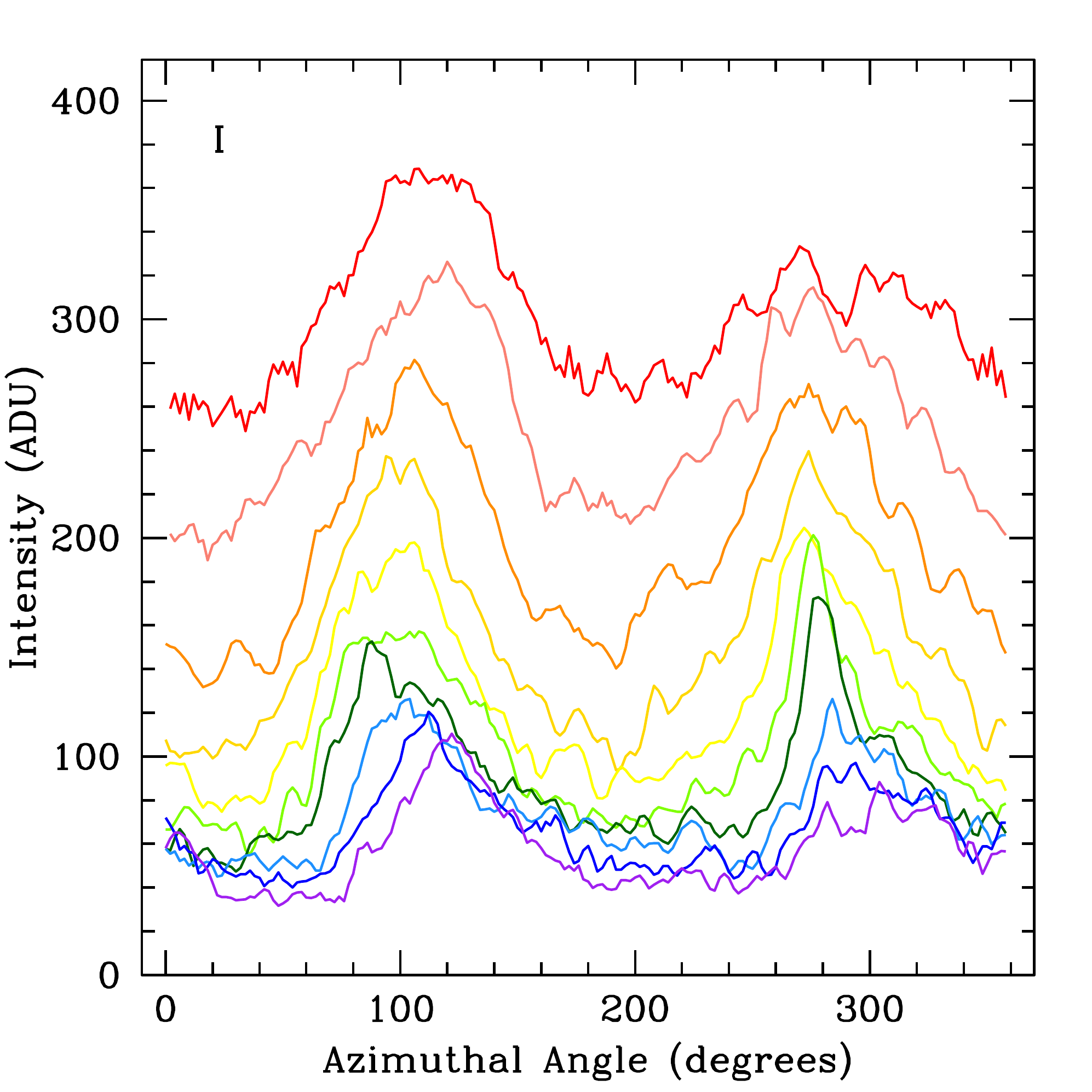}
  \includegraphics[scale=0.35]{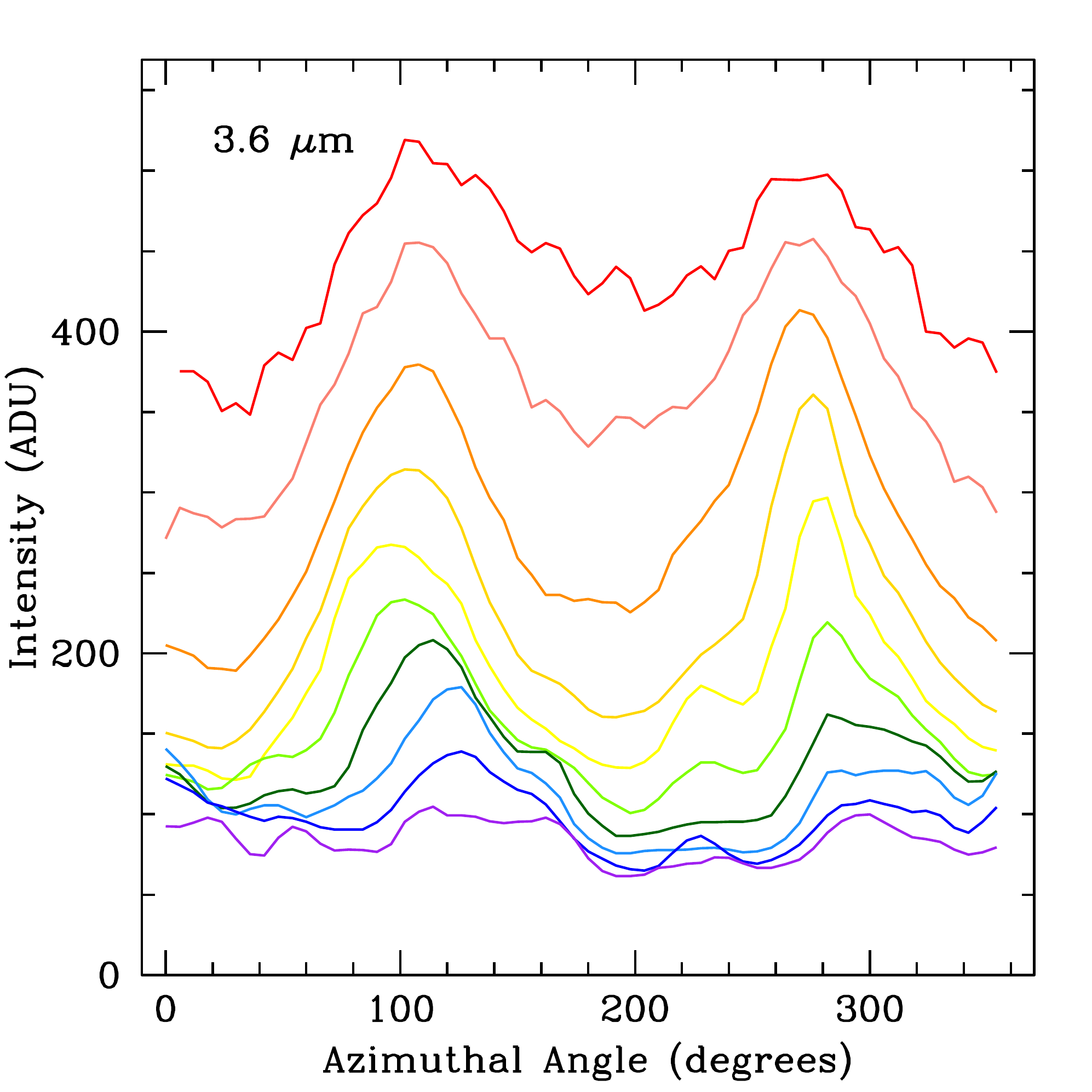}
  \caption{Azimuthal light profiles for UGC~628 in \textit{B} (top), \textit{I} (middle), and 3.6 $\mu$m (bottom). For clarity, profiles are plotted every 1.2$\arcsec$\ (every 6 pixels) for the \textit{B} and \textit{I}-bands and 1.8$\arcsec$\ (every 3 pixels) for 3.6 $\mu$m. All three bands begin at 3$\arcsec$. Profiles are plotted in a rainbow continuum, with red as the inner radius and purple the outermost.}
  \label{ugc628_az}
\end{figure}

\subsubsection{Bar Radii}
\label{sssec:ugc628_bar}

\begin{figure}
  \centering
  \includegraphics[scale=0.4]{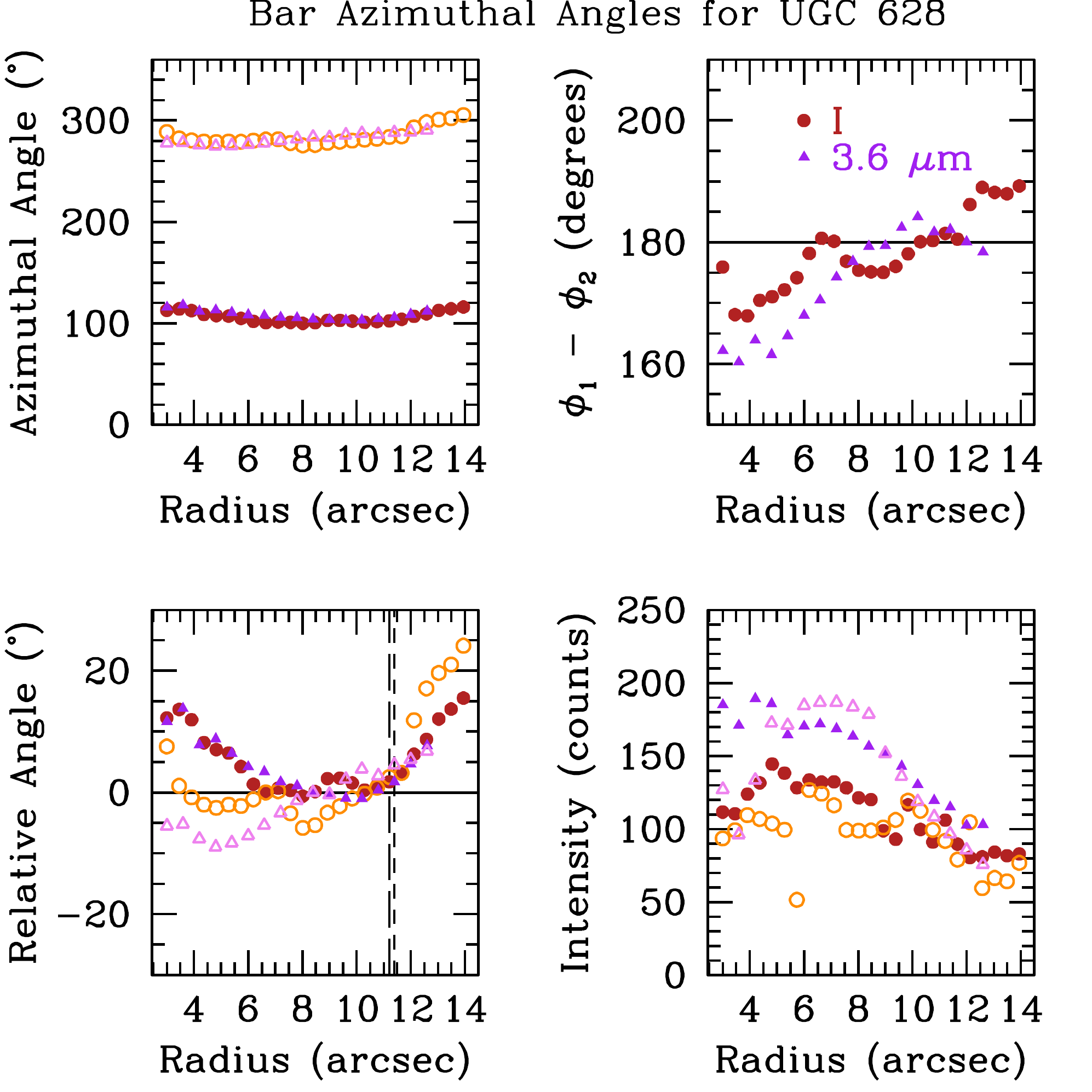}
  \caption{Bar azimuthal angle centroid positions for UGC~628. The brown and orange circles denote the location of the \textit{I}-band humps in Fig.~\ref{ugc628_az}, and the purple and pink triangles denote the location of the 3.6 $\mu$m humps. The top left panel shows the angular positions of the two humps, derived from fits with Gaussians. The top right panel shows the angular difference between the centroids of the two humps, with the horizontal line showing 180 degrees difference. The bottom left panel shows the relative motion from the bar centroid of the humps. The vertical dashed lines indicate the radius of the bar. The bottom right panel shows the intensity of the humps above the galaxy light.}
  \label{ugc628_az_bar}
\end{figure}

Our azimuthal light profiles for UGC~628 are shown in Fig.~\ref{ugc628_az}. Here, the profiles are plotted only every 1.2$\arcsec$\ for the \textit{B} and \textit{I}-bands and every 1.8$\arcsec$\ for 3.6 $\mu$m in order to more clearly see the behaviour of the profiles. The profiles are plotted in a rainbow continuum with the red profile (top line) showing the inner radius and the purple profile (bottom line) the outermost. This galaxy displays strong humps in its light profiles at inner radii, confirming the presence of a bar. In order to quantitatively track the motion of the humps, we fit Gaussians to the light profiles in order to obtain azimuthal centroids and intensities (Sec.~\ref{sssec:az_profs}), seen in Fig.~\ref{ugc628_az_bar}. 

In this figure, the brown and orange circles denote the \textit{I}-band humps, and the purple and pink triangles denote the 3.6 $\mu$m humps. We only show the region around the bar, ending around 14$\arcsec$. The top left panel shows the azimuthal position of the two humps seen in Fig.~\ref{ugc628_az}. The top right panel shows the azimuthal difference between the two humps ($\phi_{1}$\ and $\phi_{2}$), with the horizontal black line denoting 180 degrees. Humps due to a bar should remain at this value \citep{ohta1990}. We can see that between roughly 7$\arcsec$\ to 12$\arcsec$\ the humps are close to 180 degrees out of phase. We therefore take the angle of each hump at the beginning of this region to be the angle of the bar, or bar centroid. The bottom left panel shows the angle of the humps relative to the bar centroid, remaining roughly constant until $\sim$11.5$\arcsec$, where both humps in both bands begin moving towards larger azimuthal angle. We find that the humps remain roughly constant in angle before moving together towards larger azimuthal angle, where they begin tracing the path of the spiral arms. We find that there is still quite significant structure within the bar, as the centroids do not remain exactly constant in the bar region. In fact, the very inner regions ($<$\ 8$\arcsec$) do not appear constant at all, possibly indicating the presence of a bulge of some sort (see the large movement in the top right and bottom left panels of Fig.~\ref{ugc628_az_bar}). In the bottom right panel we show the peak intensity of the humps above the galaxy light, measured as the height of the hump above the `continuum' at each radius seen in the azimuthal light profiles. Here, we see that the intensity drops off significantly after the bar region, \textcolor{black}{most pronounced in the open symbols. This is a reflection of the somewhat non-symmetric nature of the bar, where the northern part of the bar (open symbols) has a more intense decrease in intensity than the southern part (closed symbols). This trend is seen in both bands.}

We determine the bar length from the azimuthal light profiles via the bottom left panel of Fig.~\ref{ugc628_az_bar}. We take the radius where the centroids begin moving together towards larger azimuthal angle (the start of the spiral arms) as an indication that the bar has ended, shown as the vertical dashed lines, short-dashed for the \textit{I}-band and long-dashed for 3.6 $\mu$m. The bar lengths from this method are 11.21$\arcsec \pm$0.92$\arcsec$\ in \textit{I} and 11.40$\arcsec \pm$1.2$\arcsec$\ in 3.6 $\mu$m. \textcolor{black}{Since the radius at which the behaviour given in \citet{ohta1990} (see Sec.~\ref{sssec:az_profs}) could be assigned to a few of the points in Fig.~\ref{ugc628_az_bar}, we assign the error to be four pixels (equivalent to 0.92$\arcsec$) in the \textit{I}-band and two pixels (equivalent to 1.2$\arcsec$) in 3.6 $\mu$m.}


\begin{figure}
  \centering
  \includegraphics[scale=0.4]{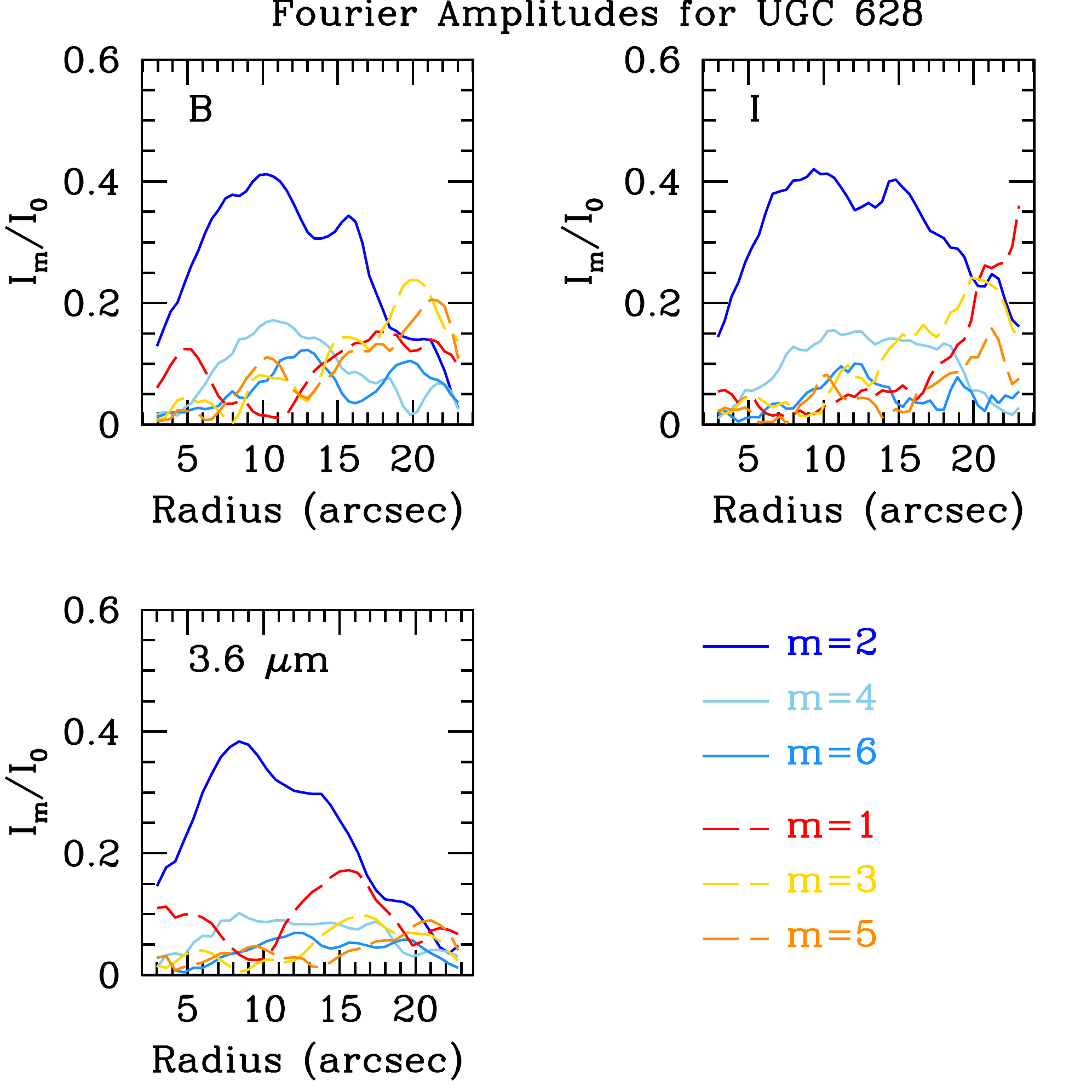}
  \caption{Fourier amplitudes for UGC~628. Bluer colours show the even modes, with the dark blue being $m = 2$. Redder colours show the odd modes, with the dark red being $m = 1$. All three bands show strong $m = 2$\ and $m = 4$\ modes within the inner regions, strongly supporting the presence of a bar.}
  \label{ugc628_amp}
\end{figure}

\begin{figure}
  \centering
  \includegraphics[scale=0.4]{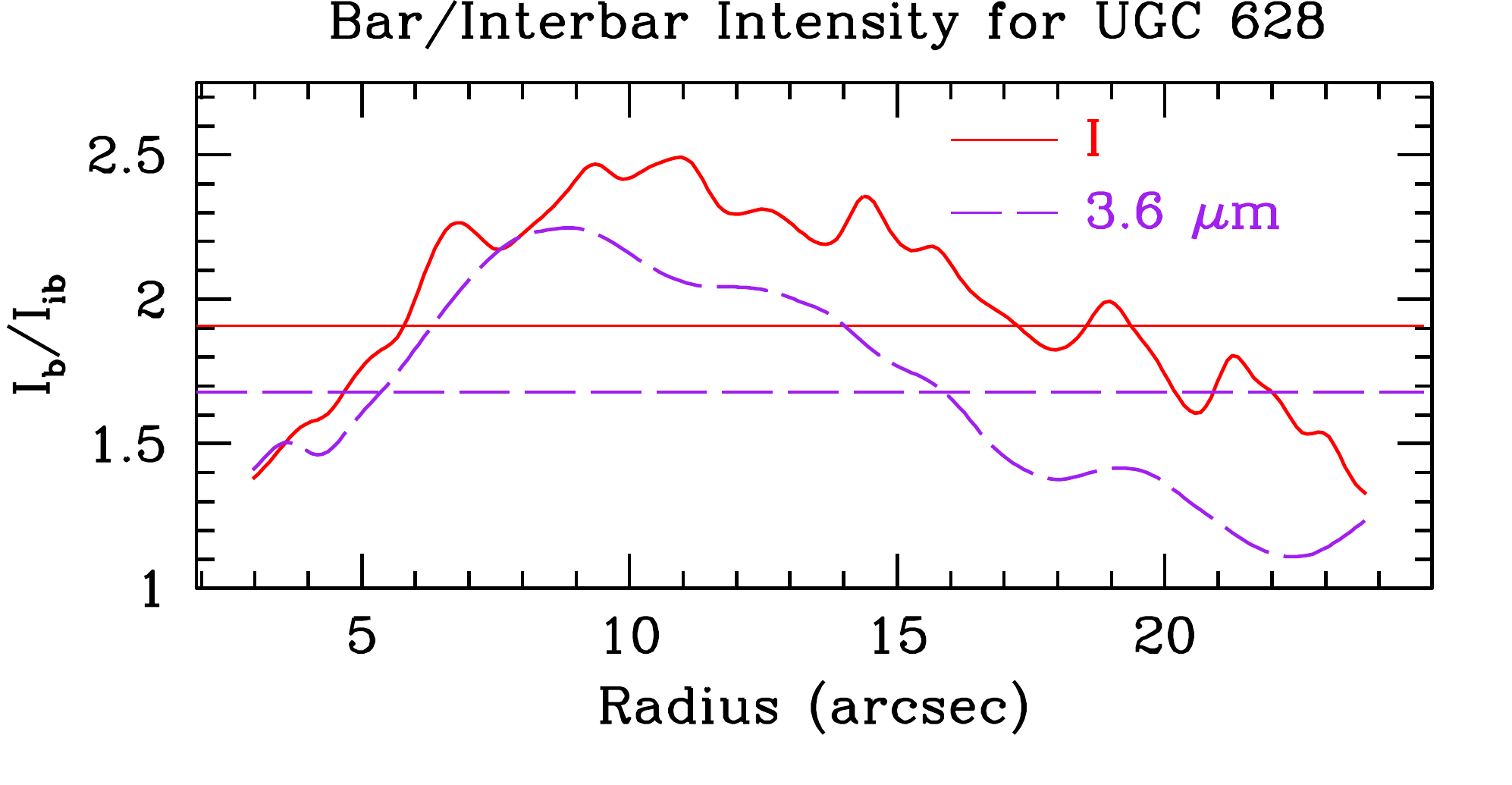}
  \caption{Bar/Interbar Fourier intensities for the \textit{I}-band (solid red) and 3.6 $\mu$m (dashed purple) for UGC~628. The horizontal lines show equation 7.}
  \label{ugc628_four}
\end{figure}

Our second bar length measure comes from our Fourier analysis. In Figures \ref{ugc628_amp} and \ref{ugc628_four} we show the relative Fourier amplitudes and bar/interbar Fourier intensities for UGC~628. All three bands show strong $m = 2$\ and $m = 4$\ modes within the inner regions, strongly supporting the presence of a bar. Using the Fourier Bar/Interbar method from \citet{aguerri2000}, we find a bar length of 16.96$\arcsec \pm$0.46$\arcsec$ for the \textit{I}-band and 15.86$\arcsec \pm$0.60$\arcsec$\ for 3.6 $\mu$m. \textcolor{black}{As this measure for the bar length is simply where the Fourier intensities cross the value given by Equation 8, the error comes from the radii spacing of our azimuthal light profiles: two pixels (equivalent to 0.46$\arcsec$) for the \textit{I}-band and one pixel (equivalent to 0.60$\arcsec$) for 3.6 $\mu$m.}


\begin{figure}
  \centering
  \includegraphics[scale=0.4]{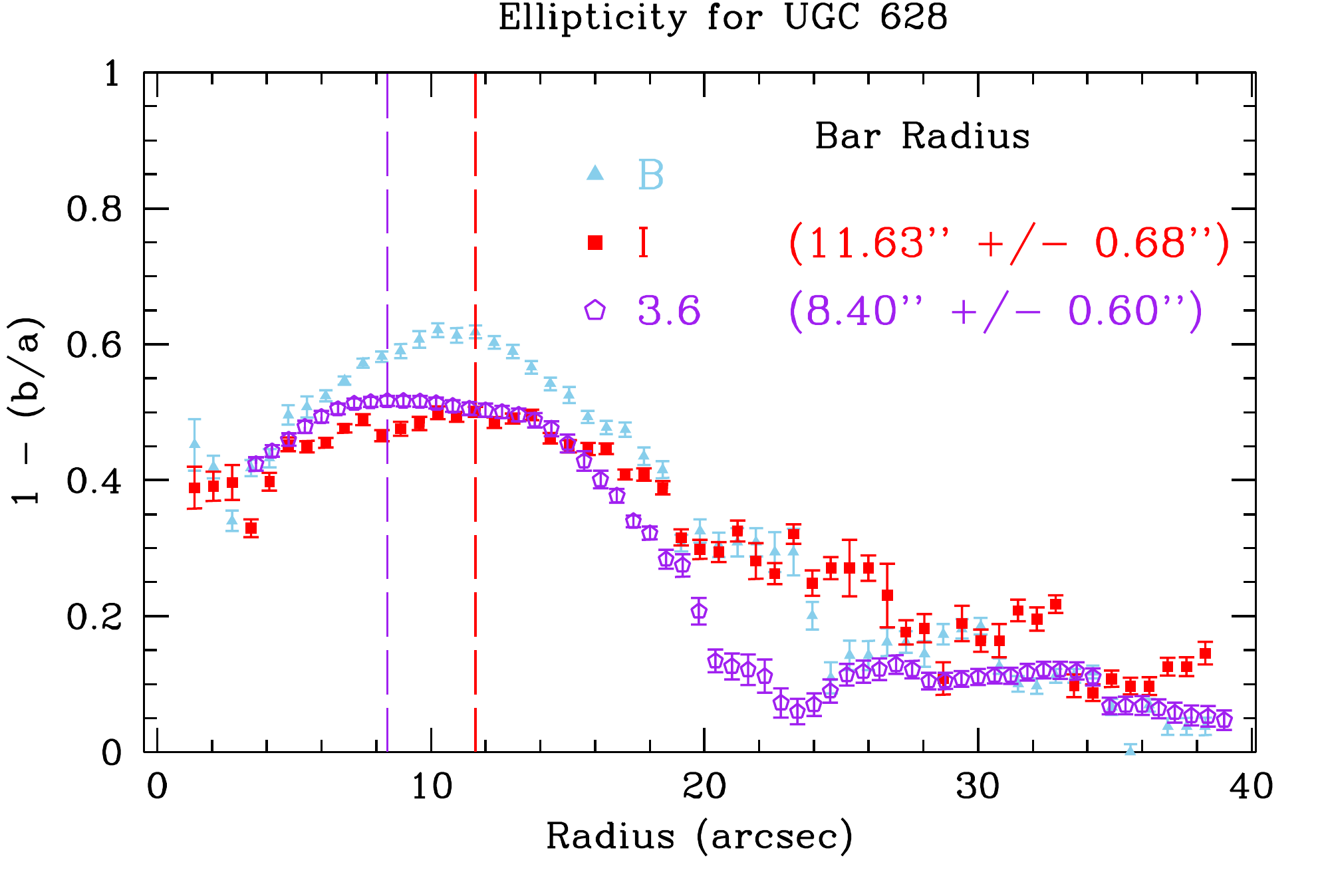}
  \caption{Ellipticity as a function of radius of elliptical isophotes for the deprojected images of UGC~628. The roughly zero ellipticity in all three bands at large radii shows our deprojection was successful. The vertical dashed lines indicate the radius of maximum ellipticity in \textit{I} and 3.6 $\mu$m, which we take as the isophotal bar radius. The \textit{B}-band is shown by the blue triangles, \textit{I}-band by the red squares, and 3.6 $\mu$m by the open purple pentagons.}
  \label{ugc628_isobar}
\end{figure}

Our third bar length measure is the radius of maximum ellipticity. We show the radial plot of deprojected ellipticity for UGC~628 for all three bands in Fig.~\ref{ugc628_isobar}. Here, we use only the \textit{I}-band and 3.6 $\mu$m to obtain bar radii, but also show the \textit{B}-band to check how succesful our deprojection was. In the outer radii, we can see the isophotes are roughly circular in all three bands, indicating our deprojection was successful. Using the radius of maximum ellipticity of elliptical isophotes, we find the \textit{I}-band bar radius to be 11.63$\arcsec \pm$0.68$\arcsec$\ and a 3.6 $\mu$m bar radius of 8.40$\arcsec \pm$0.60$\arcsec$. However, we note that both the \textit{I}-band and 3.6 $\mu$m ellipticity profiles appear quite flat within the bar radius, \textcolor{black}{making the reliability of choosing the radius of maximum ellipticity uncertain for this galaxy. In order to confirm our results, we began our starting radius for \texttt{ELLIPSE} at various radii to see what effect this had on the bar radius chosen. We found that this did not affect our bar lengths from this method. Therefore, the error on the bar length from this method is quivalent to the radial spacing from \texttt{ELLIPSE}.} 


\begin{figure}
  \centering
  \includegraphics[scale=0.7]{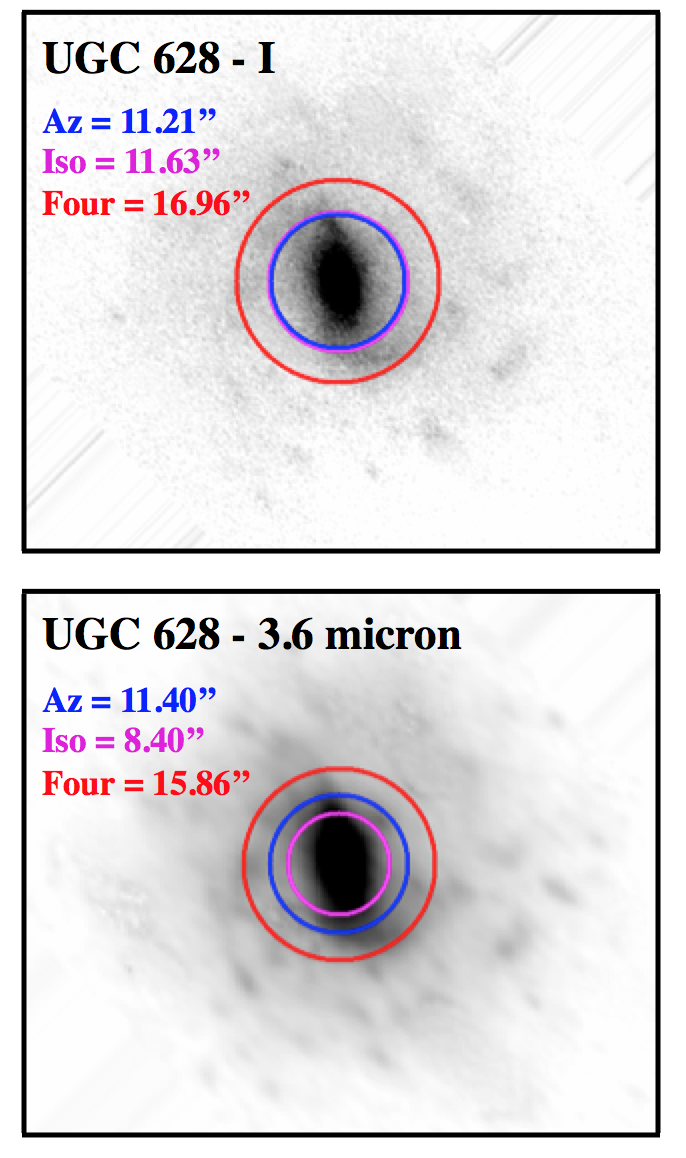}
  \caption{Bar radii for the \textit{I}-band (top) and 3.6 $\mu$m (bottom) for UGC~628 overplotted on the deprojected images. In both panels, blue circles are derived from the azimuthal light profile method, pink circles are derived from the elliptical isophote method, and red circles are derived from the Fourier bar/interbar intensities. Radii are in arcsec.}
  \label{ugc628_bar}
\end{figure}

These three techniques give us a bar radius that ranges from $\sim$8.5$\arcsec$ to $\sim$16.8$\arcsec$. At first glance, we are left with values that span a factor of two, but this is due to outliers. To compare the results visually, the bar radii are plotted over the deprojected \textit{I}-band and 3.6 $\mu$m images of UGC~628 in Fig.~\ref{ugc628_bar}. From this figure, it is clear that the Fourier method overshoots the bar radius in both bands rather significantly, as the red circle extends beyond the fat, bright bar and even includes a rather large portion of the spiral arms. Therefore, we can safely assume the Fourier bar radii are not accurate. The azimuthal and elliptical isophote bar radii agree for the \textit{I}-band, but the elliptical isophote bar radius for 3.6 $\mu$m appears to be too short, with the bar visually extending past the pink circle. Given the agreement between the results for both bands, we use the bar length derived using the azimuthal light profile method as the bar radius for UGC~628 in both \textit{I} and 3.6 $\mu$m, listed in Table~\ref{res_table}.


\subsubsection{Bar Strength}
\label{sssec:ugc628_str}

The next parameter that we have measured is the bar strength. Using Equation 9, we measure bar strengths to be 0.26 and 0.22 in the \textit{I} and 3.6 $\mu$m respectively (see Table~\ref{res_table}). As described in Sec.~\ref{sssec:four_bar}, these are lower limits on the bar strength. However, based on the relative Fourier amplitudes in Fig.~\ref{ugc628_amp}, it is most likely that these values are close to the true strength, as the $m = 2$\ mode appears to be trending towards zero near the centre. In addition, when decreasing the starting radius from 3$\arcsec$\ to 1.5$\arcsec$\ when constructing the azimuthal light profiles (see Sec.~\ref{ssec:bar_strength}) our values remained the same, suggesting our lower limits are accurate.

\subsubsection{Corotation Radii}
\label{sssec:ugc628_corot}

\begin{figure}
  \centering
  \includegraphics[scale=0.4]{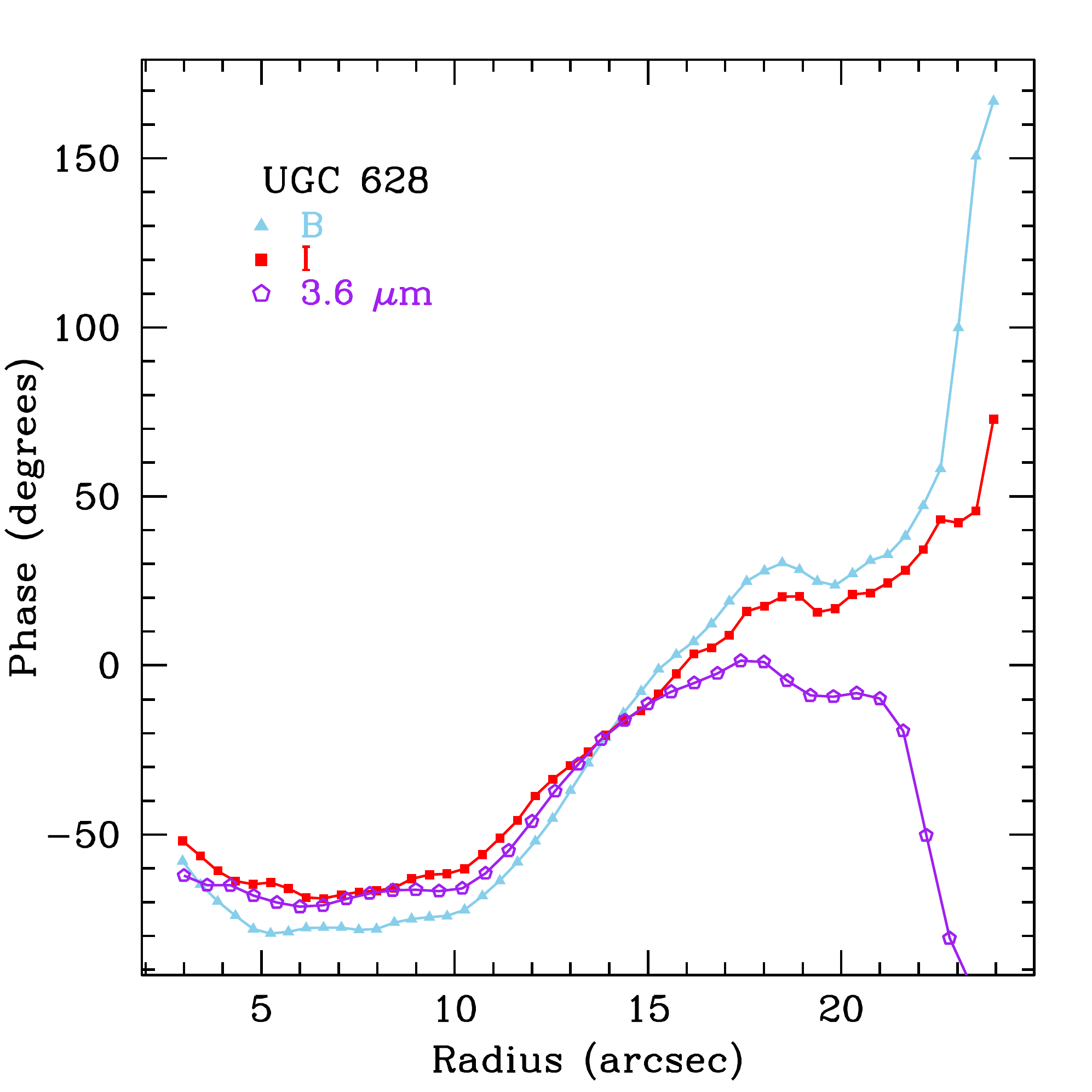}
  \caption{Phase profiles for UGC~628: \textit{B}-band (blue triangles), \textit{I}-band (red squares), and 3.6 $\mu$m (open purple pentagons).}
  \label{ugc628_phase}
\end{figure}

In Fig.~\ref{ugc628_phase} we show the phase profiles, calculated with equation 10, for all three bands for UGC~628. We find a \textit{B},\textit{I} phase crossing at 13.96$\arcsec \pm$0.46$\arcsec$ and a \textit{B},3.6 $\mu$m phase crossing at 13.96$\arcsec \pm$0.60$\arcsec$. We also find a second \textit{B},3.6 $\mu$m phase crossing at 30.36$\arcsec \pm$0.60$\arcsec$, but no second phase crossing between \textit{B} and \textit{I}. This second B,3.6 $\mu$m phase crossing is beyond the range plotted in Fig.~\ref{ugc628_phase}. Given the clear phase reversal present in both the \textit{B},\textit{I} profiles and \textit{B},3.6 $\mu$m profiles, we are confident in placing the corotation radius at 13.96$\arcsec \pm$0.46$\arcsec$ for the \textit{I}-band and 13.96$\arcsec \pm$0.60$\arcsec$ for 3.6 $\mu$m, listed in Table~\ref{res_table}. \textcolor{black}{The error on the corotation radius is equivalent to the radial spacing of the azimuthal light profiles.}

As an interesting aside, we can also use the phase profiles to determine if the spiral arms are leading or trailing (Fig.~1 in PD97). Based on the observed phase profiles (Fig.~\ref{ugc628_phase}), we find the shock front (\textit{B}-band) begins below the density wave (\textit{I}-band), intersects, and remains at greater phase for all radii. This is indicative of a Z-leading spiral pattern, consistent with our images of this galaxy.


\subsection{F568\nobreakdash-1}
\label{ssec:f56801_res}

F568\nobreakdash-1 shows clear dual spiral arm structure, with a small bar and bulge in the centre of the galaxy (Fig.~\ref{gal_ims}). As with UGC~628, we see that the spiral arms are stronger in the \textit{B}-band image. We see that the arms in the \textit{I}-band image are quite diffuse and less extensive than in \textit{B}. In both bands we see that the arms somewhat vanish before returning at a much fainter level (the south-western arm for instance). We also see that the bar in both bands is quite small angularly.

\subsubsection{Bar Radii}
\label{sssec:f56801_bar}

Our azimuthal light profiles for F568\nobreakdash-1 are shown in Fig.~\ref{f56801_az}. We find that F568\nobreakdash-1 displays a dual humped pattern in its azimuthal light profiles, indicating the presence of a bar. Curiously, we find that while the humps remain at a constant azimuthal separation in the \textit{I}-band with one another, they do so at 190 degrees as opposed to 180 degrees (top right panel of Fig.~\ref{f56801_az_bar}). In addition to the bar, our azimuthal light profiles also clearly show the presence of a dual spiral arm pattern, seen in the path of the humps after the bar radius in the lower left panel of Fig.~\ref{f56801_az_bar}. The spiral arms also appear to be 190 degrees out of phase with each other in the \textit{I}-band as well (top right panel). We do not find significant inner structure in this galaxy, with the humps remaining at quite constant angle within the bar (bottom left panel).

Using our azimuthal light profile method, we find a bar length of 4.37$\arcsec \pm$0.46$\arcsec$ in the \textit{I}-band and 4.80$\arcsec \pm$0.60$\arcsec$ in 3.6 $\mu$m for F568\nobreakdash-1, shown in Fig.~\ref{f56801_az_bar}. After the bar, there is a clear dual armed spiral pattern in this galaxy, seen in the bottom left panel where all four humps begin moving away from the centroid at the same rate. We find the intensity of the humps drops off rather significantly after the bar radius (bottom right panel). \textcolor{black}{We do note that the few points within the bar region indicate that we are pushing the limits of this method. However, given the very clear behaviour of the spiral arms seen in Fig.~\ref{f56801_az_bar}, the error on this bar radius measurement is two pixels in the \textit{I}-band (equivalent to 0.46$\arcsec$) and one pixel in 3.6 $\mu$m (equivalent to 0.60$\arcsec$).}

For our second technique, we show the relative Fourier amplitudes for F568\nobreakdash-1 in Fig.~\ref{f56801_amp}, and the Fourier bar/interbar intensities in Fig.~\ref{f56801_four}. We find strong $m = 2$\ and $m = 4$\ modes in the inner region of the galaxy, supporting the presence of a bar. We also find strong $m = 2$\ modes past $\sim$10$\arcsec$, an indicator of the grand-design spiral arms. Using the bar/interbar Fourier intensities, we find an \textit{I}-band bar radius of 7.76$\arcsec \pm$0.46$\arcsec$\ and a 3.6 $\mu$m bar radius of 7.56$\arcsec \pm$0.60$\arcsec$. Values have been trimmed past 13$\arcsec$\ in Fig.~\ref{f56801_four} as the strong even modes due to the spiral arms result in additional crossings far beyond the bar region. \textcolor{black}{As with UGC 628, the errors on this bar radius are equivalent to the radial spacing of the azimuthal light profiles.}

For our third technique, we show the radial plots of deprojected ellipticity for F568\nobreakdash-1 in Fig.~\ref{f56801_isobar}. After $\sim$26$\arcsec$, the \textcolor{black}{average} ellipticity of the isophotes is roughly zero in the three bands, indicating our deprojection was successful. \textcolor{black}{The ellipticity values in the \textit{I}-band oscillate between 0 and 0.2, likely due to the very faint outer disc. However, the average ellipticity is quite similar to the other two bands.} Using the radius of maximum ellipticity for elliptical isophotes, we find an \textit{I}-band bar radius of 8.44$\arcsec \pm$0.91$\arcsec$\ and a 3.6 $\mu$m bar radius of 6.60$\arcsec \pm$1.20$\arcsec$. As with UGC~628, the radius of maximum ellipticity is hard to measure for this galaxy due to the somewhat flat behaviour within the bar radius and beginning of the spiral arms. \textcolor{black}{The error is set to the radial spacing of \texttt{ELLIPSE}.}

These three techniques give us bar radii that range from $\sim$4.4$\arcsec$\ to $\sim$8.4$\arcsec$. Like with UGC~628 these values span a factor of two due to outliers. The bar radii are visually plotted over the deprojected \textit{I}-band and 3.6 $\mu$m images in Fig.~\ref{f56801_bar}. Here, the azimuthal light profile bar radius is consistent between the two bands, but the radius of maximum ellipticity differs between the two bands. The Fourier method gives consistent results between both bands, but like with UGC~628 it overshoots the bar radius and contains part of the spiral arms. Therefore, we take the bar radius derived from our azimuthal light profile method to be the bar radius in both \textit{I} and 3.6 $\mu$m, listed in Table~\ref{res_table}.

\subsubsection{Bar Strength}
\label{sssec:f56801_str}

We find lower limits on the bar strength to be 0.09 and 0.10 in \textit{I} and 3.6 $\mu$m, respectively. However, when decreasing the starting the starting radius from 3$\arcsec$\ to 1.5$\arcsec$\ we find our bar strengths increase to 0.13 and 0.14 in \textit{I} and 3.6 $\mu$m respectively (Table~\ref{res_table}). Due to the small angular size of the bar, we are not as confident that these are representative of the true strength as we were with UGC~628.

\subsubsection{Corotation Radii}
\label{sssec:f56801_corot}

We show the phase profiles for F568\nobreakdash-1 in Fig.~\ref{f568-1_phase}. We find three sets of clear phase crossings for F568\nobreakdash-1. The first \textit{B},\textit{I} phase crossing occurs at 5.86$\arcsec \pm$0.46$\arcsec$ and the first \textit{B},3.6 $\mu$m at 7.06$\arcsec \pm$0.60$\arcsec$. The second \textit{B},\textit{I} phase crossing occurs at 9.66$\arcsec \pm$0.46$\arcsec$ and the second \textit{B},3.6 $\mu$m phase crossing occurs at 9.66$\arcsec \pm$0.60$\arcsec$. The third occurs at 13.26$\arcsec \pm$0.46$\arcsec$ for \textit{B} and \textit{I} and at 13.46$\arcsec \pm$0.60$\arcsec$ for \textit{B} and 3.6 $\mu$m. Although the \textit{B} and \textit{I}-band phase profiles are quite close in value near the first phase crossing, this is simply due to the wide range of angles plotted (-150$\degr$ to 350$\degr$) causing the lines to appear close together and is similar to what is seen in PD97. In addition, the 3.6 $\mu$m profile exhibits clear phase reversal as well, leading us to conclude that the first phase crossing is the radius of corotation for both bands, listed in Table~\ref{res_table}. \textcolor{black}{As with UGC 628, the error on the corotation radius is equivalent to the radial spacing of the azimuthal light profiles.}

We also used these phase profiles to characterise the spiral pattern in F568\nobreakdash-1. We find the phase profiles of the shock front and density wave to follow that of an S-leading spiral pattern. This is consistent with our images of this galaxy.

\subsection{F568\nobreakdash-3}
\label{ssec:f56803_res}

F568\nobreakdash-3 poses a few complications when examining its morphology. Firstly, when looking at Fig.~\ref{gal_ims} we see that there is a very rough dual arm structure, with one clear, strong arm on the southern side of the galaxy and a weaker arm on the northern side. In addition, it also appears as though there may be a third weaker arm that begins on the eastern side of the bar. Whether this is a true third arm or whether it connects to the northern arm is unclear. \textcolor{black}{Secondly, the bar structure is not as straightforward as it may appear when Fig.~\ref{gal_ims} is scaled differently. As shown in Fig.~\ref{f56803_rescale}, there appears to be either a second bar or an inner spiral structure resembling a lightning bolt.} This inner structure is quite small, contained within the inner 3$\arcsec$\ and thus not probed by our azimuthal light profiles. What this means is that the main bar is not exactly symmetric, as it begins at the end of the inner bar. Finally, the southern arm appears to be quite a bit brighter than the northern arm, which may possibly influence our determination of the bar radius. 

\subsubsection{Bar Radii}
\label{sssec:f56803_bar}

The complicated morphology described above can be seen in our azimuthal light profiles for this galaxy in Fig.~\ref{f56803_az}. Here, the left hump is the southern arm of the bar/galaxy and shows the rather non-Gaussian shape of the arm, as well as how the two humps are not symmetric. We can see the two humps are not roughly 180 degrees out of phase (top right panel of Fig.~\ref{f56803_az_bar}), but closer to 170 degrees. The southern hump also does not appear Gaussian at all, preventing us from obtaining accurate centroids for this hump. However, the northern hump is well behaved, seen in the bottom left and right panels of Fig.~\ref{f56803_az_bar}. Here, we see that the northern humps remain quite stationary for the duration of the bar before dramatically moving. We also find that the intensity within the bar region is rather constant, and drops off significantly outside. 

Using our azimuthal light profile method, we find a bar length of 8.93$\arcsec \pm$0.92$\arcsec$ in the \textit{I}-band and 9.60$\arcsec \pm$1.2$\arcsec$ in 3.6 $\mu$m for F568\nobreakdash-3, shown in Fig.~\ref{f56803_az_bar}. Here we see that there is at least one clear spiral arm, noticeable as the movement away from the centroid after the denoted bar radii. The humps at smaller azimuthal angle appear to remain roughly constant after the bar radius, likely indicating the bright spiral arm seen in Fig.~\ref{gal_ims}. \textcolor{black}{Given that only one of the spiral arms is well behaved in Fig.~\ref{f56803_az_bar}, we assign an error of four pixels (equivalent to 0.92$\arcsec$) in the \textit{I}-band and two pixels (equivalent to 1.2$\arcsec$) in 3.6 $\mu$m.}

We show the relative Fourier amplitudes for F568\nobreakdash-3 in Fig.~\ref{f56803_amp}, and the Fourier bar/interbar intensities in Fig.~\ref{f56803_four}. We find strong $m = 2$\ and $m = 4$\ modes in the inner region, but also quite strong $m = 1$\ modes. This is likely due to the complicated morphology, and the very bright, southern arm of the galaxy, reinforced by the increase in the $m = 1$\ mode past $\sim$15$\arcsec$. Using the bar/interbar fourier intensities we find an \textit{I}-band bar radius of 13.96$\arcsec \pm$0.46$\arcsec$\ and a 3.6 $\mu$m bar radius of 18.16$\arcsec \pm$0.60$\arcsec$. \textcolor{black}{Again, the errors on the bar length from this method are equivalent to the radial spacing of our azimuthal light profiles.}

We show the radial plots of deprojected ellipticity for F568\nobreakdash-3 in Fig.~\ref{f56803_isobar}. We find the outer radii are consistent with ellipticities of zero, albeit with a large amount of scatter. Using the radius of maximum ellipticity for elliptical isophotes, we find an \textit{I}-band bar radius of 7.75$\arcsec \pm$0.68$\arcsec$\ and a 3.6 $\mu$m bar radius of 9.00$\arcsec \pm$1.20$\arcsec$. \textcolor{black}{The errors on this measurement are equivalent to the radial spacing from \texttt{ELLIPSE}.}

For F568\nobreakdash-3, we have bar radii that range from $\sim$7.8$\arcsec$\ to $\sim$18.2$\arcsec$, shown visually in Fig.~\ref{f56803_bar}. Once again, this large discrepancy can be attributed to the Fourier method. We find agreement between the azimuthal light profile and ellipticity methods, around $\sim$8$\arcsec$\ in \textit{I} and $\sim$9$\arcsec$\ in 3.6 $\mu$m, with the ellipticity method finding a slightly shorter bar in both bands. Again, we find the Fourier method to overshoot the bar radius, quite significantly in 3.6 $\mu$m. Here it almost entirely encompasses the bright portion of the southern arm. We take the bar radius from our azimuthal light profile method to be the bar radius for both bands, listed in Table~\ref{res_table}.

\subsubsection{Bar Strength}
\label{sssec:f56803_str}

We find bar strength lower limits of 0.18 in both \textit{I} and 3.6 $\mu$m. When decreasing the starting radius from 3$\arcsec$\ to 1.5$\arcsec$\ when constructing the azimuthal light profiles we find the bar strength increases only by 0.01 in both bands to 0.19 (Table~\ref{res_table}). Based on the behaviour of the relative Fourier amplitudes, we find these bar strengths are indicative of the true bar strength for F568\nobreakdash-3. 

\subsubsection{Corotation Radii}
\label{sssec:f56803_corot}

Our phase profiles for F568\nobreakdash-3 are shown in Fig.~\ref{f568-3_phase}. We find a first \textit{B},\textit{I} phase crossing at 10.06$\arcsec \pm$0.46$\arcsec$, and a second at 13.86$\arcsec \pm$0.46$\arcsec$. We find only one \textit{B},3.6 $\mu$m phase crossing at 13.86$\arcsec \pm$0.60$\arcsec$. While we do not find a phase crossing near 10$\arcsec$ between \textit{B} and 3.6 $\mu$m, it is a clear phase reversal for the \textit{B} and \textit{I}-bands, and cannot be ignored. A possible reason this phase crossing does not appear for the 3.6 $\mu$m is due to the large pixel scale. Regardless, if we only had the \textit{B} and \textit{I}-bands, the phase profile behaviour would be clear enough to trust the intersections. This gives a bar corotation radius of 10.06$\arcsec \pm$0.46$\arcsec$ for the \textit{I}-band. Because we do not see the phase crossing in 3.6 $\mu$m near 10$\arcsec$, we must report the corotation radius in 3.6 $\mu$m as 13.86$\arcsec \pm$0.60$\arcsec$. \textcolor{black}{Again, the errors on the corotation radius are equivalent to the radial spacing of the azimuthal light profiles.}

While there is only one \textit{clear} arm present in this galaxy, we also attempted to characterise the spiral pattern in F568\nobreakdash-3. We find the phase profiles of the shock front and density wave to follow that of an S-trailing spiral pattern. This is consistent with our images of this galaxy.

\subsection{F563\nobreakdash-V2}
\label{ssec:f563v2_res}

F563\nobreakdash-V2 shows a very clear bar in its morphology, but not clear spiral arms. There appears to possibly be one arm on the southern edge of the galaxy, however (see Fig.~\ref{gal_ims}). The structure in the \textit{I}-band image is quite diffuse, with not much of the disc visible outside of the bar. There may possibly be a second arm visible on the northern edge of the galaxy in the \textit{B}-band image. \textcolor{black}{This lack of well-defined arm/disc structure complicates the analysis of this galaxy, particularly the deprojection, as discussed in Sec.~\ref{ssec:reduc}.}

\subsubsection{Bar Radii}
\label{sssec:f563v2_bar}

Our azimuthal light profiles for this galaxy are shown in Fig.~\ref{f563v2_az}, showing the very bright bar. We find significant deviation from 180 degree separation between the two humps (top right panel of Fig.~\ref{f563v2_az_bar}). This is due to the northern hump in both bands, which remains at a constant angle for the range plotted. We therefore only show the movement of the southern hump, seen in the bottom left panel of Fig.~\ref{f563v2_az_bar}. Here, we can see the movement of the azimuthal centroids in both bands away from the bar, showing the path of the single arm in this galaxy. We also see the dramatic decrease in intensity after the bar radius (bottom right panel).

Using our azimuthal light profile method, we find a bar length of 6.65$\arcsec \pm$0.46$\arcsec$ in the \textit{I}-band and 7.20$\arcsec \pm$0.60$\arcsec$ in 3.6 $\mu$m for F563\nobreakdash-V2. \textcolor{black}{Even though we are only examining the azimuthal behaviour of one spiral arm, the very well behaved nature of the arm gives an error of two pixels in the \textit{I}-band (equivalent to 0.46$\arcsec$) and one pixel in 3.6 $\mu$m (equivalent to 0.60$\arcsec$).}

We show the relative Fourier amplitudes for F563\nobreakdash-V2 in Fig.~\ref{f563v2_amp}, and the Fourier bar/interbar intensities in Fig.~\ref{f563v2_four}. We find strong $m = 2$\ and $m = 4$\ modes in the inner regions, reinforcing the presence of a bar. Outside the bar, we see a dramatic increase in the strength of the odd modes, reinforcing the single arm in this galaxy. Using the bar/interbar Fourier intensities, we find a bar radius in the \textit{I}-band of 7.96$\arcsec \pm$0.46$\arcsec$\ and a 3.6 $\mu$m bar radius of 8.56$\arcsec \pm$0.60$\arcsec$. \textcolor{black}{As with the other three galaxies, the error on this bar length is equivalent to the radial spacing of our azimuthal light profiles.}

We show the radial plots of \textcolor{black}{deprojected} ellipticity for F563\nobreakdash-V2 in Fig.~\ref{f563v2_isobar}.  Using the radius of maximum ellipticity of elliptical isophotes, we find an \textit{I}-band bar radius of 5.02$\arcsec \pm$0.91$\arcsec$\ and a 3.6 $\mu$m bar radius of 9.60$\arcsec \pm$2.40$\arcsec$. \textcolor{black}{The errors on these measurements is equivalent to the radial spacing from \texttt{ELLIPSE}.}

We find bar lengths that range from $\sim$5$\arcsec$\ to $\sim$9.6$\arcsec$\ for F563\nobreakdash-V2, shown in Fig.~\ref{f563v2_bar}. The wide is due to the bar length from the elliptical isophote method. We find the azimuthal light profile to provide consistent bar radii between the two bands, 6.7$\arcsec$\ and 7.2$\arcsec$. The ellipticity method does not provide consistent results, with the 3.6 $\mu$m bar radius much longer than the \textit{I}-band radius, 5$\arcsec$\ vs. 9.6$\arcsec$. The Fourier method provides consistent results, both between bands, and roughly with our azimuthal light profile method, 8$\arcsec$\ and 8.5$\arcsec$. This is the sole exception for our sample, likely due to the lack of clear spiral arms in this galaxy. Again, we take the azimuthal light profile method bar radius to be the bar length for both bands, listed in Table~\ref{res_table}.

\subsubsection{Bar Strength}
\label{sssec:f563v2_str}

We find lower limits on the bar strengths to be 0.26 in \textit{I} and 0.23 in 3.6 $\mu$m. When decreasing the starting radius from 3$\arcsec$\ to 1.5$\arcsec$\ we find the bar strengths increased by 0.03 in both bands to 0.29 and 0.26 in \textit{I} and 3.6 $\mu$m respectively (Table~\ref{res_table}). Based on this and the behaviour of the relative Fourier amplitudes, we are confident that our lower limits are accurate.

\subsubsection{Corotation Radii}
\label{sssec:f563v2_corot}

We show the phase profiles for F563\nobreakdash-V2 in Fig.~\ref{f563-v2_phase}. We find a \textit{B},\textit{I} phase crossing at 15.86$\arcsec \pm$0.46$\arcsec$ and a \textit{B},3.6 $\mu$m phase crossing at 15.86$\arcsec \pm$0.60$\arcsec$. Due to the complicated morphology of this galaxy, the phase crossings are not as well defined as the previous galaxies. We report the bar corotation radius to be at 15.86$\arcsec \pm$0.46$\arcsec$ for the \textit{I}-band and at 15.86$\arcsec \pm$0.60$\arcsec$ for 3.6 $\mu$m. \textcolor{black}{Again, the errors on the corotation radius are equivalent to the radial spacing of the azimuthal light profiles.}

As F563\nobreakdash-V2 displays very tenuous spiral structure, we did not attempt to characterise the spiral pattern in this galaxy.

\section{Discussion}
\label{sec:discussion}

Here we discuss the results for our bar lengths and strengths for our galaxies, as well as describing how $\mathrm{R_{CR}}$\ can be used to obtain relative bar pattern speeds. We discuss our bar lengths in context of our three techniques, as well as between the two photometric bands used (Sec.~\ref{ssec:dis_len}). We discuss the accuracy of the lower limit on our bar strengths using the relative Fouirer amplitudes (Sec.~\ref{ssec:dis_stren}). Finally, we go into detail about the relative bar pattern speeds of our galaxies in context with their dark matter haloes (Sec.~\ref{ssec:dis_speed}). We compare our results to those of HSBs throughout as well.

\subsection{Bar Lengths}
\label{ssec:dis_len}

We have applied three different techniques for measuring the lengths of bars in our four galaxies. We find agreement between our azimuthal light profile bar radii measurements between the \textit{I}-band and 3.6 $\mu$m for all four galaxies. We also find that the bar length from the radius of maximum ellipticity mostly agrees with our azimuthal bar radius, although for F568\nobreakdash-1 this is not the case. Finally, we find that the Fourier method consistently overshoots the bar radius when compared with the other two methods, sometimes quite significantly. For UGC~628, for example, the Fourier bar radius is larger than our first corotation radius. In general, it appears that the Fourier method includes spiral arms as well as the bar, biasing the result to larger radii.

\begin{figure}
  \centering
  \includegraphics[scale = 0.4]{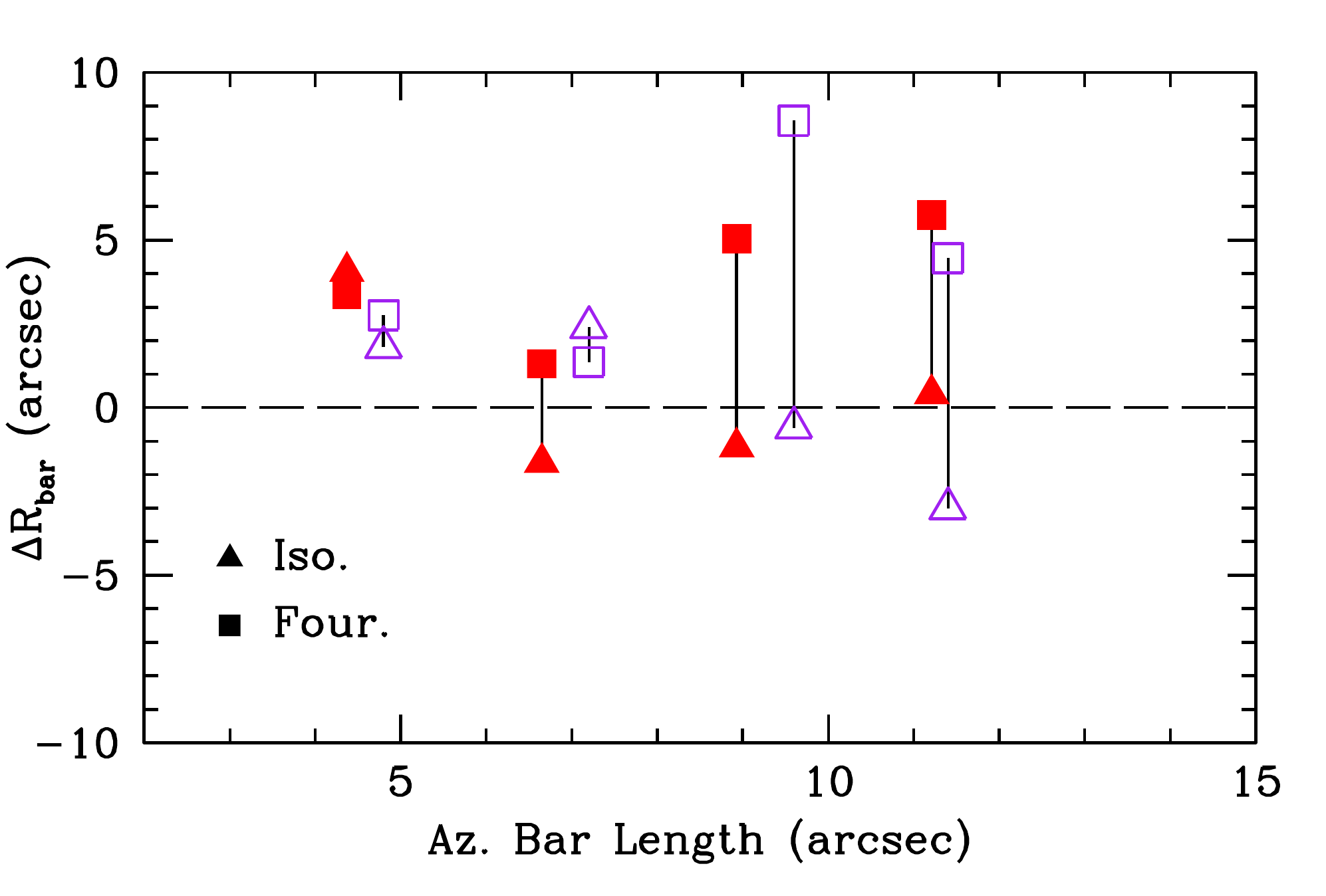}
  \caption{Bar radii measurement comparison between the three techniques for the entire sample. The x-axis is the bar radius from the azimuthal light profile method, and the y-axis is the difference from the other two techniques ($\Delta{\mathrm{R_{bar}}}$), the radius of maximum ellipticity and the Fourier bar/interbar intensities (i.e. Fourier - azimuthal). Triangle points are values from the elliptical isophote method and squares are from the Fourier method. Red, closed points are from the \textit{I}-band and purple, open points are from 3.6 $\mu$m. Vertical lines connect points for the same galaxy in each band.}
  \label{bar_len}
\end{figure}

We visually show how the bar length measurements compare for the whole sample in Fig.~\ref{bar_len}. Here, the x-axis is the bar length measured from our azimuthal light profile method and the y-axis is the difference from the other two techniques ($\Delta{\mathrm{R_{bar}}}$), the radius of maximum ellipticity and the Fourier bar/interbar intensities (i.e. $\mathrm{R_{bar,Four}} - \mathrm{R_{bar,az}}$). Triangle points are values from the elliptical isophote method and squares are from the Fourier method. The horizontal dashed line shows zero, and the vertical lines connect bar radii measurements for a given band. Red, closed points are \textit{I}-band values and purple, open points are 3.6 $\mu$m values.

We find that 3.6 $\mu$m (purple) azimuthal light profile bar radius measurements are consistently larger than the \textit{I}-band values (red) for each galaxy, while not differing too extremely. We expect them to be similar, as bars are stellar features and both our photometric bands trace this galaxy component. We also find that the elliptical isophotal bar measurements are equally smaller and larger than the azimuthal light profile method, and that the Fourier bar radius measurements are consistently larger. \textcolor{black}{\citet{aguerri2009} also found the Fourier method to overshoot the bar radius.}


When looking at the bar radii overplotted the deprojected images (Fig.~\ref{ugc628_bar}\ for instance), we find that our azimuthal light profile method provides the best measure of the bar radius for each galaxy in both bands. Based off of these figures, the elliptical isophote method provides a good measure for the bar radius as well, with F568\nobreakdash-1 being the exception. We believe the reason for this is due to transition from bar to spiral arms in this galaxy biasing the isophotes towards the bright arms.

As only F563\nobreakdash-V2 has comparable bar radii between the azimuthal light profile method and the Fourier method and has only one arm, we think that the dual spiral arms of the other three galaxies are biasing the bar radius measurement to larger values. Thus, it is possible given either angularly larger targets or better angular resolution that our bar radius measurements from the Fourier method would be more accurate for UGC~628, F568\nobreakdash-1, and F568\nobreakdash-3. We feel justified in saying this, as the Fourier method has been used to find accurate bar lengths in the literature, albeit for angularly large HSBs \citep[see][for instance]{aguerri2000}.

\begin{figure}
  \centering
  \includegraphics[scale=0.4]{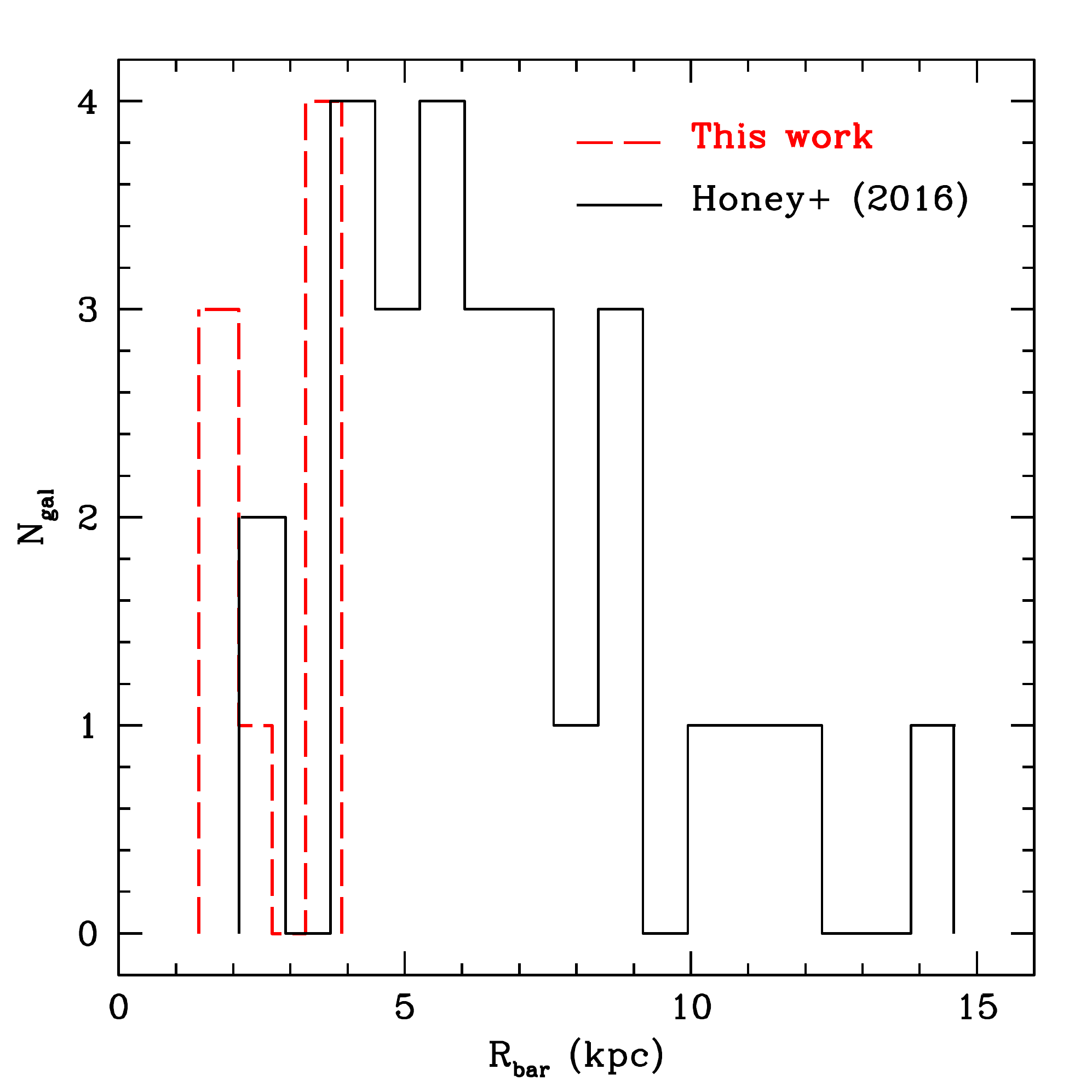}
  \caption{Histogram of bar lengths from \citet{honey2016} (solid black) and this work (dashed red). Here we include both our \textit{I}-band and 3.6 $\mu$m bar lengths. Our bar lengths clearly fall on the shorter side compared with the LSBs in \citet{honey2016}.}
  \label{lsb_len}
\end{figure}

We also find that the bar lengths in physical units (kpc) for our sample are on the lower end of those found by \citet{honey2016} (2.5 - 14.3 kpc), shown in Fig.~\ref{lsb_len}. Here we show histograms of the bar lengths (kpc) for \citet{honey2016} (solid black) and our \textit{I}-band and 3.6 $\mu$m values (dashed red). \textcolor{black}{Compared to HSB galaxies, however, we find that our bar lengths are of comparable size. For example, the majority of the bars in the HSB samples of \citet{erwin2005}, \citet{marinova2007}, and \citet{aguerri1998,aguerri2009} are $\leq$3.5 kpc, $\leq$5 kpc, and $\leq$5 kpc, respectively. In Fig.~\ref{str_len} we plot our bar lengths with the bar lengths of the HSBs in \citet{aguerri1998}.}


\subsection{Bar Strengths}
\label{ssec:dis_stren}

\begin{figure}
  \centering
  \includegraphics[scale=0.4]{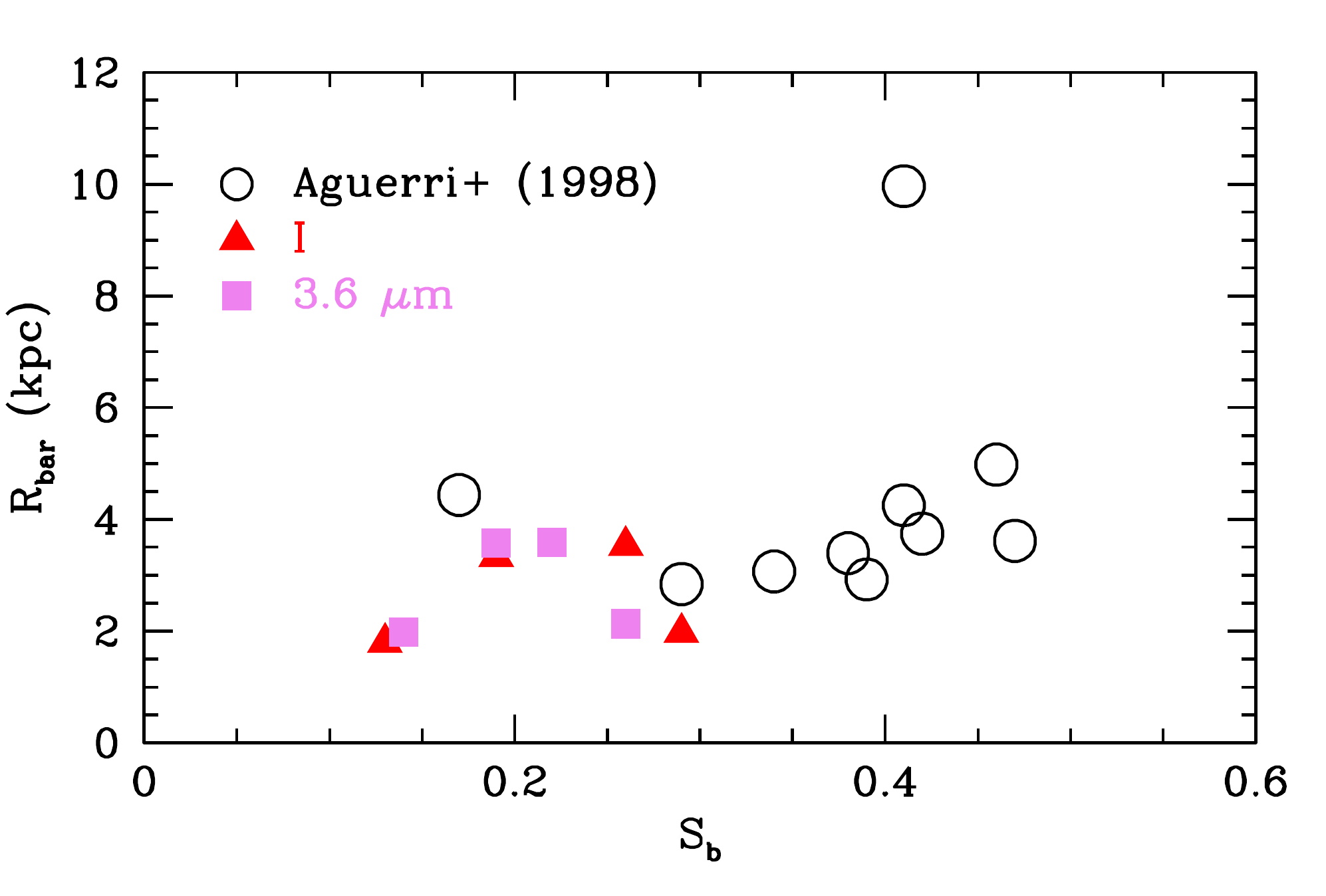}
  \caption{Comparison of bar strengths and bar lengths (kpc) between our \textit{I}-band (red triangles) and 3.6 $\mu$m (purple squares) values and those from HSBs in \citet{aguerri1998} (open black circles).}
  \label{str_len}
\end{figure}

We have found lower limits on the strengths of our bars and find that they range from $\sim$0.13 to $\sim$0.29. We find that our lower limit of bar strengths for our sample, while on the weaker side, are consistent with bar strengths found for HSBs. \citet{aguerri2000} and \citet{laurikainen2002} list bar strengths that range from $\sim$0.1 to $\sim$0.6 for example. This is also consistent with \citet{honey2016} who found that bar lengths and strengths for LSBs are similar to those in HSBs, although a different measure of bar strength was used than in our work. \textcolor{black}{We show this visually in Fig.~\ref{str_len}} where we compare our \textit{I}-band (red triangles) and 3.6 $\mu$m (purple squares) bar lengths and strengths with those from HSBs in \citet{aguerri1998} (open black circles). We find that our bar strength values form a continuum with the \citet{aguerri1998} HSB galaxies (the outlier at $\mathrm{R_{bar}} \approx$\ 10 kpc is not addressed).


Our bar strength values found for UGC~628, 0.26 and 0.22, are roughly consistent with those found by \citet{chequers2016}\ at late simulation time, $\sim$0.30, although our values are somewhat weaker. Based on the behaviour of the relative Fourier amplitudes, our bar strengths are likely indicative of the true strength.

F568\nobreakdash-1 has the lowest bar strengths of our sample, 0.13 and 0.14. Looking at Fig.~\ref{gal_ims}, this is not surprising, as the bar is quite small, and quickly turns into sprial arms. As previously mentioned in Sec~\ref{ssec:bar_strength}, bar strength was originally a visual classification based on how bright/pronounced the bar was. However, we are unable to probe down to small enough radii to be confident that our strengths are representative for this galaxy.

Our bar strengths found for F568\nobreakdash-3, 0.19 in both bands, are likely indicative of the true bar strength, as the bar is long enough for the missing, inner portion of the relative Fourier amplitudes to not be a large factor. It may seem surprising that the bar strength for this galaxy is as low as it is given the bar in Fig.~\ref{gal_ims}. However, F568\nobreakdash-3 seems to possess either a double bar, or inner spiral structure within the bar, which may be the cause of the lower strength.

As we are not missing a significant portion of the bar, our bar strengths for F563\nobreakdash-V2 are likely indicative of the true value, 0.29 and 0.26. However, we do note that when decreasing the starting radius of the azimuthal light profiles we found an increase of 0.03 in both bands. This leaves us with a relatively strong bar, which is not surprising given the \textit{I}-band image in Fig.~\ref{gal_ims}.

\subsection{Relative Bar Pattern Speeds}
\label{ssec:dis_speed}

The third bar parameter we measured was the radius of corotation ($\mathrm{R_{CR}}$). This parameter can be used to characterise the pattern speed of the bar, $\Omega_{p}$, which can give important information about the mass distribution within the host galaxy. $\Omega_{p}$\ is simply the rotational frequency of the bar itself \citep{binney2008}. The only method of directly measuring the bar pattern speed was introduced in \citet[][hereafter the TW method]{tremaine1984}. The TW method requires spectroscopy to map the velocity of orbits in the disc, either with multiple long-slit observations across the galaxy or with integral field unit (IFU) spectroscopy to create a velocity field. In addition, these velocity maps must be weighted by the underlying light distribution, measured either with the stellar continuum in the spectroscopy or with NIR imaging. This method has usually been applied to one, bright, grand-design spiral at a time \citep[i.e][]{kent1987,merrifield1995,gerssen1999}, although applications to dwarf galaxies exist as well \citep{bureau1999,banerjee2013}. However, \citet{aguerri2015} recently applied this method to 15 bright and strongly barred galaxies from the CALIFA \citep{sanchez2012} IFU survey.
 
While it is very desirable to obtain actual pattern speed measurements, the current number of LSBs with bar pattern speed measurements is only one: UGC~628. The bar pattern speed of this galaxy was measured by applying the TW method to an H$\alpha$\ velocity field \citep{chemin2009} as well as with numerical simulations \citep{chequers2016}. The lack of pattern speed measurements for LSBs is due to the large observing time required to obtain high signal to noise spectroscopy for LSBs. As a result, LSBs are largely absent from IFU surveys, such as the aforementioned CALIFA survey. However, large numbers of galaxies can be characterised by measuring the dimensionless parameter $\mathcal{R} = \mathrm{R_{CR}}/\mathrm{R_{bar}}$, or the `relative bar pattern speed', allowing us to probe LSBs in a more general context. Both the corotation radius ($\mathrm{R_{CR}}$) and bar radius ($\mathrm{R_{bar}}$) have been measured for our sample using photometry alone.

$\mathcal{R}$ can be measured simply with photometry (Sec.~\ref{sec:measurements}), and can be used to infer properties of the host galaxy. Bars are classified as `fast' if $\mathcal{R} < 1.4$ or `slow' if $\mathcal{R} > 1.4$ \citep{athanassoula1992, elmegreen1996, debattista2000}. \textcolor{black}{This designation arises from the prediction that galaxies dominated by baryons (as opposed to dark matter) in their central regions should have $\mathcal{R} = 1.2\pm0.2$\ \citep{athanassoula1992}}. In other words, corotation should occur very close to the end of the bar. Based on studies of HSBs, there seems to be little dependence on $\mathcal{R}$ with either morphology \citep{aguerri1998, aguerri2015} or redshift \citep{perez2012}, although none of these studies go to Hubble types later than SBbc. For those galaxies that have had $\mathcal{R}$\ measured, the almost unanimous result are `fast' bars, confirming the prediction of $\mathcal{R} = 1.2\pm0.2$. 

\textcolor{black}{Due to the relative lack of studies of barred LSBs, it is uncertain what values of $\mathcal{R}$\ are/will be observed for LSBs.  However, we have a few insights based on simulations and theory as to what to expect.  While there is debate about the likelihood of bars forming and surviving in very centrally dense haloes, the general consensus is that only slow, weak bars can form and that they will exist as transient features \citep[][but see \citealt{valenzuela2003}]{mihos1997, elzant2002, mayer2004}. It is expected that centrally dense dark matter haloes should be hosts to slow bars ($\mathcal{R}\ > 1.4$) because the bar should be dynamically slowed down by the dense halo \citep{weinberg1985, debattista2000}.}

As previously \textcolor{black}{introduced} in Sec.~\ref{sec:introduction}, \textcolor{black}{the kinematics of LSB galaxies are well-studied and have been used to measure the overall density and the density profile of the dark matter haloes in which they are embedded.} In general, two types of density profiles for the haloes are considered in these studies: \textcolor{black}{cuspy} profiles and \textcolor{black}{cored} profiles. \textcolor{black}{The most common form of a cuspy halo is the NFW profile \citep{navarro1996}.} It is characterised by a density power law $\rho \sim r^{\alpha}$\ and dark matter concentration $c$. \textcolor{black}{Haloes in cold dark matter simulations} generally have a slope of $\alpha = -1$, or steeper (i.e. `cusp'). \textcolor{black}{These haloes are very centrally dense.} \textcolor{black}{For cored haloes, the pseudo-isothermal density profile is the most commonly used.} These have a constant density centre, $\rho \sim r^{0}$ (i.e. `core'). \textcolor{black}{The central density of cored haloes is much lower than cuspy haloes. Cored haloes are motivated by observations rather than coming from simulations or theory.} 

\textcolor{black}{Observations of LSBs are typically better fit by cored haloes rather than cuspy haloes \citep[e.g.,][]{deblok1996a, deblok1996b, mcgaugh2000, swaters2003}. In addition, the overall halo density for LSBs is also lower than expected from simulations \citep[see Fig. 1b in ][for example]{kuzio2011a}} The conflict between kinematic observations being better fit by cores and simulations creating cusps is referred to as the `cusp-core' problem and is well documented in the literature \citep[see the review,][]{deblok2010}.

All four of our galaxies have kinematic data in the form of rotation curves, and have had mass models that confirm their dark matter domination \textcolor{black}{down to small radii} \citep{deblok2001a, mcgaugh2001, kuzio2006, kuzio2008}. Rotation curves of LSBs are better explained by pseudo-isothermal haloes rather than NFW haloes, suggesting the density profile of the dark matter is `cored'. \textcolor{black}{These studies find the data to be better explained by cored haloes rather than very centrally dense cuspy haloes.  Now that we have measured the length and corotation radius of the bar, we can use the ratio of these two parameters to determine the relative bar pattern speed ($\mathcal{R}$) and probe the dark matter halo in a new way.}

\begin{figure*}
  \centering
  \includegraphics[scale=0.8]{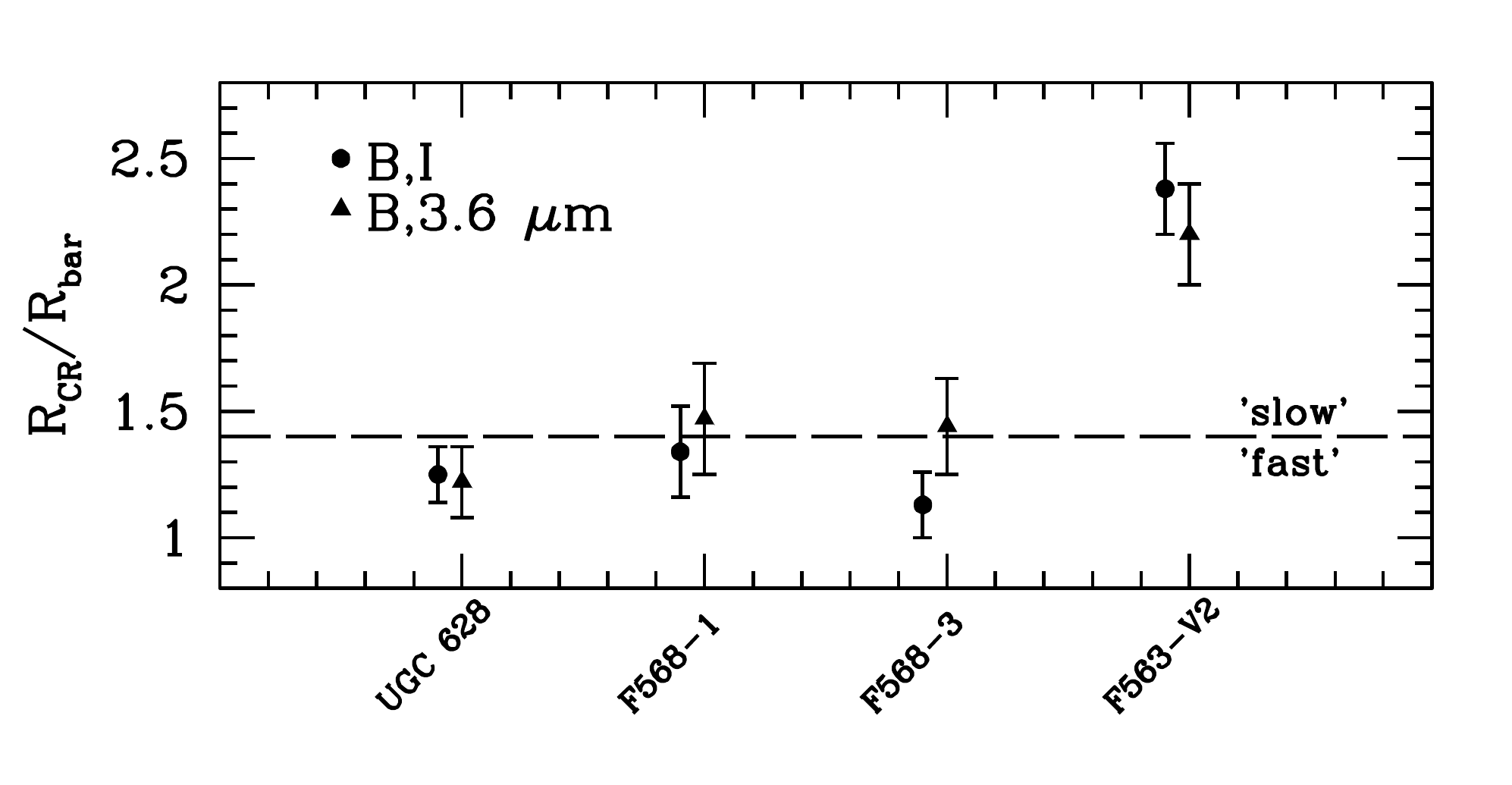}
  \caption{Relative bar pattern speeds for our sample. From left to right: UGC~628, F568-1, F568-3, F563-V2. Filled circles show the \textit{B},\textit{I} values and filled triangles show the \textit{B},3.6 $\mu$m values. Horizontal dashed line indicates the distinction between `fast' and `slow' bars. If LSBs reside in \textcolor{black}{centrally dense (i.e. cuspy) haloes} as expected, they should reside in the upper portion of this plot. `Normal' HSBs almost exclusively live in the bottom portion of the figure, as they are baryon dominated in their inner regions.}
  \label{speed}
\end{figure*}

Using the azimuthal bar radius and \textit{first} phase crossing after the bar radius using the PD97 method, our relative bar pattern speeds ($\mathcal{R}$) are shown in Table~\ref{res_table}. \textcolor{black}{Here we have propagated the bar length and corotation radius errors through to determine the error on $\mathcal{R}$.} The relative bar pattern speeds for our sample are shown in Fig.~\ref{speed}. With the exception of F568\nobreakdash-3, we find excellent agreement between the relative bar pattern speeds of both photometric bands. We note that we are comparing two different corotation radii for F568\nobreakdash-3 between the \textit{I}-band and 3.6 $\mu$m. Excluding F563\nobreakdash-V2, our relative bar pattern speeds span values from 1.13 to 1.47, \textcolor{black}{suggesting that these galaxies are not embedded in dark matter haloes with high central densities}. While a few values straddle the line between fast and slow with large errors, the values are still comparable to those found for HSBs (i.e. fast) with F563\nobreakdash-V2 being the clear exception. Since we find such a slow bar in F563\nobreakdash-V2, we are confident that we are able to differentiate between fast and slow bars.


\begin{figure}
  \centering
  \includegraphics[scale=0.4]{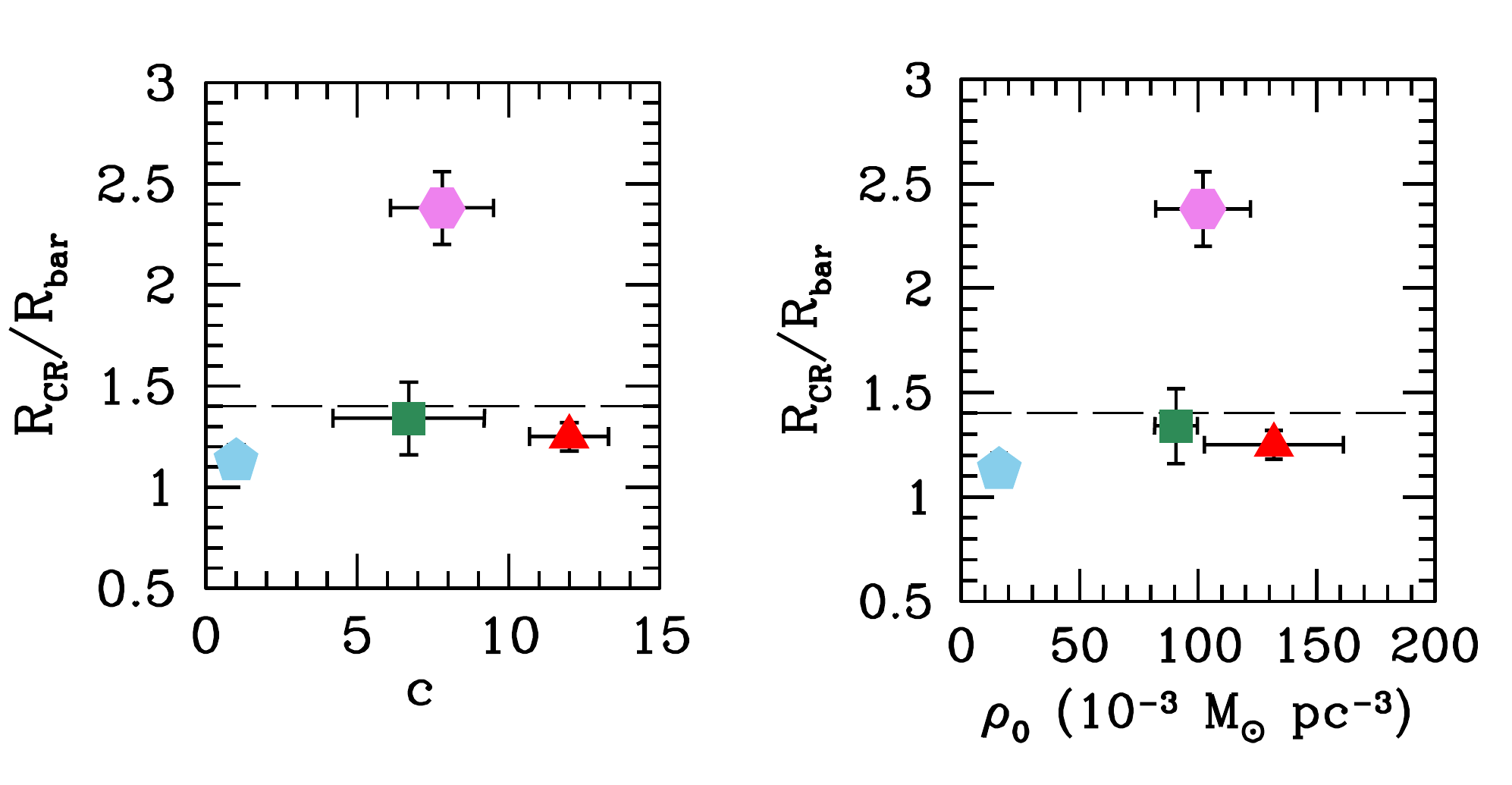}
  \caption{Relative bar pattern speed plotted versus NFW concentration \textit{c} (left) and pseudo-isothermal central density $\rho_{0}$\ (right), both halo fits from population synthesis models. Red triangles are UGC~628, green squares are F568-1, blue pentagons are F568-3, and purple hexagons are F563-V2. Halo parameters for UGC~628 are taken from \citet{deblok2002}; parameters for F568-1 are from \citet{deblok2001b}; parameters for F568-3 and F563-V2 are from \citet{kuzio2008}. NFW parameters for F568-3 are not fits, but held fixed. The horizontal dashed line shows $\mathcal{R} = 1.4$, the deliminator between fast and slow bars.}
  \label{halo_speed}
\end{figure}

As previously mentioned in Sec.~\ref{ssec:sample}, our four galaxies have kinematic data in the literature and were part of dark matter halo studies. In Fig.~\ref{halo_speed} we show the relative bar pattern speeds versus the NFW concentration $c$\ and pseudo-isothermal central density $\rho_{0}$. Red triangles in Fig.~\ref{halo_speed} are UGC~628, green squares are F568\nobreakdash-1, blue pentagons are F568\nobreakdash-3, and purple hexagons are F563\nobreakdash-V2. Halo parameters for UGC~628 are taken from \citet{deblok2002}, parameters for F568\nobreakdash-1 are from \citet{deblok2001b}, and parameters for F568\nobreakdash-3 and F563\nobreakdash-V2 are from \citet{kuzio2008}. The NFW fits to F568\nobreakdash-3 \citep{kuzio2008} were held fixed unlike the other three galaxies, because a fit to the data could not be made. \textcolor{black}{In each of these studies, R-band photometry from \citet{deblok1995} and \citet{swaters1999} and stellar M/L ratios from population synthesis models have been used to determine the contribution of the stars to the total mass of each system.} The horizontal dashed line shows $\mathcal{R} = 1.4$. In our small sample we do not find any correlation between the dark matter halo parameters and the relative bar pattern speeds.


\begin{figure}
  \centering
  \includegraphics[scale=0.4]{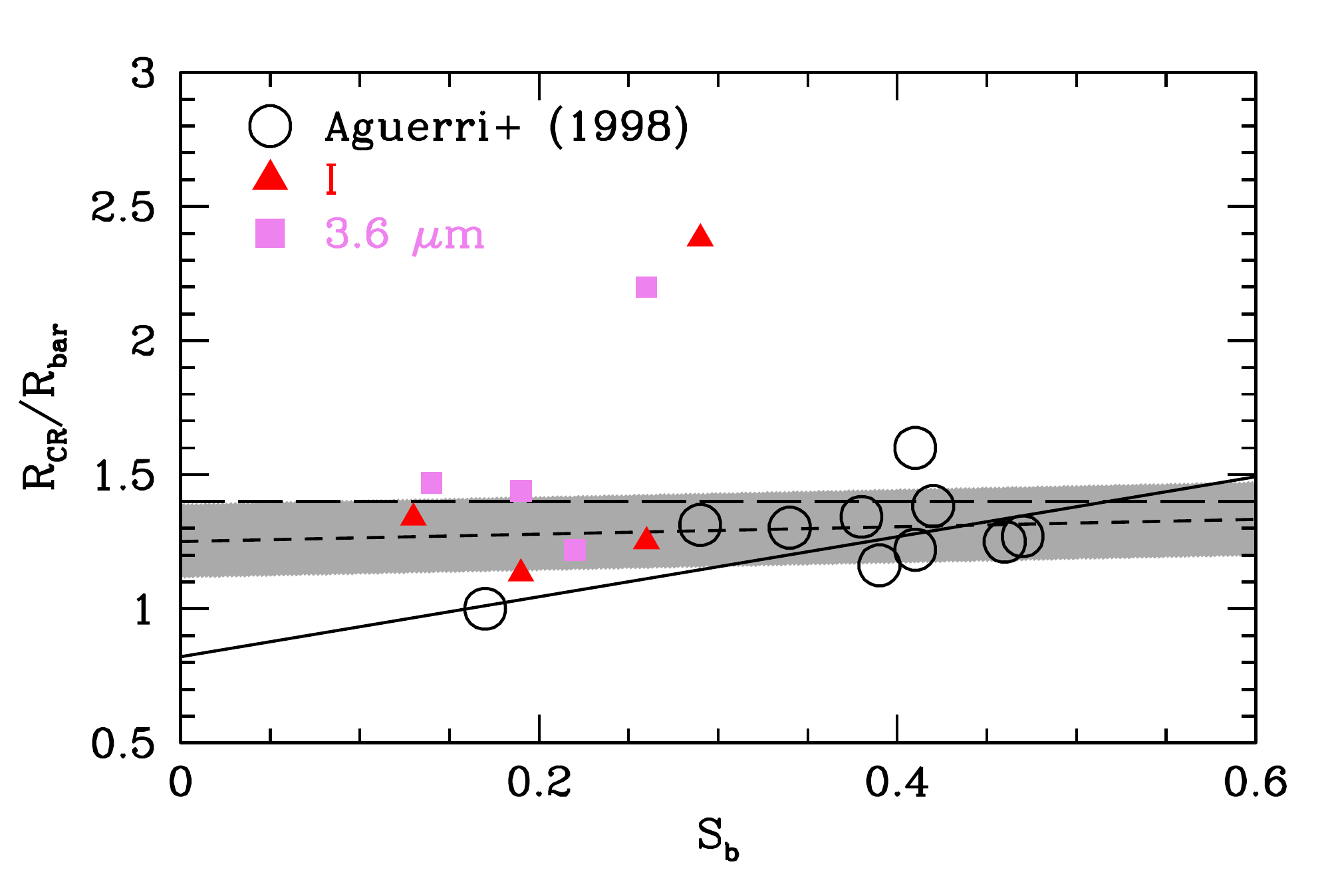}
  \caption{Same as Fig.~\ref{str_len}, but with relative bar pattern speed instead of length. The solid line is the fit to the HSBs, found by \citet{aguerri1998}, and the horizontal long dashed line shows the deliminator between fast and slow bars. The short dashed line is a fit to the HSBs and LSBs, excluding F563-V2 (the two points above the rest): $\mathcal{R} = 1.25 + 0.14S_{b}$. \textcolor{black}{The shaded gray region denotes the scatter about the fit: $\sigma$ = 0.137.}}
  \label{str_speed}
\end{figure}

We show our relative bar pattern speeds \textcolor{black}{versus strength} compared with the HSB galaxies from \citet{aguerri1998} in Fig.~\ref{str_speed}. The black line indicates the fit from \citet{aguerri1998}, $\mathcal{R} = 0.82 + 1.12S_{b}$. We find that our galaxies all fall above this relation. Excluding F563\nobreakdash-V2, the two points far above the rest, it appears as though there is no correlation with relative bar pattern speed and bar strength for both LSBs and HSBs. In fact, were it not for the HSB at $S_{B} \approx 0.18$, the \citet{aguerri1998} sample would be roughly flat as well. To test this, we fit both the LSB and HSB datasets (excluding F563\nobreakdash-V2) and found $\mathcal{R} = 1.25 + 0.14S_{b}$\ \textcolor{black}{with a scatter $\sigma = 0.137$ shown as the short dashed line and gray shaded band in Fig.~\ref{str_speed}.}


In the following sections we discuss the relative bar pattern speeds, $\mathcal{R}$, for each galaxy individually in context with any previous work, as well as with previous dark matter studies.

\subsubsection{Comments on UGC~628}
\label{sssec:ugc628_dis}

Our relative bar pattern speeds for UGC~628 are $\mathcal{R}_{I} = 1.25\pm0.11$\ and $\mathcal{R}_{3.6} = 1.22\pm0.14$, clearly indicating a fast bar, suggesting the dark matter halo is not \textcolor{black}{as centrally dense as would be expected for a cuspy halo}. \citet{deblok2002} find that both NFW halo and pseudo-isothermal fits to the H$\alpha$\ rotation curve of UGC~628 provide similar results. In addition, they find that UGC~628 has a steep, cuspy inner slope ($\alpha \sim$\ -1.3). However, \citet{deblok2005} \textcolor{black}{show how cuspy haloes do not provide realistic or physically meaningful fits to the data for UGC 628 (and many other LSBs).}

Our relative pattern speeds are in conflict with \citet{chemin2009} and \citet{chequers2016}, who both find a slow bar ($\mathcal{R} \sim$ 2). As previously mentioned, \citet{chemin2009} applied the TW method to an H$\alpha$\ velocity field to measure the bar pattern speed, and \citet{chequers2016} used an N-body simulation to model the bar pattern speed. We think this conflict ultimately stems from the measure of the bar radius. Both of these previous works place the bar radius at $\sim$14.5$\arcsec$\ in the \textit{R}-band, far longer than our value of $\sim$11.5$\arcsec$. Looking at Fig.~\ref{ugc628_bar}, a bar radius of $\sim$14.5$\arcsec$\ would be more similar to the bar radii from the Fourier intensities method and would contain a portion of the spiral arms. In addition, our radius of corotation is $\sim$14$\arcsec$, falling \textit{inside} the bar for these other two works. However, corotation cannot occur within the bar, since these orbits would be unstable and eventually cause the bar to be destroyed \citep{contopoulos1980}. We also note that we do find the radius of corotation reported by \citet{chemin2009} in 3.6 $\mu$m, $\sim$30$\arcsec$. If we use this value for R$_{\mathrm{CR}}$, we find $\mathcal{R} >$\ 2, which brings us into agreement with \citet{chemin2009} and \citet{chequers2016}.

We also emphasize that the result from the \citet{chequers2016} modeling \textit{assumes} a \textcolor{black}{centrally-dense,} cuspy NFW dark matter halo for their simulation, so it is not unexpected that a slow bar was found. This assumption is contrary to previous works that found UGC~628 to not be well fit by an NFW halo \citep[e.g. poor fits in][]{deblok2001a, deblok2005}. Based on our results, we reclassify UGC~628 as having a fast bar. \citet{chequers2016} find that despite the slow pattern speed, their model of UGC~628 implies that the galaxy is not dark matter dominated in the inner radii, as found by \citet{chemin2009}. This would be consistent with our finding of a fast bar in this galaxy.

\subsubsection{Comments on F568\nobreakdash-1}
\label{sssec:f56801_dis}

Because we are confident that the first phase crossing for both \textit{I} and 3.6 $\mu$m are real, we have relative bar pattern speeds of $\mathcal{R}_{I} = 1.34\pm0.18$\ and $\mathcal{R}_{3.6} = 1.47\pm0.22$, respectively. This leaves us with results that span both the fast and slow regimes for F568\nobreakdash-1. This is likely due to the large errors found for this galaxy, resulting from the angularly small bar radius. 

In terms of mass profiles for this galaxy, \citet{deblok2001a} find that the pseudo-isothermal halo profile is the best fit to the rotation curve for F568\nobreakdash-1, and that the NFW profiles all over predict the rotation curve in the inner and outer regions. \citet{swaters2000} also found a relatively flat inner density slope, $\alpha \sim 0.3$, suggesting a cored halo. According to \citet{deblok2001a}, the maximum disc profile is the best scenario for the NFW fits, albeit still worse than the pseudo-isothermal haloes. In addition, \citet{fuchs2003} showed that the maximum disc hypothesis holds for dual-armed spiral LSBs; F568\nobreakdash-1 was included in this work. \citet{lee2004} find that a bottom heavy IMF in combination with very recent star formation can explain the results from \citet{fuchs2003}, lowering the required mass of the dark matter halo significantly.

\textcolor{black}{When considering the errors, the relative bar pattern speeds calculated for F568\nobreakdash-1 span the range of 1.15 to 1.69, bracketing the $\mathcal{R} = 1.4$ dividing line between the fast and slow regimes.} The only way for us to obtain unambiguously high relative bar pattern speeds would be to ignore the first phase crossing for both \textit{I} and 3.6 $\mu$m and take the second intersection of 9.66$\arcsec$\ in both bands to be the corotation radius. This would give relative bar pattern speeds of $\mathcal{R}_{I} = 2.21\pm0.11$\ and $\mathcal{R}_{3.6} = 2.01\pm0.28$. Finding pattern speeds that are not clearly slow or clearly fast is not uncommon for the PD97 method; \citet{sierra2015} for instance find roughly ten of the 57 galaxies in their sample have relative bar pattern speeds that span both fast and slow within their errors. All galaxies in \citet{sierra2015}, \textcolor{black}{however, are baryon-dominated HSBs that display little need for dark matter until large radii (the flat part of their rotation curves).  In contrast, LSBs require dark matter to become a significant mass component much farther in.  That the bars in UGC 628, F568-1, and F568-3 (see below) do not reside well inside the slow regime is perhaps suggestive that these galaxies do not contain as much dark matter near their centers as would be expected for systems inside centrally dense cuspy dark matter haloes.}

\subsubsection{Comments on F568\nobreakdash-3}
\label{sssec:f56803_dis}

For F568\nobreakdash-3, we are left with two very different results, both of which depend on the corotation radius that is used. If we take the first phase intersection between the \textit{B} and \textit{I}-bands, then we find that $\mathcal{R}_{I} = 1.13\pm0.13$, clearly indicating a fast bar. However, this phase intersection is not seen in 3.6 $\mu$m. If we take the corotation radius as the second phase intersection between \textit{B} and \textit{I}, which is in agreement with the \textit{B},3.6 $\mu$m phase crossing, then we have $\mathcal{R}_{I} = 1.55\pm0.17$, or a slow bar. To further complicate things, the phase intersection between \textit{B} and 3.6 $\mu$m gives $\mathcal{R}_{3.6} = 1.44\pm0.19$, spaning both fast and slow.

Because the phase reversal between the \textit{B} and \textit{I}-bands is quite clear, it is very likely real, as justified in Sec.~\ref{sssec:f56803_corot}. We therefore trust that our results for the \textit{B} and \textit{I} phase crossing are more certain than the \textit{B} and 3.6 $\mu$m phase crossings. Thus we conclude that F568\nobreakdash-3 is more likely to have a fast bar, suggesting the dark matter halo is not \textcolor{black}{centrally dense}. F568\nobreakdash-3 is also present in the \citet{fuchs2003} and \citet{lee2004} samples, suggesting that the stellar component in this galaxy may be more massive than previously thought. This is also supported by the complex morphology that is present in the inner radii for this galaxy. If the dark matter halo for F568\nobreakdash-3 were \textcolor{black}{centrally dense}, then it would be very unlikely to find the inner spiral structure and bar as these \textit{global} features would be smoothed out by the dark matter halo \citep{mihos1997}.

The NFW profile rotation curve fitting from \citet{deblok2001a} and \citet{kuzio2008} for this galaxy were unable to converge, forcing the authors to hold best guess values fixed by hand. In addition, \citet{swaters2000} found a flat inner density slope, $\alpha \sim 0.2$, inconsistent with a cuspy halo.

\subsubsection{Comments on F563\nobreakdash-V2}
\label{sssec:f563v2_dis}

Based off of the bar radii and corotation radii from both bands, F563\nobreakdash-V2 clearly has a slow bar. For the \textit{I}-band we have $\mathcal{R}_{I} = 2.38\pm0.18$, and $\mathcal{R}_{3.6} = 2.20\pm0.20$\ for 3.6 $\mu$m. Our azimuthal light profiles reveal there to be only one strong arm, with possibly a fainter arm as well. \textcolor{black}{As described in Sec.~\ref{ssec:reduc}, this prevented  us from being able to use ELLIPSE to determine the deprojection  and we relied instead on values from the literature.} In addition, the sole-arm could pose issues for the PD97 method, which selects the $m = 2$\ mode for the Fourier transform. This likely explains the large errors on our $\mathcal{R}$\ values. Due to this, the phase profiles for this galaxy are not as clear as the rest of our sample. Somewhat supporting this, F563\nobreakdash-V2 is a clear outlier \textcolor{black}{when looking at Fig.~\ref{str_speed}}, seen as the two points above the rest. This figure suggests that there is no real correlation between bar strength and pattern speed.

Based off of the rotation curve fitting from \citet{deblok2001a}, the NFW profile does much worse than the pseudo-isothermal profile. This is supported both by the overestimating of the inner portions of the rotation curve as well as the relatively flat inner slope of the density profile, $\alpha \sim$\ 0, from \citet{swaters2000}, suggestive of a cored halo. However, our relative bar pattern speed for F563\nobreakdash-V2 would be in agreement with a \textcolor{black}{very centrally dense} dark matter density profile.

\section{Conclusions}
\label{sec:conclusions}

We have used multi-band photometry to characterise bars in four low surface brightness galaxies. In general, we find that techniques used for high surface brightness galaxies can be successfully applied to LSBs. We list our conclusions here.

\begin{enumerate}
\item Bar morphology in LSBs is quite complicated, allowing for structures such as double bars or inner spirals within the bar region, as well as being generally lopsided features. 
\item We find measuring bar properties in LSBs with poorly-defined spiral arms is more difficult than in those with a clear dual-arm structure because of the uncertainties associated with the disc deprojection.
\item \textcolor{black}{The bars in our sample have comparable lengths to those found in late-type HSB galaxies. We require a larger sample to know if this holds for LSBs in general or not.}
\item The bars in our LSBs appear to have weaker bars than HSBs on average, but not drastically so.
\item Excluding F563\nobreakdash-V2, our barred LSBs appear to have relative bar pattern speeds that are comparable to HSBs, with the majority of our measurements consistent with fast bars. This implies that the dark matter haloes of our LSBs \textcolor{black}{are unlikely to be centrally dense (i.e. cuspy)}. Since we find such a slow bar in F563\nobreakdash-V2, however, we are confident in differentiating between fast and slow bars for our sample.
\item We reclassify UGC~628 as having a fast bar due to both our clear phase intersections and shorter bar radius than previous works.
\item We find no correlation between bar strength and relative bar pattern speed when including measurements from HSBs.
\end{enumerate}

Our initial findings that LSB bars are fast is intriguing, but we require a much larger sample before we can say anything about the general population as a whole. Observations of nearly two-dozen more barred LSBs are currently underway and the results from those studies will be combined with our findings here to give a more clear and wide-ranging picture of LSB galaxies.

\section*{Acknowledgements}

\textcolor{black}{We would like to thank the referee for their constructive comments which helped to clarify and strengthen our paper. We would also like to thank Dr. Marc Seigar and Dr. Amber Sierra for their very helpful discussions regarding the methods used in this paper, as well as Russet McMillan from Apache Point Observatory for her help in taking calibration data for the ARCTIC imager.}

This research has made use of the NASA/ IPAC Infrared Science Archive, which is operated by the Jet Propulsion Laboratory, California Institute of Technology, under contract with the National Aeronautics and Space Administration.


\appendix

\section{F568\nobreakdash-1}

\begin{figure}
  \centering
  \includegraphics[scale=0.35]{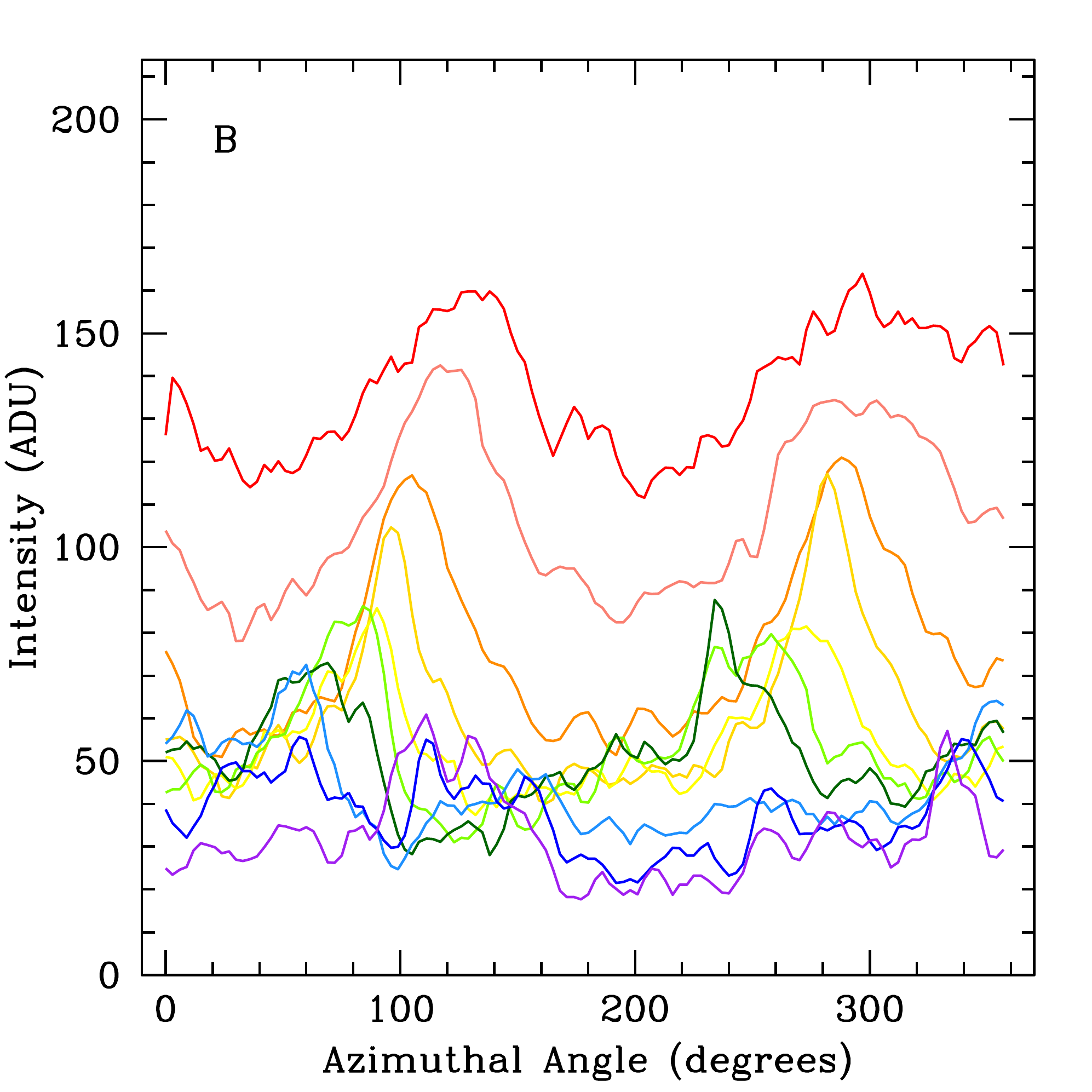}
  \includegraphics[scale=0.35]{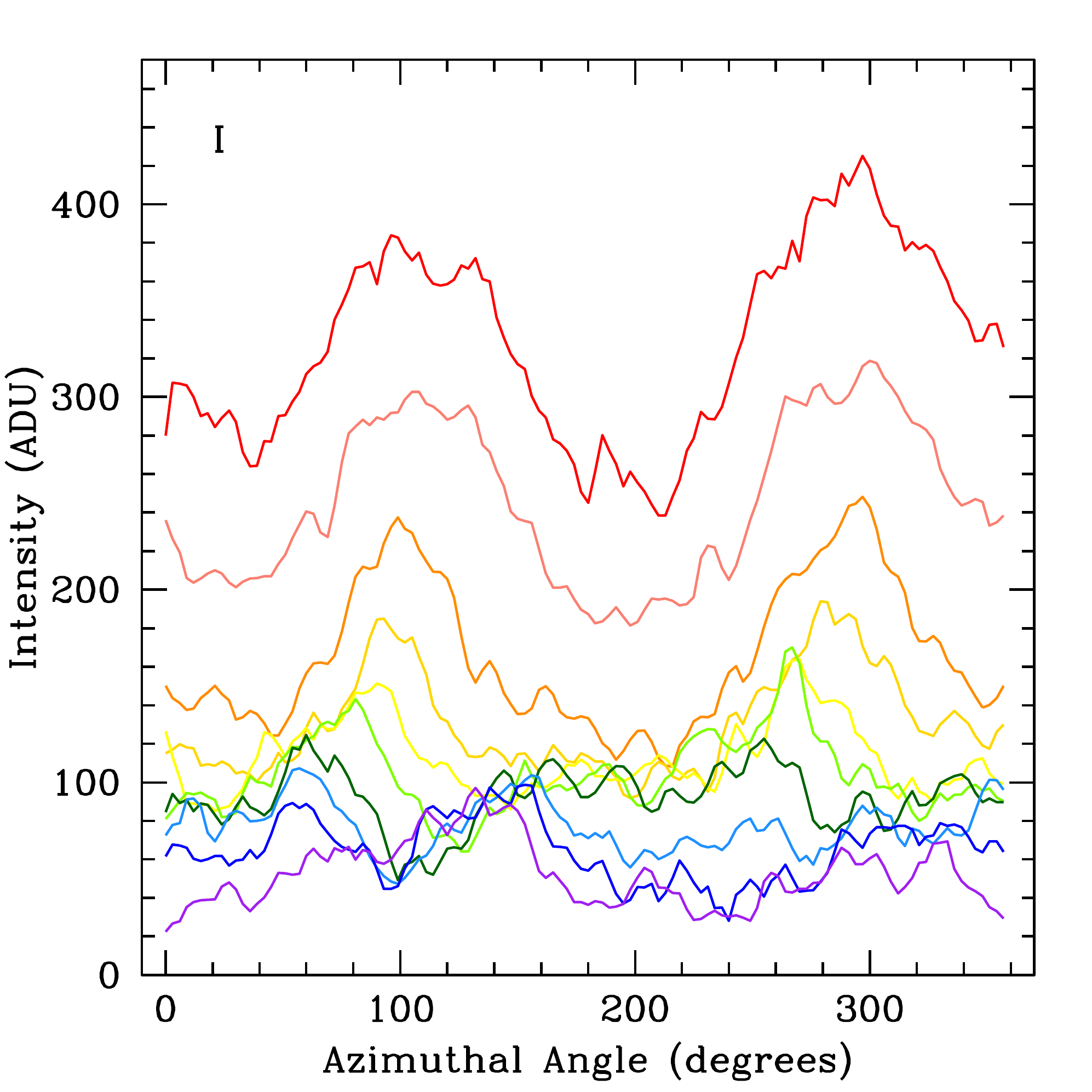}
  \includegraphics[scale=0.35]{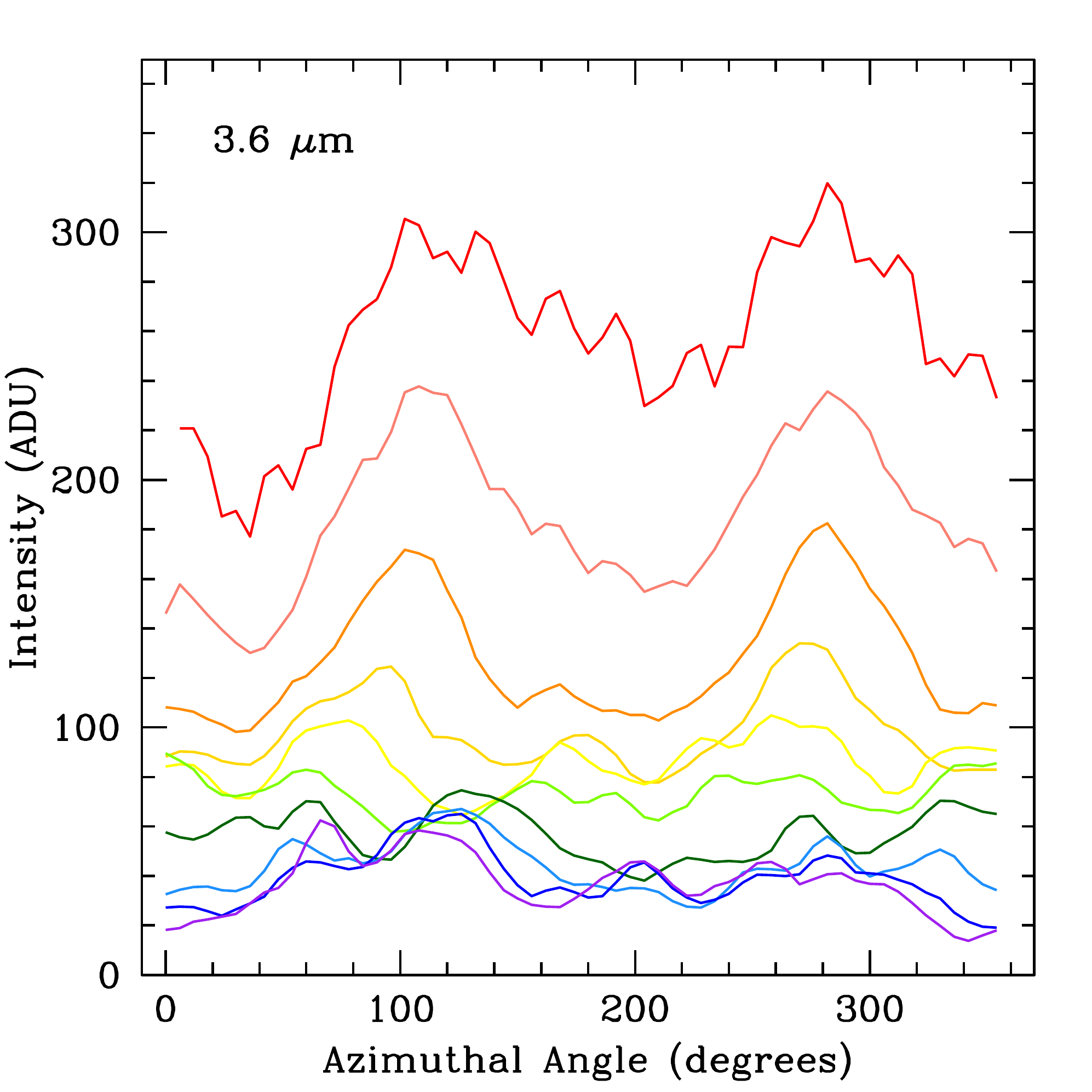}
  \caption{Same as Fig.~\ref{ugc628_az}, but for F568-1.}
  \label{f56801_az}
\end{figure}

\begin{figure}
  \centering
  \includegraphics[scale=0.4]{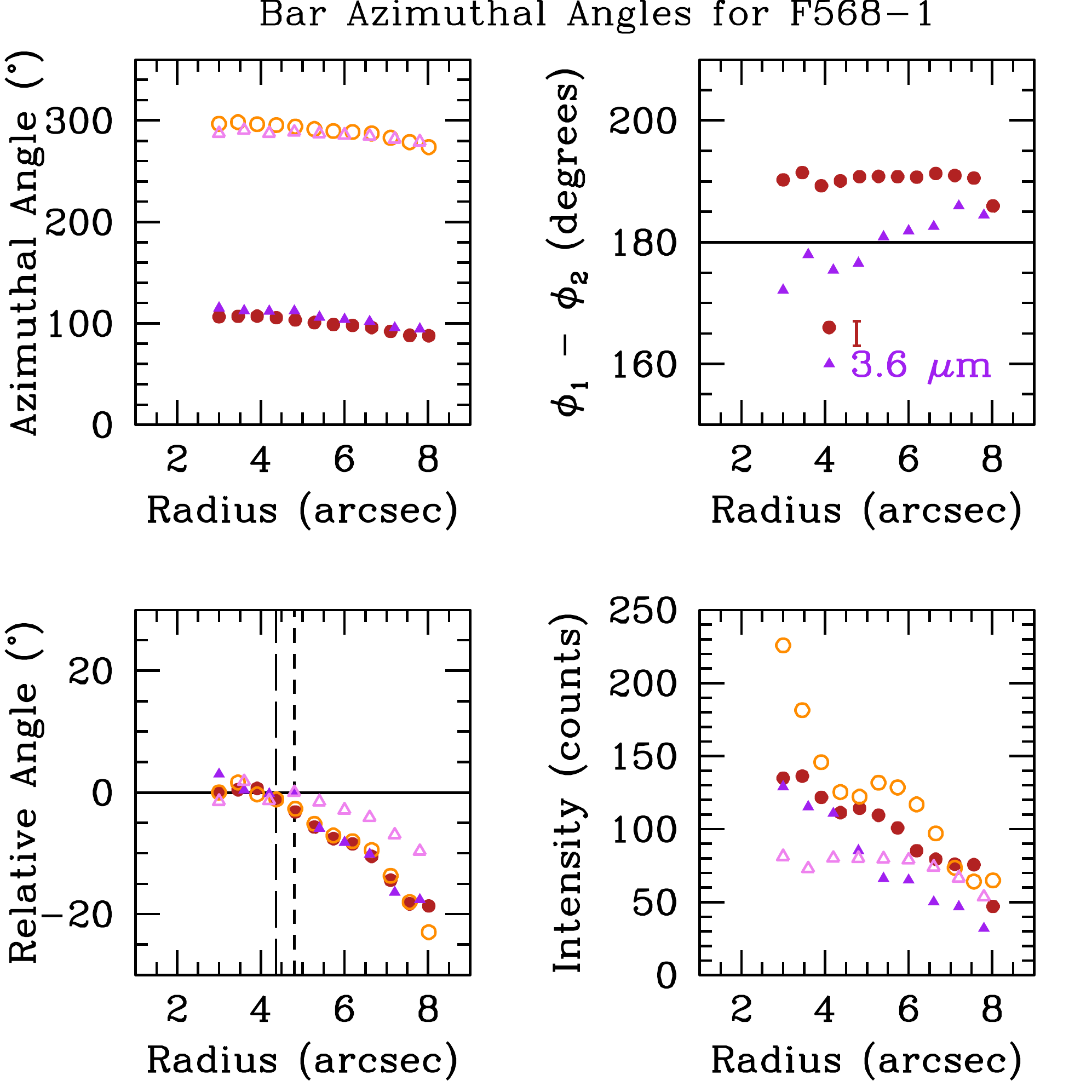}
  \caption{Same as Fig.~\ref{ugc628_az_bar}, but for F568-1.}
  \label{f56801_az_bar}
\end{figure}

\begin{figure}
  \centering
  \includegraphics[scale=0.4]{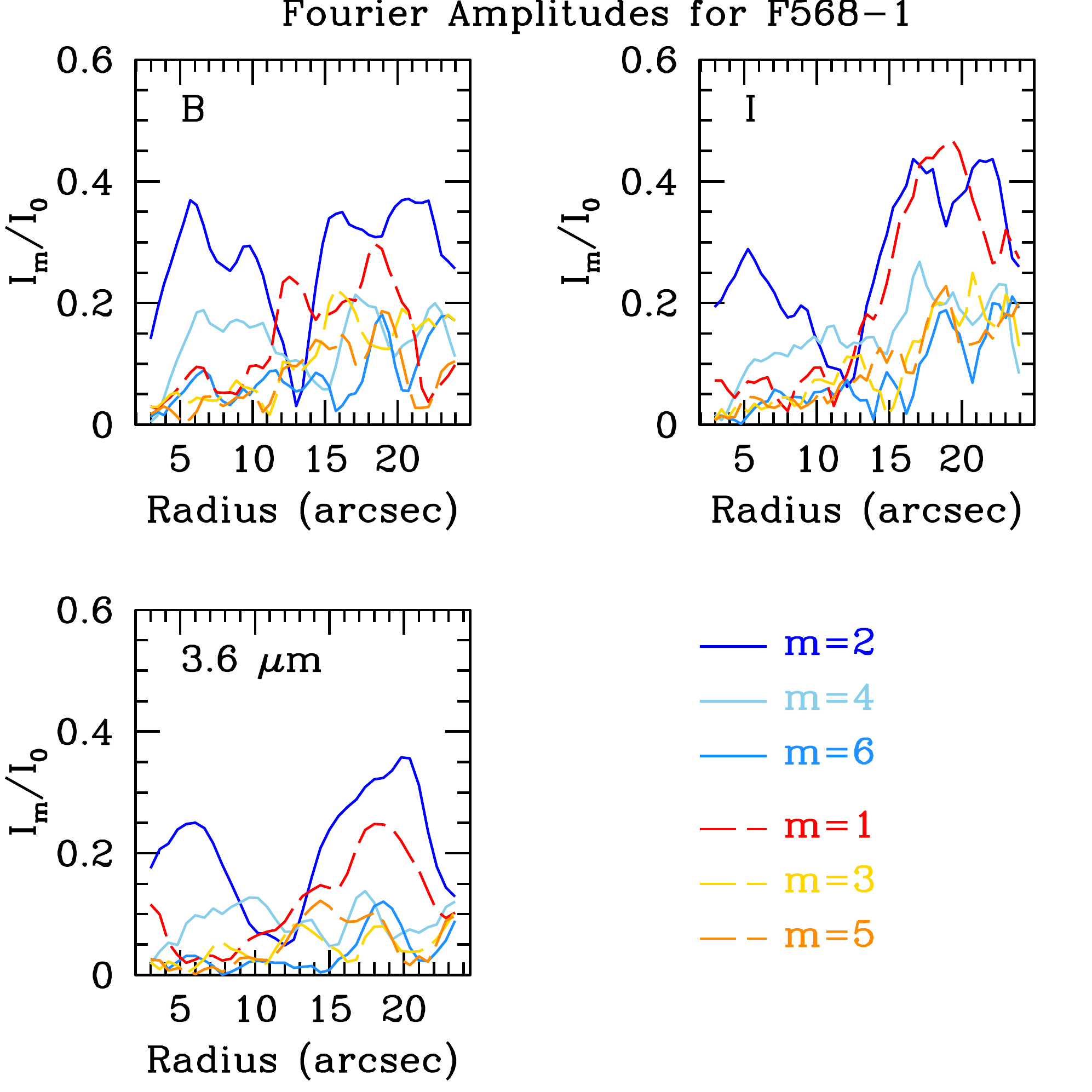}
  \caption{Same as Fig.~\ref{ugc628_amp}, but for F568-1.}
  \label{f56801_amp}
\end{figure}

\begin{figure}
  \centering
  \includegraphics[scale=0.4]{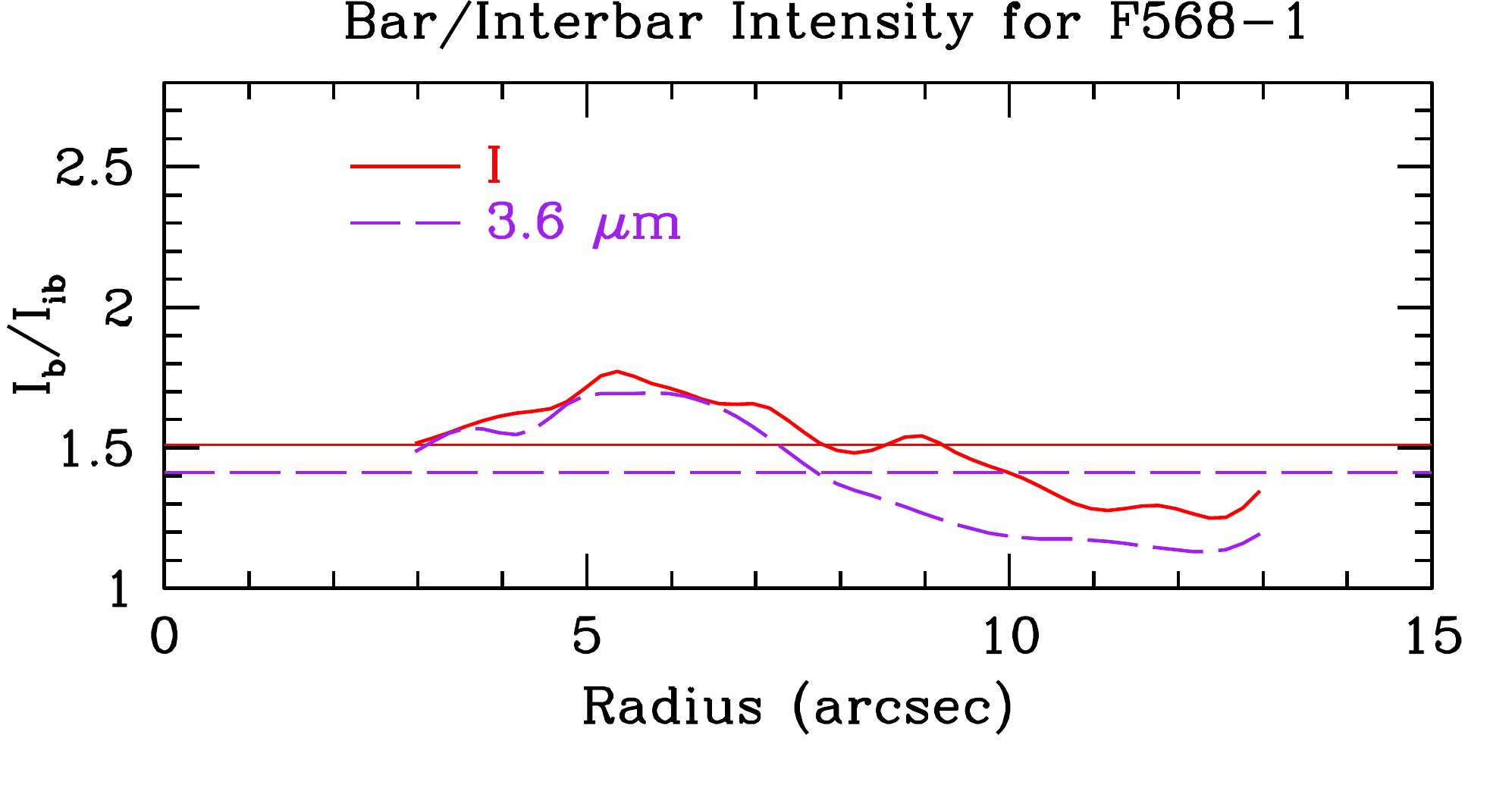}
  \caption{Same as Fig.~\ref{ugc628_four}, but for F568-1.}
  \label{f56801_four}
\end{figure}

\begin{figure}
  \centering
  \includegraphics[scale=0.4]{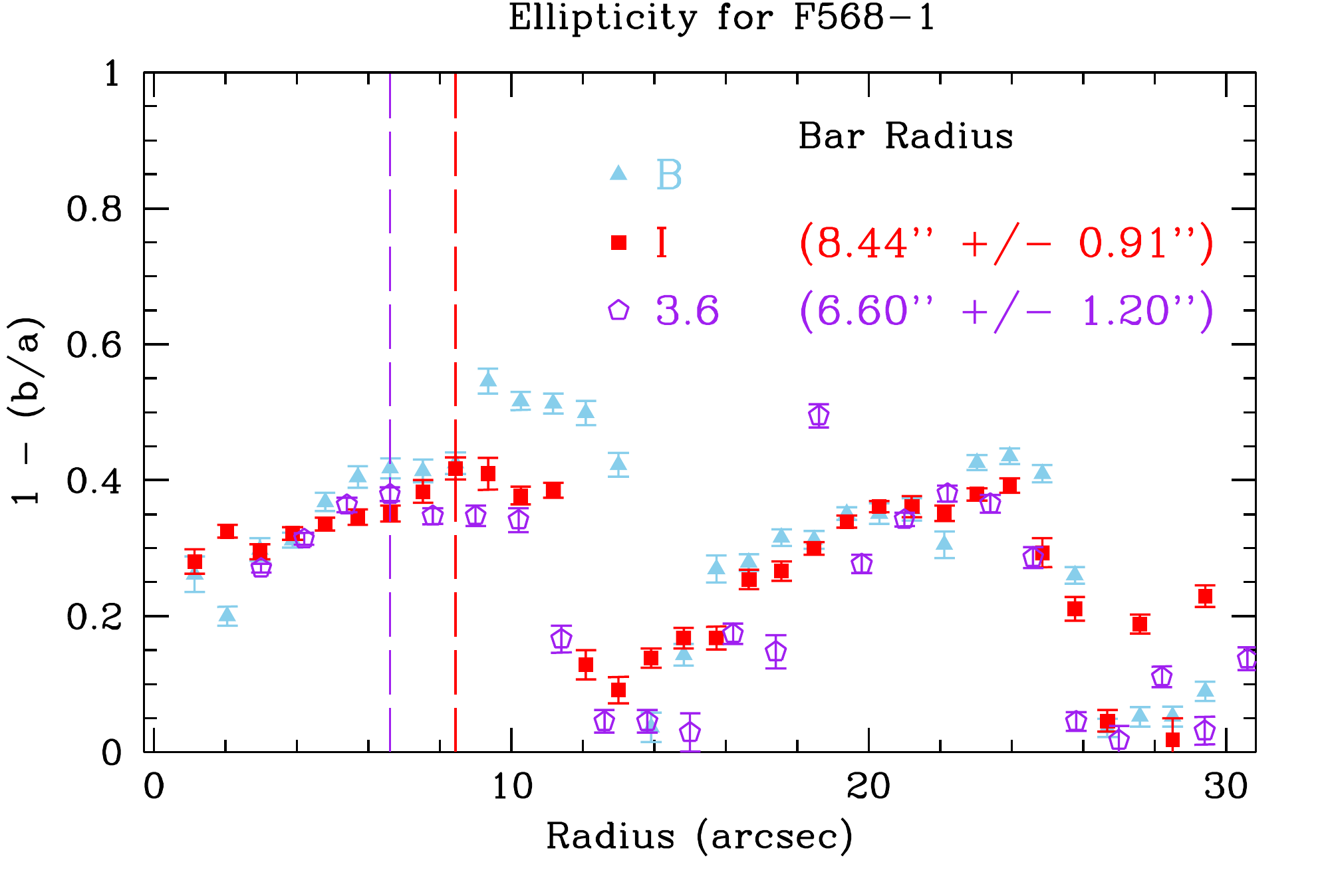}
  \caption{Same as Fig.~\ref{ugc628_isobar}, but for F568-1.}
  \label{f56801_isobar}
\end{figure}

\begin{figure}
  \centering
  \includegraphics[scale=0.7]{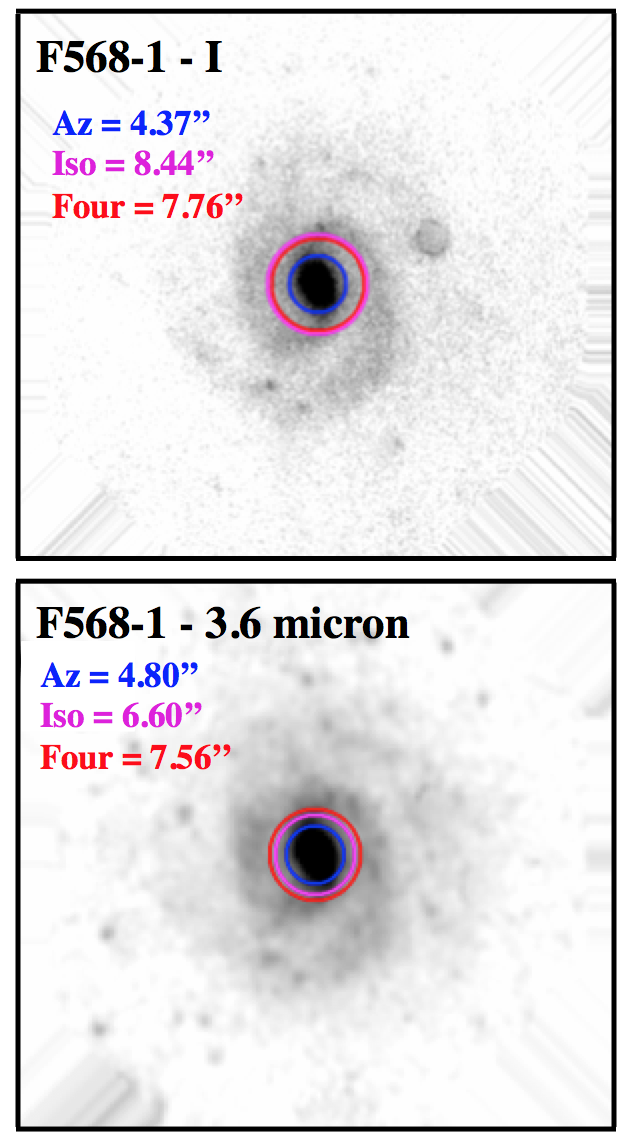}
  \caption{Same as Fig.~\ref{ugc628_bar}, but for F568-1.}
  \label{f56801_bar}
\end{figure}

\begin{figure}
  \centering
  \includegraphics[scale=0.4]{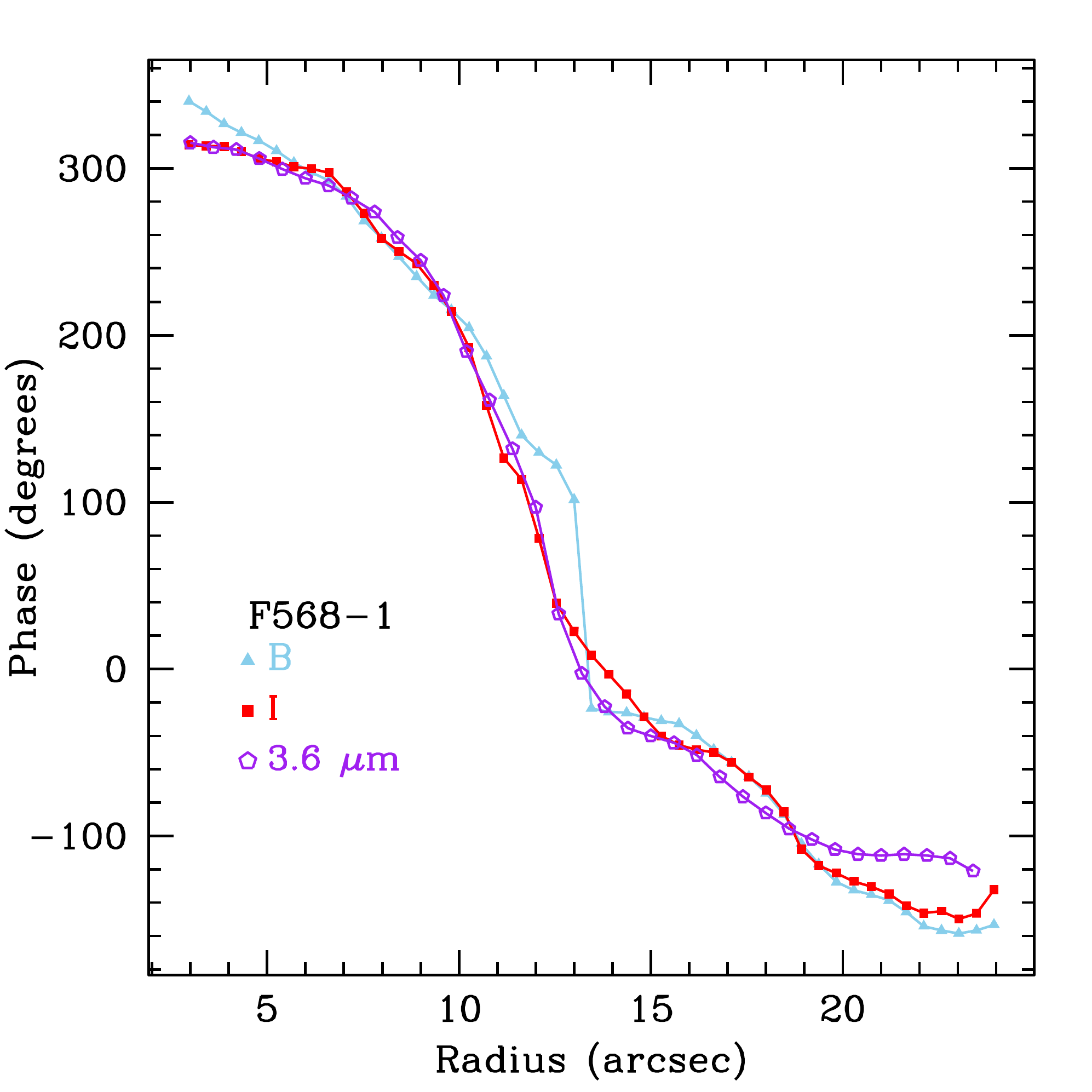}
  \caption{Same as Fig.~\ref{ugc628_phase}, but for F568-1.}
  \label{f568-1_phase}
\end{figure}

The azimuthal light profiles for F568\nobreakdash-1 are shown in Fig.~\ref{f56801_az}. The azimuthal centroids of the bar are shown in Fig.~\ref{f56801_az_bar}. The relative Fourier amplitudes are shown in Fig.~\ref{f56801_amp} and the Fourier bar/interbar intensities are shown in Fig.~\ref{f56801_four}. The radial plots of ellipticity are shown in Fig.~\ref{f56801_isobar}. The various bar radii overplotted on the deprojected \textit{I} and 3.6 $\mu$m images are shown in Fig.~\ref{f56801_bar}. The phase profiles are shown in Fig.~\ref{f568-1_phase}.


\section{F568\nobreakdash-3}

\begin{figure}
  \centering
  \includegraphics[scale=0.4]{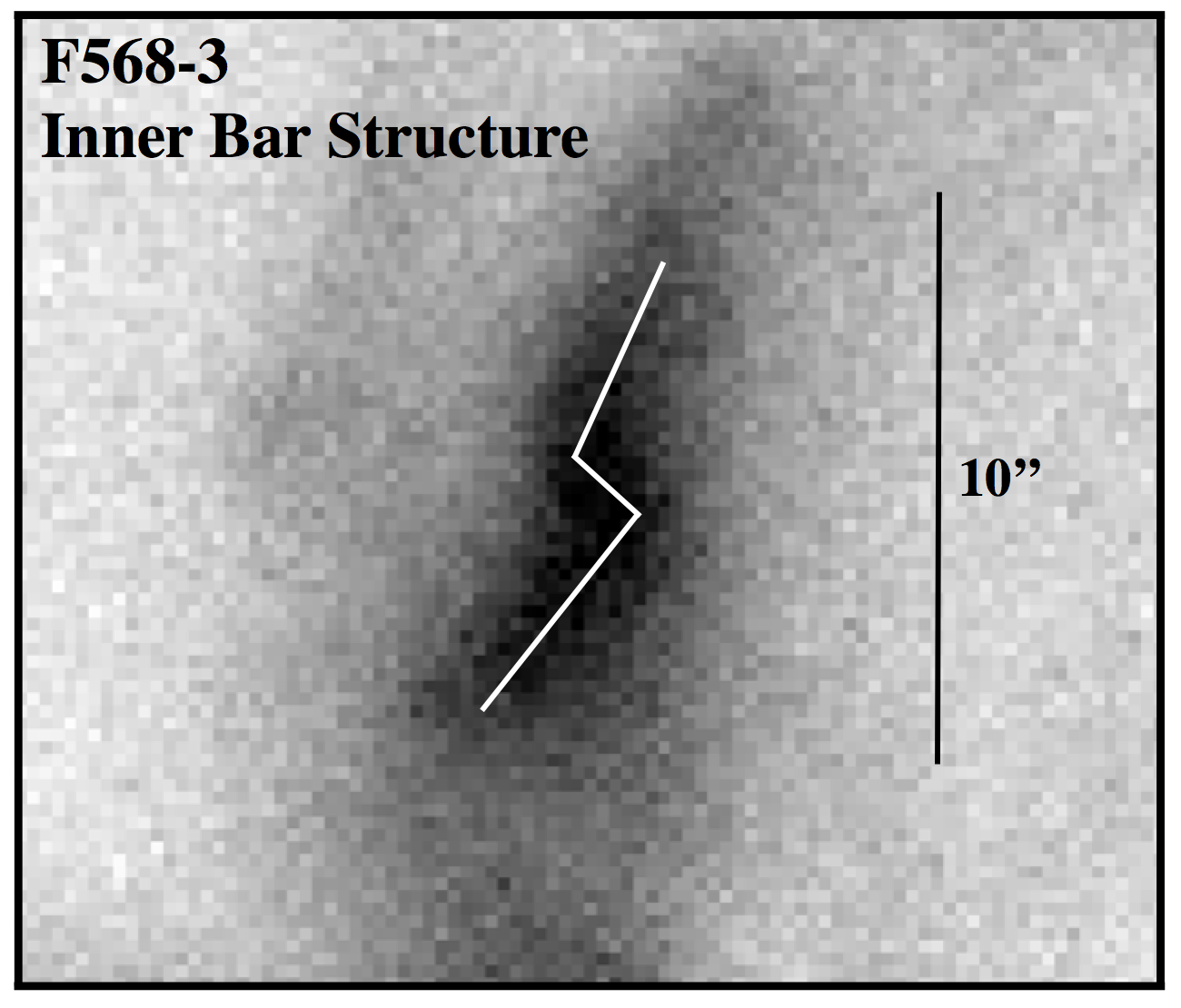}
  \caption{Re-scaled \textit{B}-band image of F568-3 showing the inner bar structure. White lines have been drawn to show the asymmetric nature of the bar.}
  \label{f56803_rescale}
\end{figure}

\begin{figure}
  \centering
  \includegraphics[scale=0.35]{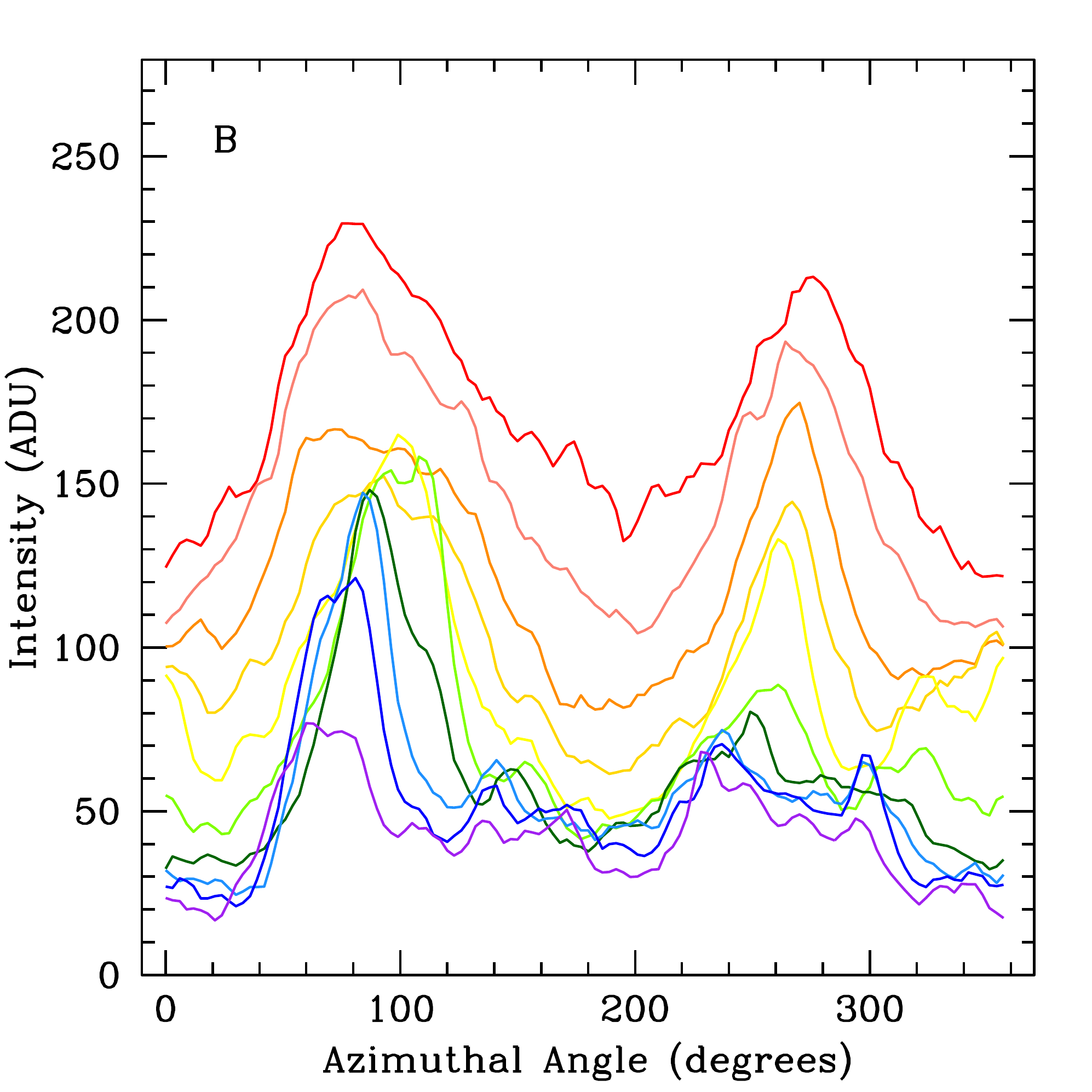}
  \includegraphics[scale=0.35]{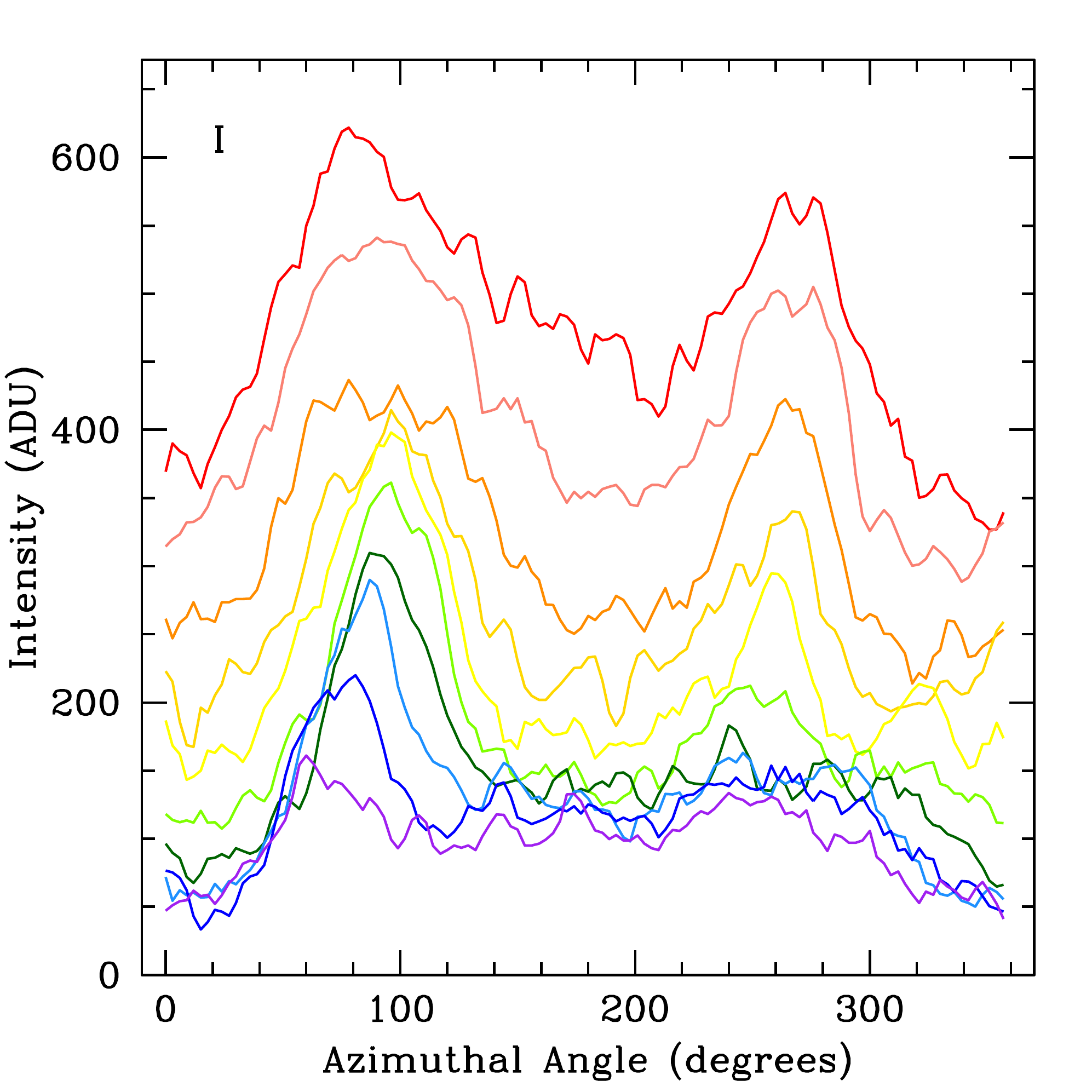}
  \includegraphics[scale=0.35]{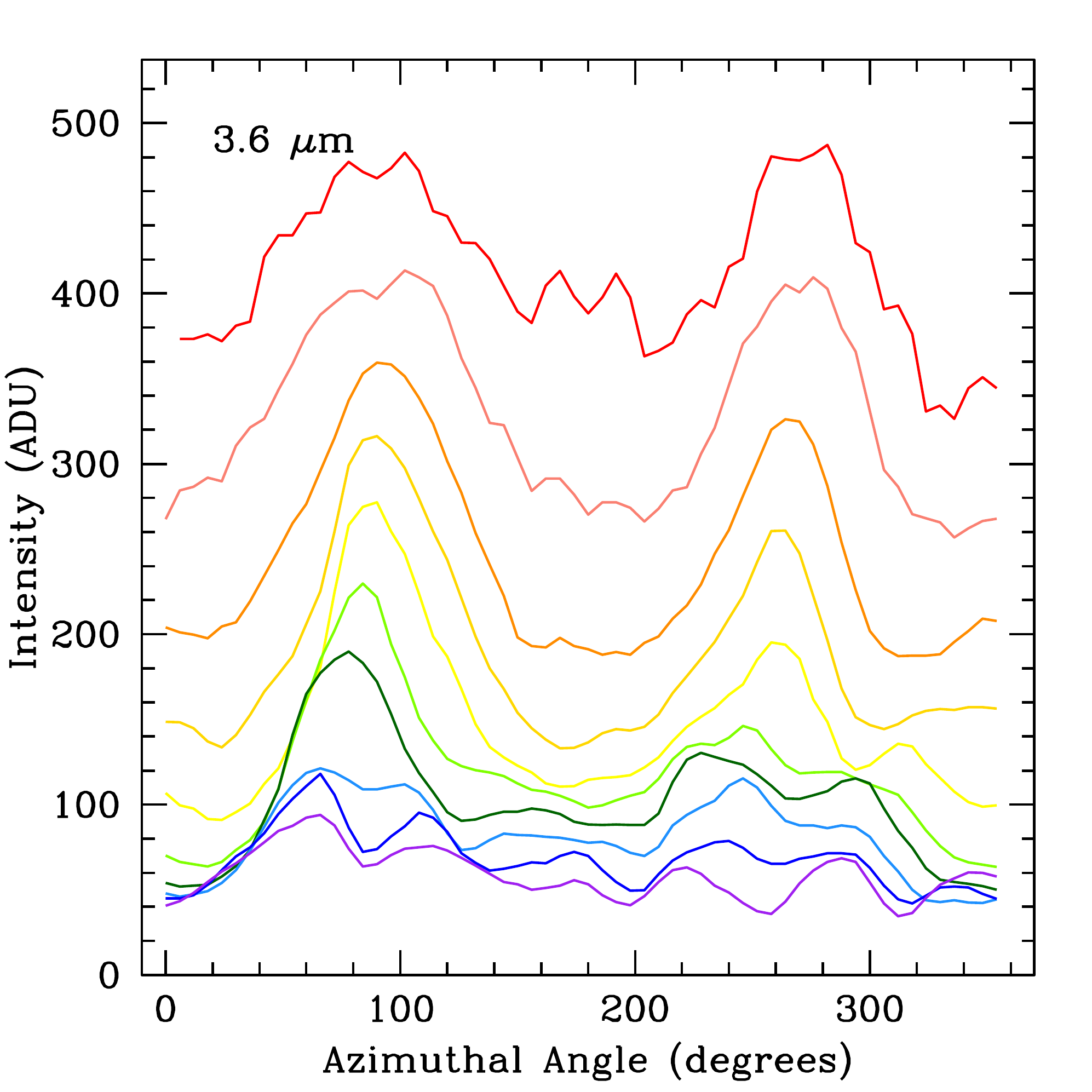}
  \caption{Same as Fig.~\ref{ugc628_az}, but for F568-3.}
  \label{f56803_az}
\end{figure}

\begin{figure}
  \centering
  \includegraphics[scale=0.4]{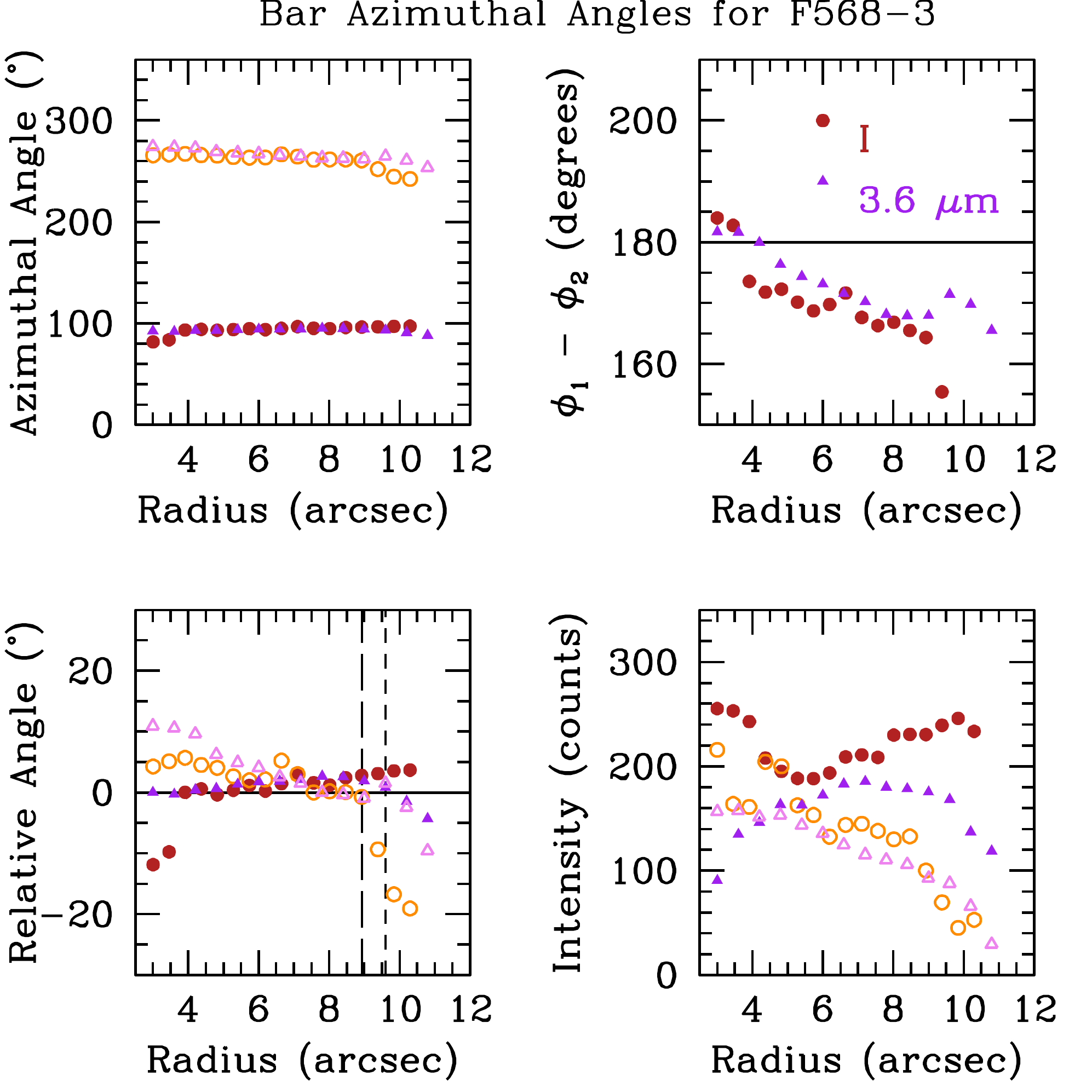}
  \caption{Same as Fig.~\ref{ugc628_az_bar}, but for F568-3.}
  \label{f56803_az_bar}
\end{figure}

\begin{figure}
  \centering
  \includegraphics[scale=0.4]{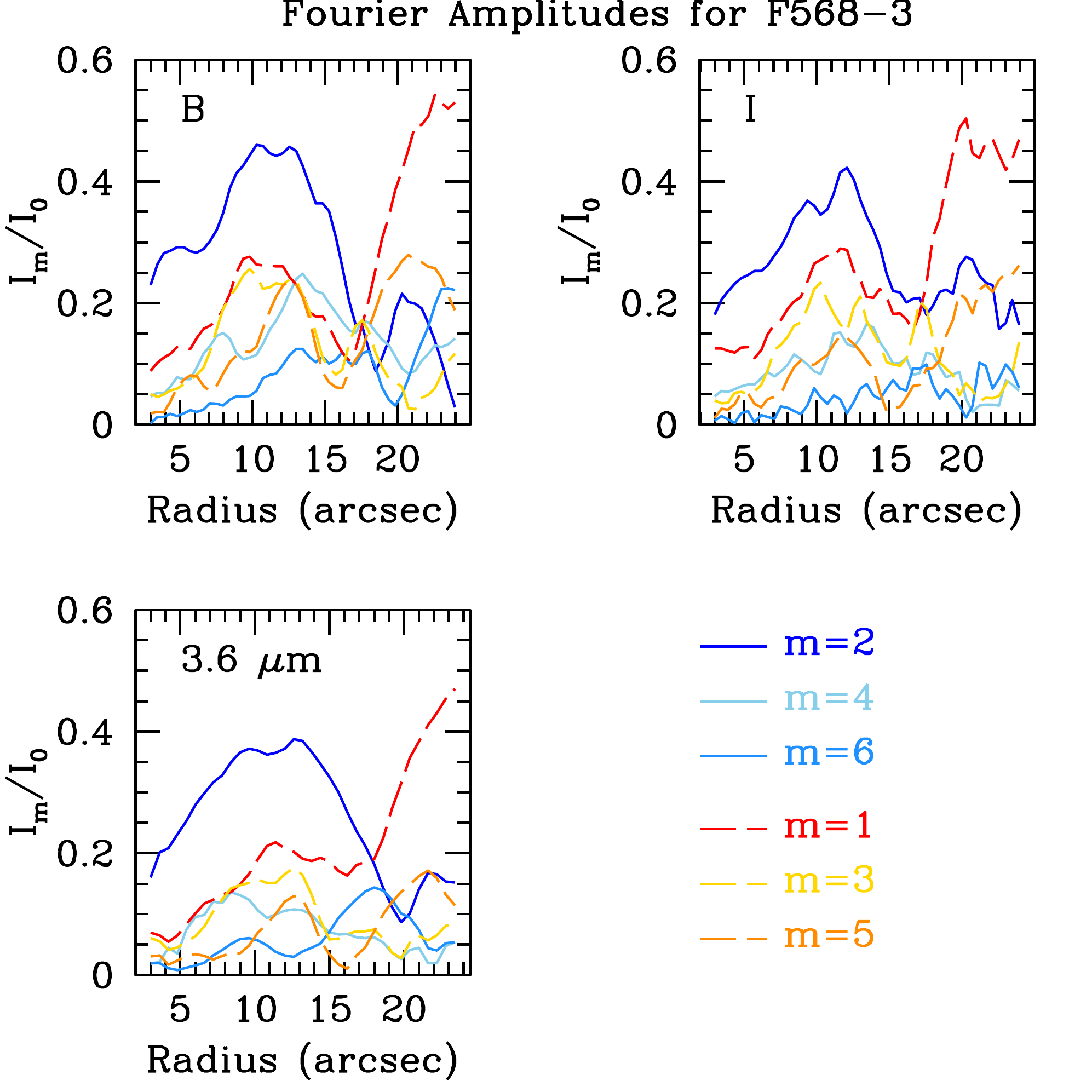}
  \caption{Same as Fig.~\ref{ugc628_amp}, but for F568-3.}
  \label{f56803_amp}
\end{figure}

\begin{figure}
  \centering
  \includegraphics[scale=0.4]{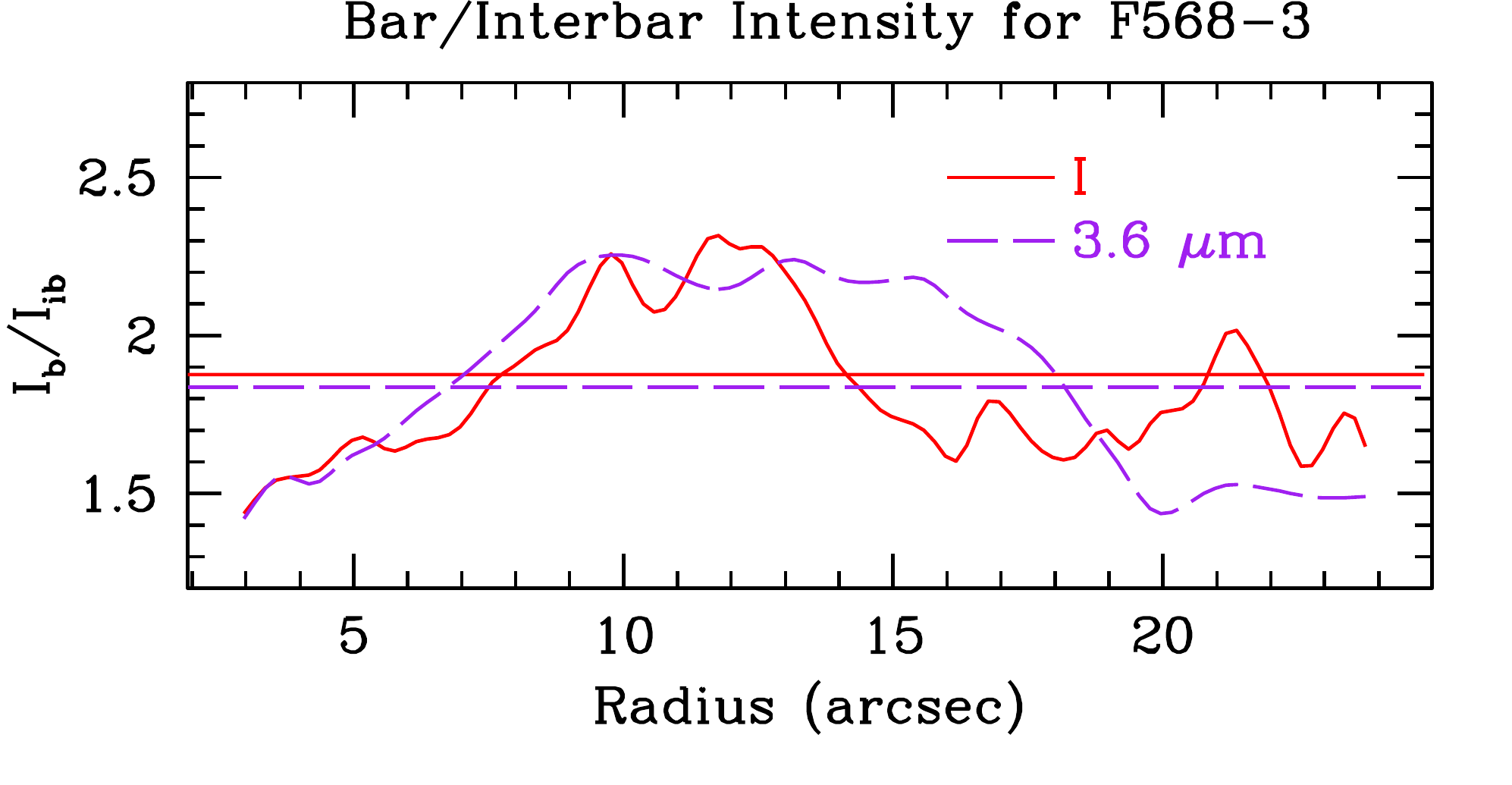}
  \caption{Same as Fig.~\ref{ugc628_four}, but for F568-3.}
  \label{f56803_four}
\end{figure}

\begin{figure}
  \centering
  \includegraphics[scale=0.4]{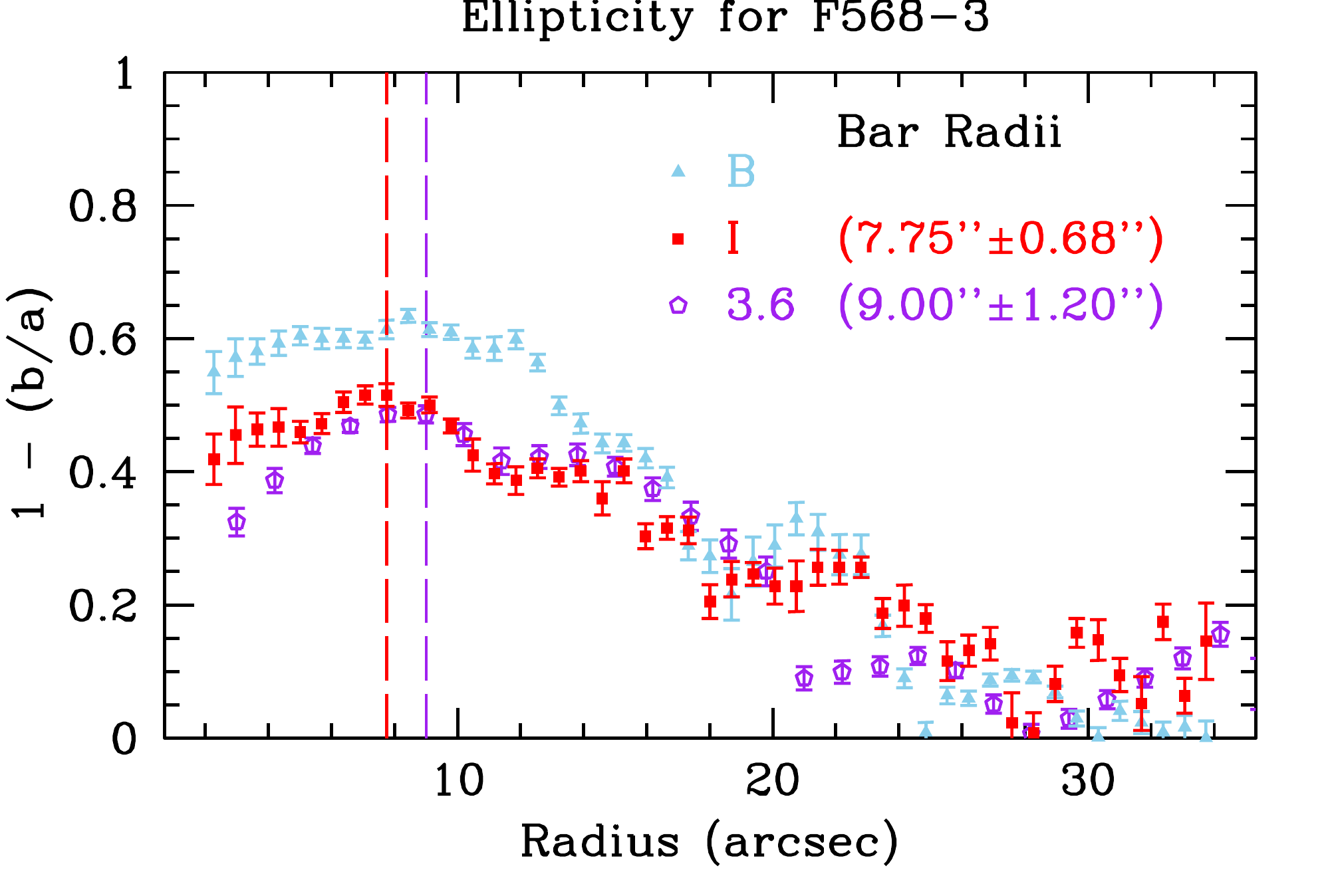}
  \caption{Same as Fig.~\ref{ugc628_isobar}, but for F568-3.}
  \label{f56803_isobar}
\end{figure}

\begin{figure}
  \centering
  \includegraphics[scale=0.7]{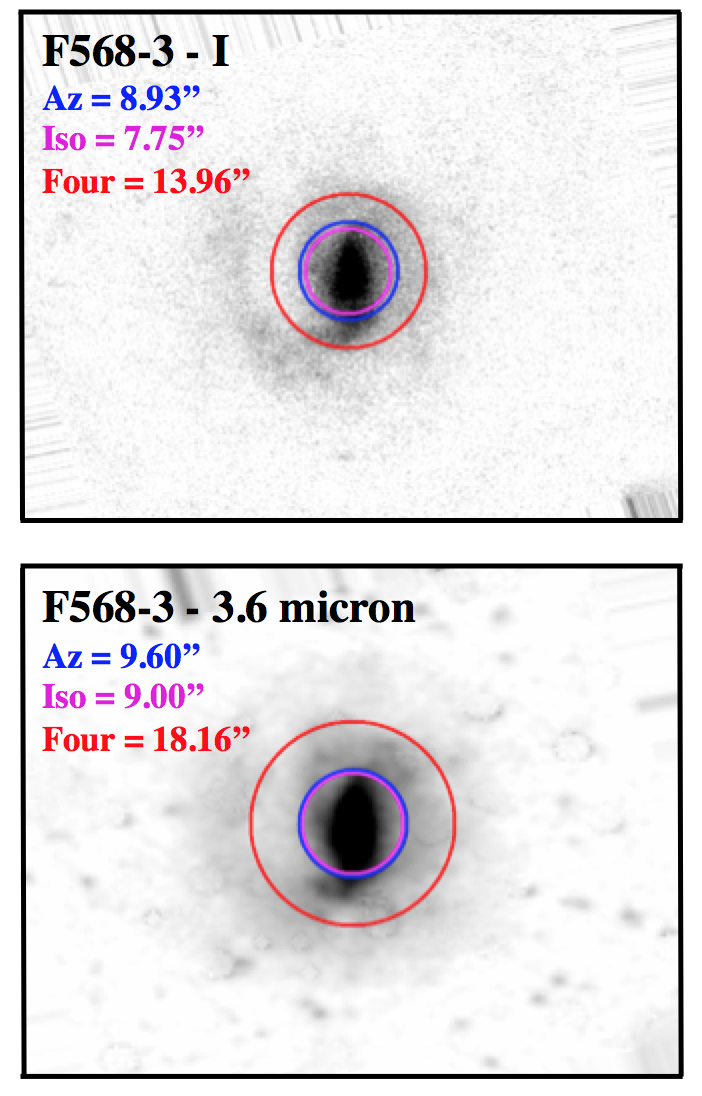}
  \caption{Same as Fig.~\ref{ugc628_bar}, but for F568-3.}
  \label{f56803_bar}
\end{figure}

\begin{figure}
  \centering
  \includegraphics[scale=0.4]{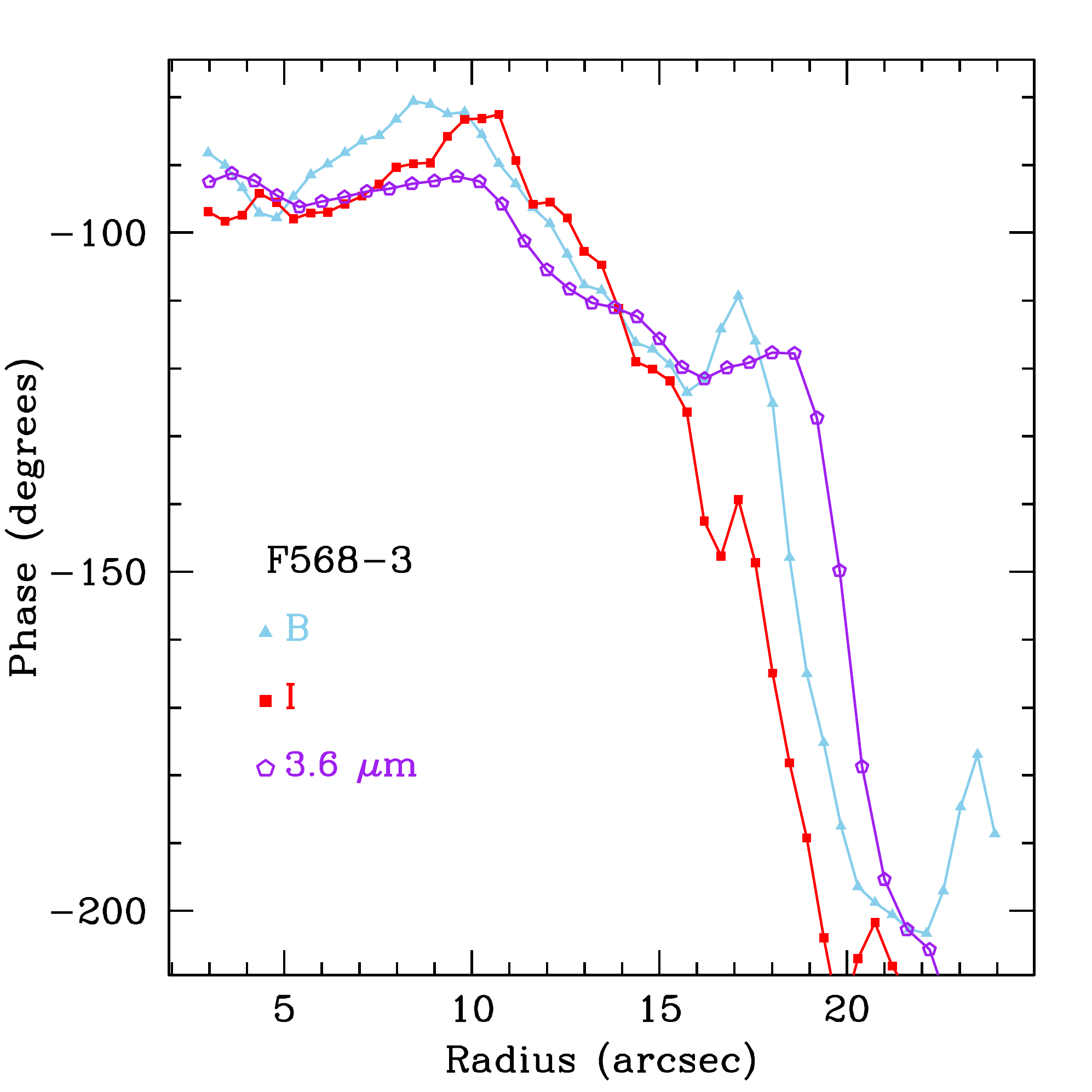}
  \caption{Same as Fig.~\ref{ugc628_phase}, but for F568-3.}
  \label{f568-3_phase}
\end{figure}

\textcolor{black}{We show a zoomed-in and re-scaled image of the \textit{B}-band image of F568-3 in Fig.~\ref{f56803_rescale} to highlight the inner bar structure. The white lines are drawn to better show the asymmetric nature of the bar.} The azimuthal light profiles for F568\nobreakdash-3 are shown in Fig.~\ref{f56803_az}. The azimuthal centroids of the bar are shown in Fig.~\ref{f56803_az_bar}. The relative Fourier amplitudes are shown in Fig.~\ref{f56803_amp} and the Fourier bar/interbar intensities are shown in Fig.~\ref{f56803_four}. The radial plots of ellipticity are shown in Fig.~\ref{f56803_isobar}. The various bar radii overplotted on the deprojected \textit{I} and 3.6 $\mu$m images are shown in Fig.~\ref{f56803_bar}. The phase profiles are shown in Fig.~\ref{f568-3_phase}.

\section{F563\nobreakdash-V2}

\begin{figure}
  \centering
  \includegraphics[scale=0.35]{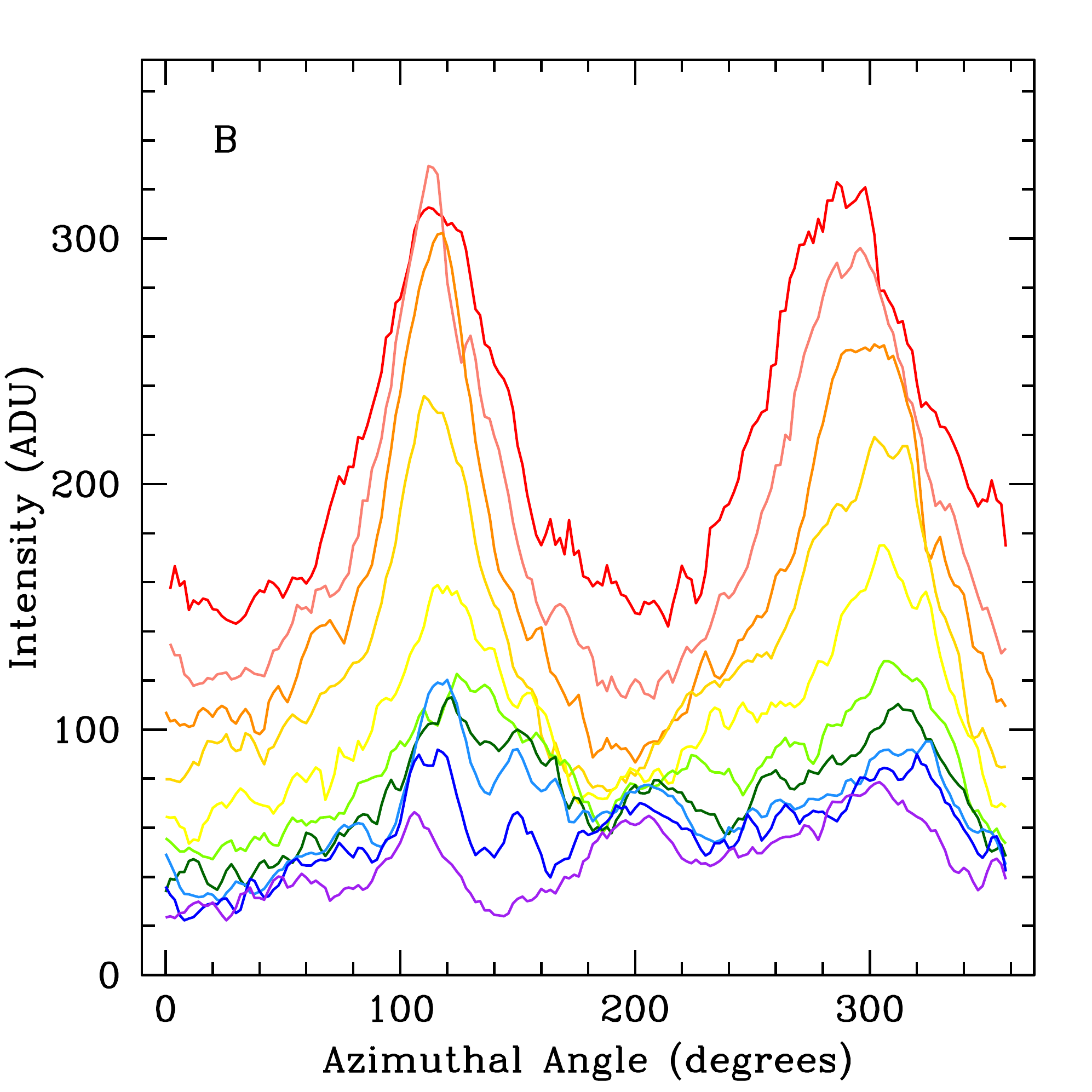}
  \includegraphics[scale=0.35]{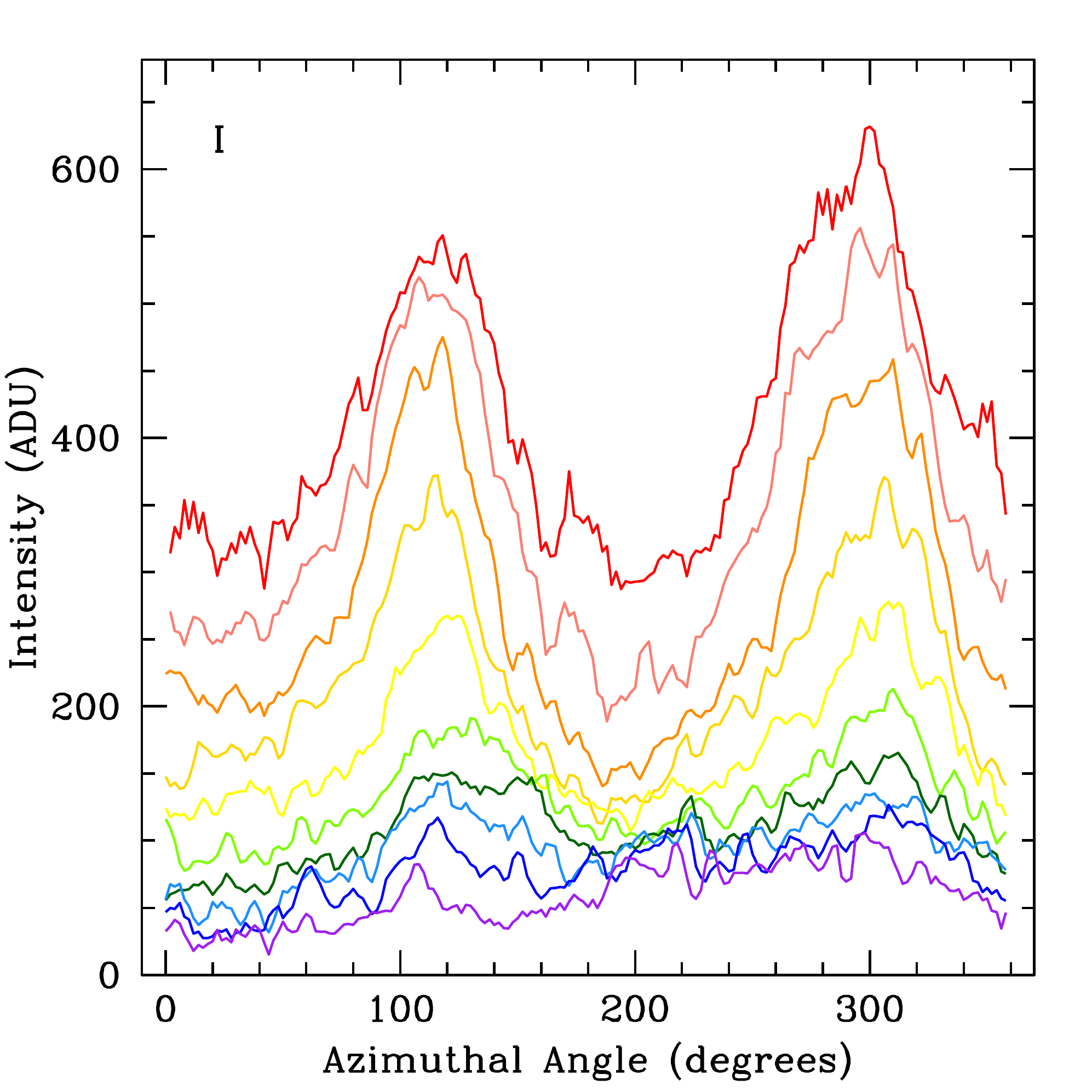}
  \includegraphics[scale=0.35]{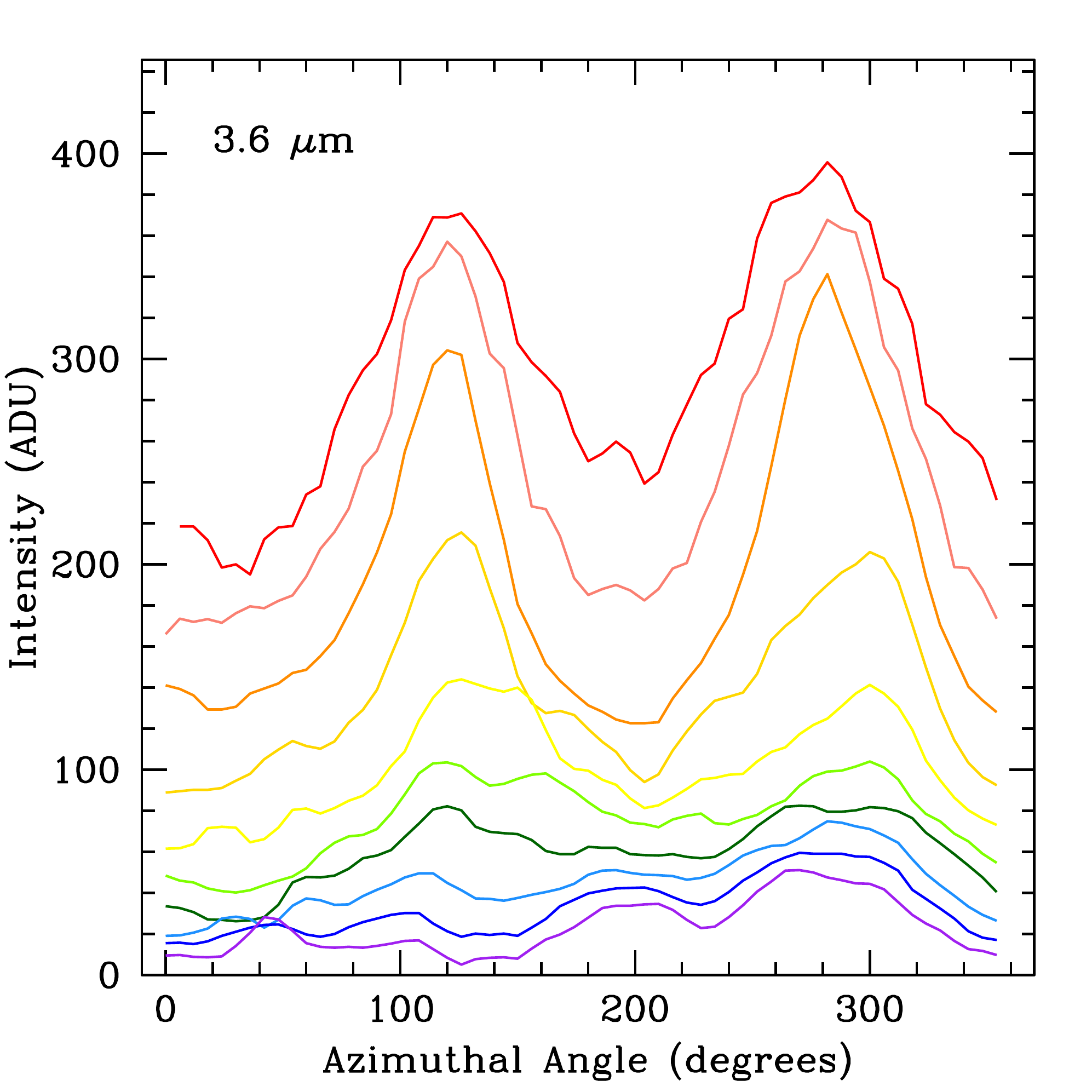}
  \caption{Same as Fig.~\ref{ugc628_az}, but for F563-V2.}
  \label{f563v2_az}
\end{figure}

\begin{figure}
  \centering
  \includegraphics[scale=0.4]{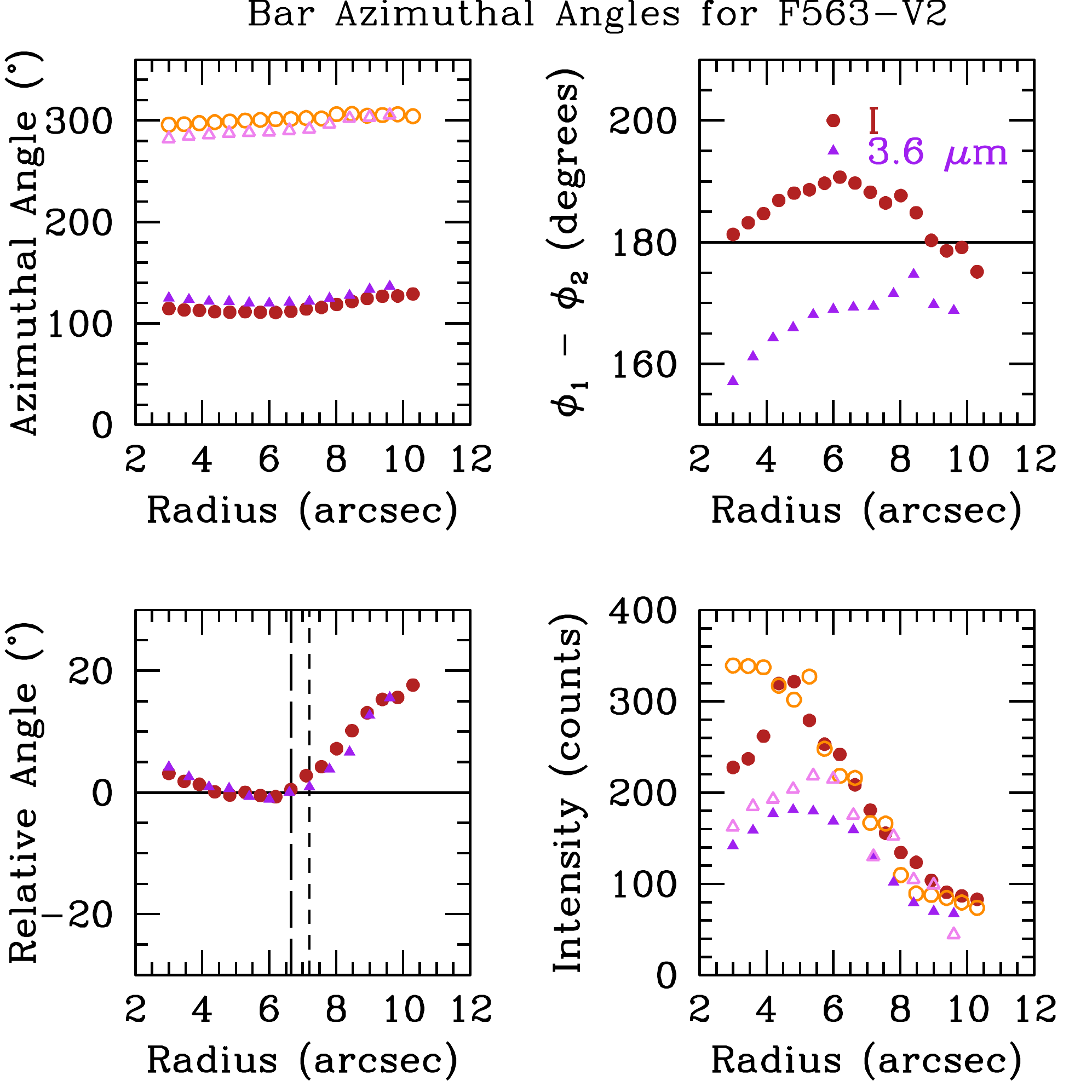}
  \caption{Same as Fig.~\ref{ugc628_az_bar}, but for F563-V2.}
  \label{f563v2_az_bar}
\end{figure}

\begin{figure}
  \centering
  \includegraphics[scale=0.4]{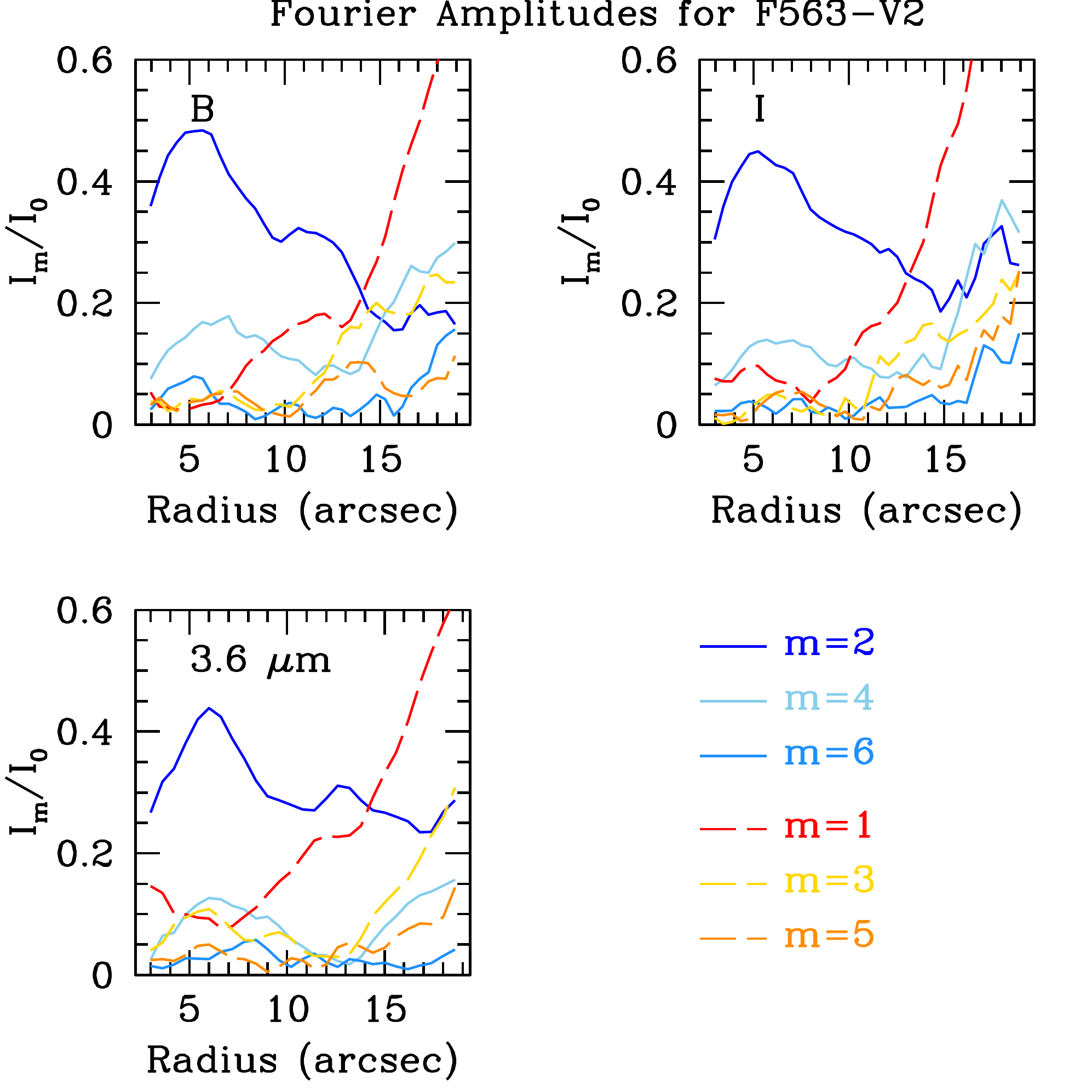}
  \caption{Same as Fig.~\ref{ugc628_amp}, but for F563-V2.}
  \label{f563v2_amp}
\end{figure}

\begin{figure}
  \centering
  \includegraphics[scale=0.4]{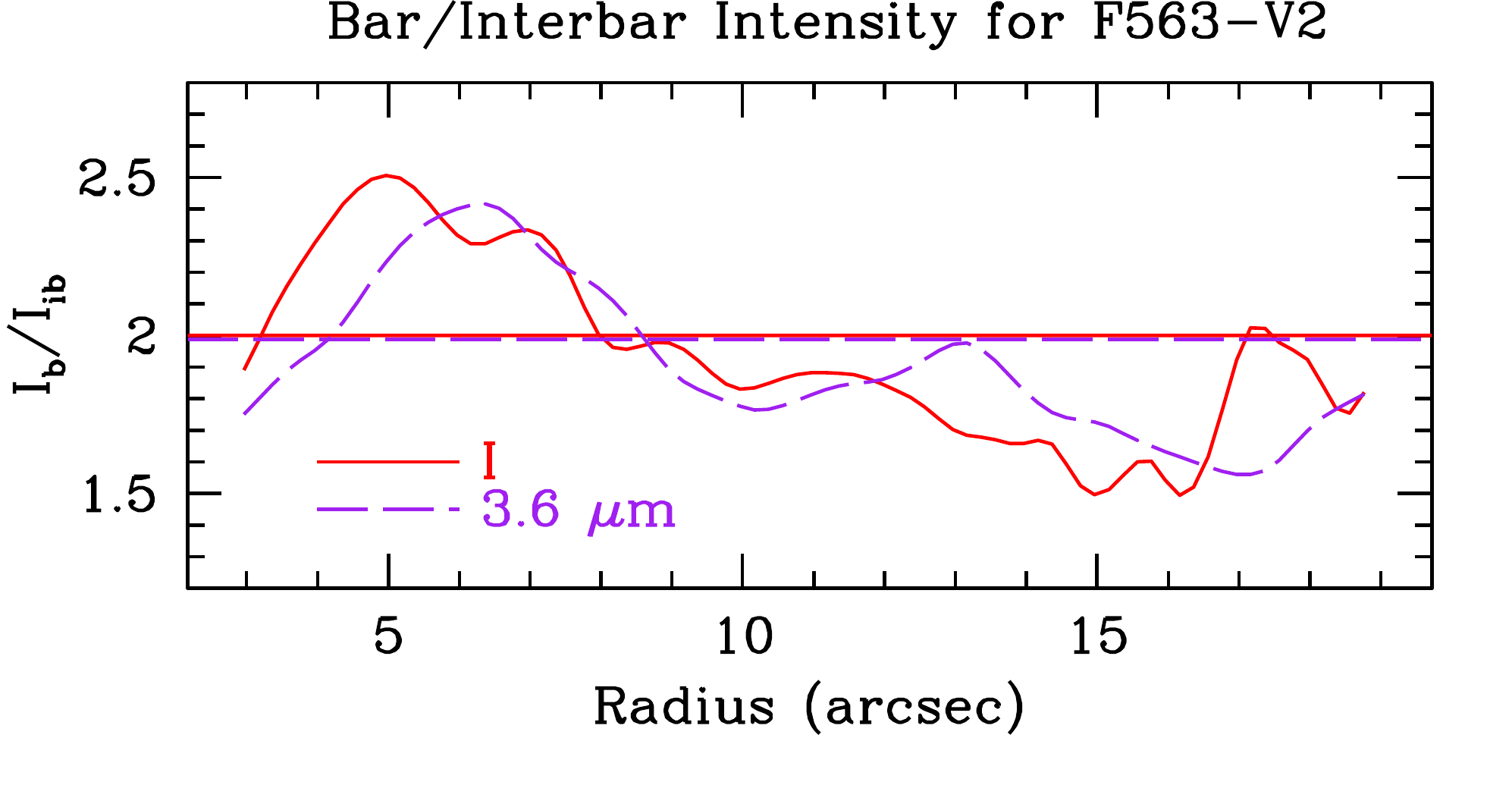}
  \caption{Same as Fig.~\ref{ugc628_four}, but for F563-V2.}
  \label{f563v2_four}
\end{figure}

\begin{figure}
  \centering
  \includegraphics[scale=0.4]{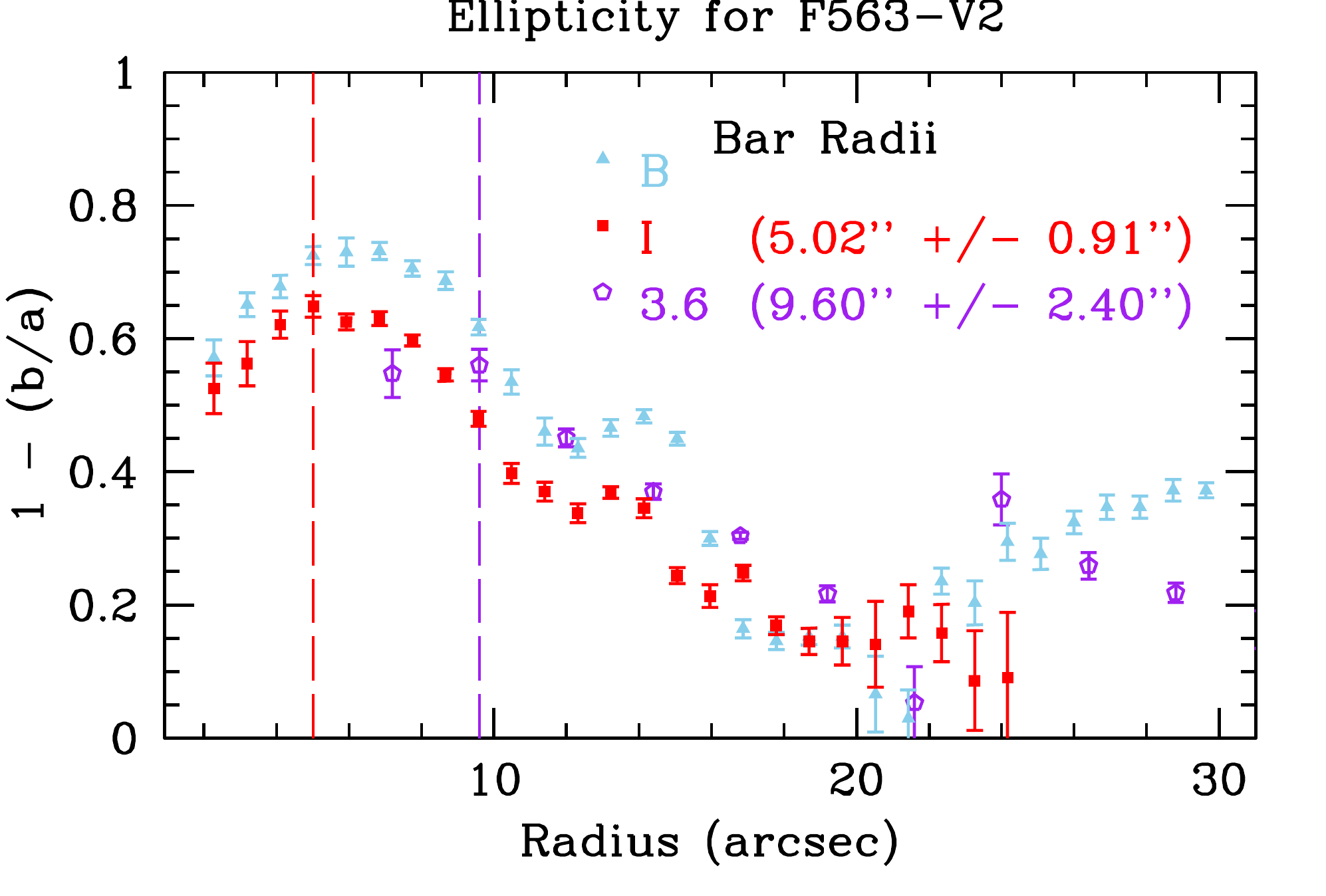}
  \caption{Same as Fig.~\ref{ugc628_isobar}, but for F563-V2.}
  \label{f563v2_isobar}
\end{figure}

\begin{figure}
  \centering
  \includegraphics[scale=0.7]{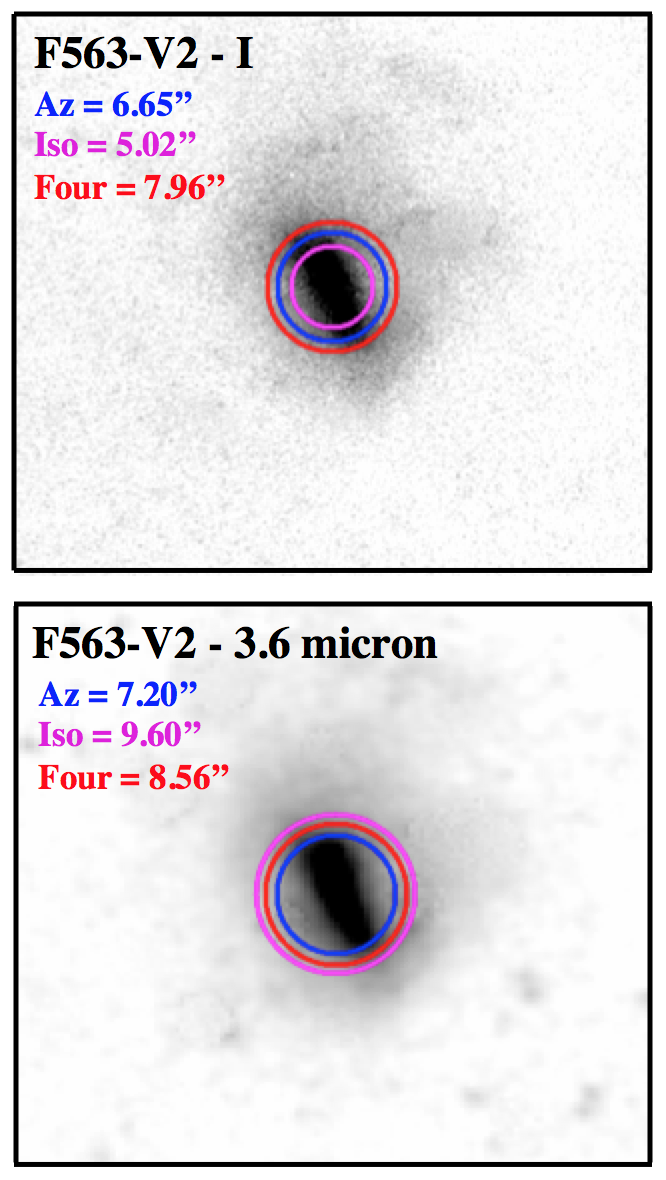}
  \caption{Same as Fig.~\ref{ugc628_bar}, but for F563-V2.}
  \label{f563v2_bar}
\end{figure}

\begin{figure}
  \centering
  \includegraphics[scale=0.4]{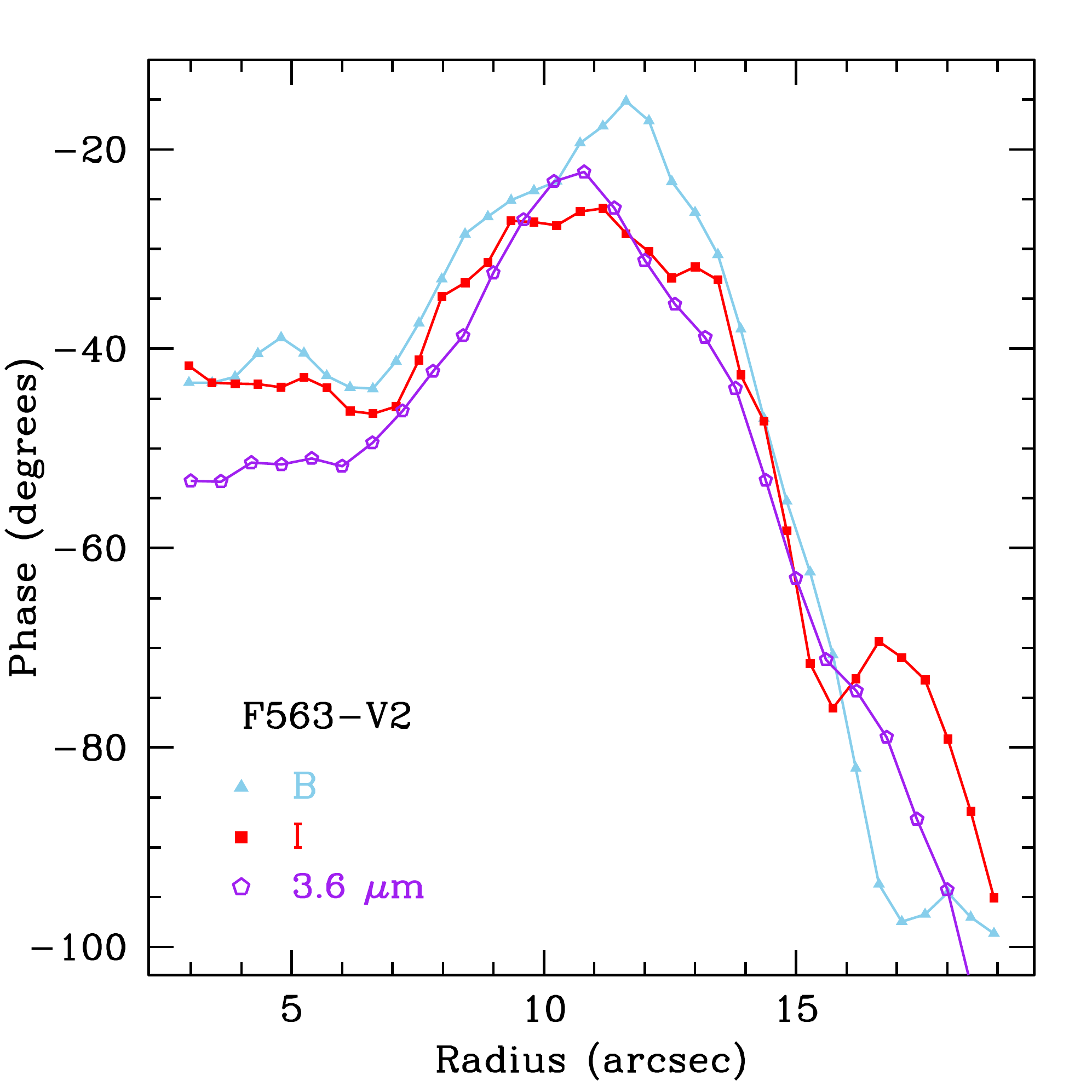}
  \caption{Same as Fig.~\ref{ugc628_phase}, but for F563-V2.}
  \label{f563-v2_phase}
\end{figure}

The azimuthal light profiles for F563\nobreakdash-V2 are shown in Fig.~\ref{f563v2_az}. The azimuthal centroids of the bar are shown in Fig.~\ref{f563v2_az_bar}. The relative Fourier amplitudes are shown in Fig.~\ref{f563v2_amp} and the Fourier bar/interbar intensities are shown in Fig.~\ref{f563v2_four}. The radial plots of ellipticity are shown in Fig.~\ref{f563v2_isobar}. The various bar radii overplotted on the deprojected \textit{I} and 3.6 $\mu$m images are shown in Fig.~\ref{f563v2_bar}. The phase profiles are shown in Fig.~\ref{f563-v2_phase}.


\bsp	
\label{lastpage}
\end{document}